\pdfoutput=1
\documentclass[a4paper,11pt]{article}
\usepackage[utf8]{inputenc}
\usepackage{jcappub}
\usepackage[utf8]{inputenc}
\usepackage{amsmath,amssymb}
\usepackage{graphicx}
\usepackage{multicol}
\usepackage{hyperref}
\usepackage{physics}
\usepackage{lipsum}
\usepackage{caption}
\usepackage{subcaption}
\usepackage{accents}
\usepackage{caption}
\usepackage{subcaption}
\usepackage{bm}
\usepackage{comment}
\usepackage{ wasysym }
\usepackage{cancel}
\usepackage{tensor}
\usepackage{soul}
\newcommand{\m}{m_{\rm{P}}}

\newcommand{\Fone}{F^{(1)}}
\newcommand{\Ftwo}{F^{(2)}}
\newcommand{\Tzero}{T^{(0)}}
\newcommand{\Tone}{T^{(1)}}

\newcommand{\fzero}{f^{(0)}}
\newcommand{\fone}{f^{(1)}}
\newcommand{\ftwo}{f^{(2)}}
\newcommand{\rhom}{\rho_{\rm m}}

\hypersetup{
	colorlinks=true,       
	linkcolor=blue,        
	citecolor=blue,        
	filecolor=blue,     
	urlcolor=blue         
}

\allowdisplaybreaks

\usepackage{pgfplots}
\usepgfplotslibrary{fillbetween}
\usetikzlibrary{patterns}
\pgfplotsset{compat=1.8}

\title{\boldmath Scalar-Induced Gravitational Waves in Palatini $f(R)$ Gravity}
\author[a]{Samuel S\'anchez L\'opez}
\author[b]{and Jos\'e Jaime Terente D\'iaz}

\emailAdd{ssanchezlopez@imsc.res.in}
\emailAdd{jterente@uc.pt}

\vspace{5cm}

\affiliation[a]{The Institute of Mathematical Sciences, CIT Campus, Chennai 600113, India.}
\affiliation[b]{CFisUC, Departamento de Física, Universidade de Coimbra, Rua Larga P-3004-516, \\ Coimbra, Portugal.}

\abstract{Primordial scalar perturbations that reenter the horizon after inflation may induce a second-order Gravitational Wave spectrum with information about the primordial Universe on scales inaccessible to Cosmic Microwave Background experiments. In this work, we develop a general framework for the study of Scalar-Induced Gravitational Waves in Palatini  $f(R)$ gravity, a theory that was proven to successfully realise inflation and quintessence, and consider the case of the Starobinsky-like model as an example. A regime of radiation domination with a subdominant matter component is assumed, allowing for a well-motivated perturbative approach to the gravity modifications. We calculate the kernel function and the density spectrum numerically and find accurate analytical expressions. The spectral density, which may be tested across a wide range of frequencies by upcoming Gravitational Wave experiments, is shown to differ from the General Relativity and metric $f(R)$ gravity predictions under certain conditions. We comment on previous results in the literature regarding the metric formulation and make special emphasis on the potential of these distinctive features of the spectrum to probe the two formalisms of gravity.}

\begin{document}

\maketitle
\section{Introduction}
While the expansion history of the Universe is well-known since the recombination epoch, we have scarce observational evidence of the physics on much earlier stages \cite{Baumann:2022mni,CosmoVerse:2025txj}. The highly dense primordial plasma hinders any possibility of extracting essential information from light sources which span different frequencies in the electromagnetic spectrum \cite{Baumann:2018muz}. Fortunately, the past decade saw the first direct observation of gravitational waves (GWs) from the merge of two black holes orbiting around each other \cite{LIGOScientific:2016aoc,LIGOScientific:2016sjg}, as well as the first multi-messenger observation of a binary neutron star system's merger \cite{LIGOScientific:2017vwq,LIGOScientific:2017ync,LIGOScientific:2017zic}. These two collective achievements ushered in a new era of cosmological and astrophysical probes, with far-reaching implications on theoretical models of the early and late universe, and set the stage for the exploration of primordial processes that preceded the time of neutrino decoupling.

Besides, Pulsar Timing Array (PTA) collaborations NANOGrav \cite{NANOGrav:2023gor,NANOGrav:2023hde,NANOGrav:2023hvm}, EPTA/InPTA \cite{EPTA:2023fyk,EPTA:2023sfo,EPTA:2023xxk}, PPTA \cite{Reardon:2023gzh,Reardon:2023zen,Zic:2023gta} and CPTA \cite{Xu:2023wog} provided mounting evidence in the present decade for a stochastic GW background around the nHz frequencies ($\sim 10^{-9}$~-~$10^{-7}$~Hz). The source of this signal is not clear but binaries of supermassive black holes are the most probable candidates \cite{Huang:2023chx,Depta:2023uhy,Gouttenoire:2023nzr}. In spite of this likely astrophysical origin, there is a wide variety of cosmological mechanisms generating GWs \cite{Guzzetti:2016mkm,Caprini:2018mtu} with typical frequencies within the prospects of future GWs experiments. Amongst these, the so-called `Scalar-Induced Gravitational Waves' (SIGWs) \cite{Domenech:2021ztg,Tomita:1967wkp,Domenech:2024rks} are quite appealing\footnote{Additionally, one can also consider `vector-' or even `tensor-' induced GWs. A recent example of work on first-order tensor perturbations as sources of GWs is Ref.~\cite{Picard:2023sbz}, or \cite{Wu:2024qdb} in the context of PTA observations and GWs induced by scalar-tensor perturbation interactions. The distinctive signatures of these latter GWs were also studied in Ref.~\cite{Bari:2023rcw} during radiation domination. Magnetically-generated first-order vector perturbations of the metric during kination domination in reheating acting as sources of GWs were examined in Ref.~\cite{Bhaumik:2025kuj}.} given the (otherwise inaccessible) comoving scales that are sensitive to the physics of the final stages of inflation, which is the most compelling paradigm of the early Universe \cite{Starobinsky:1979ty,Starobinsky:1980te,Sato:1980yn,Guth:1980zm,Linde:1981mu}. Since the amplitude of the scalar spectrum of primordial perturbations was measured at a pivot scale $k_{*} = 0.05~\textrm{Mpc}^{-1}$ \cite{Achucarro:2022qrl,Planck:2018jri}, the SIGWs give an absolute lower limit to the total spectrum of GWs \cite{Baumann:2007zm}, and serve to probe distinctive features of inflation such as sizeable scalar non-Gaussinities \cite{Cai:2018dig,Adshead:2021hnm,Acquaviva:2002ud,Perna:2024ehx,Iovino:2024sgs}. There is also recent work on these primordial non-Gaussian effects on scalar-tensor induced GWs as well \cite{Picard:2024ekd}, and on the clustering properties of Primordial Black Holes (PBHs) \cite{He:2024luf}, which are the result of the collapse of large primordial perturbations. This is however one of many ongoing lines of research which aim to harness the power of the spectrum of SIGWs to constrain the amount of PBH production \cite{PhysRevD.83.083521,Saito:2008jc,Loc:2024qbz}.

Although the study of SIGWs gained decisive momentum in the early 90s \cite{Matarrese:1992rp,Matarrese:1993zf,Matarrese:1997ay}, it was not until very recently that modified gravity effects were considered in the analysis of the induced GW spectrum \cite{Domenech:2024drm,Zhou:2024doz,Kugarajh:2025rbt}. Such additional terms are expected to arise at the quantum level \cite{Callan:1970ze,Allen:1983dg,Hrycyna:2015vvs} and must be taken into account in the effective description of the gravitational sector \cite{Antoniadis:2022cqh,Lima:2016npg}. In the context of $f(R)$ gravity (for some excellent reviews we suggest Refs.~\cite{DeFelice:2010aj,Sotiriou:2008rp,Olmo:2011uz,Clifton:2011jh}), two of the aforementioned references followed different methodologies to calculate the energy density spectrum of SIGWs,\footnote{Having mentioned the PBHs, an example of relatively recent work on the effects of an $f(R)$ gravity theory at the level of the gravitational potential of Poisson-distributed PBHs can be found in Ref.~\cite{Papanikolaou:2021uhe}. Later in that work, the authors focus on the Starobinsky-like model chiefly. Also, we may mention the recent Ref.~\cite{Bruton:2025dyr} where the contribution from the scalar mode is explored in the GW emission from BH pairs under certain conditions.} and considered well-known functions of the Ricci scalar $R$. Particularly, Ref.~\cite{Kugarajh:2025rbt} resorted to a perturbative way of treating the modified gravity corrections to the General Relativistic case, which was originally developed by one of its authors in Ref.~\cite{Bertacca:2011wu}. This allowed for an analytical approach to the study of the behaviour of cosmological perturbations when the gravitational sector includes further terms in addition to the Einstein-Hilbert one of General Relativity (GR), while, at the same time, these are sufficiently weak in order for the background evolution not to deviate significantly from the cosmological model of consensus; namely, the $\Lambda$CDM model \cite{Joyce:2014kja}. This proved to be useful to account for regimes where, in the study of structure formation, the quasi-static approximation \cite{Esposito-Farese:2000pbo,Boisseau:2000pr,delaCruz-Dombriz:2008ium} does not hold, and to derive the kernel function and time-averaged power spectrum of SIGWs for modes reentering at the time of radiation domination. In this latter respect, the function $f(R)$ of Starobinsky gravity \cite{Starobinsky:1980te} was assumed, which includes an $R^2$ term and a dimensionful $\alpha$ coupling parameter. Using this as expansion parameter of the perturbative series, they obtained a distinctive total SIGW spectrum (with the modified gravity corrections implemented already) whose amplitude was of the same order of magnitude as the one prescribed by GR, but with slightly smaller amplitude, depending on the comoving scales (and hence the GW experiment) under consideration.

However, these preceding studies implicitly assumed the metric formalism of gravity, which constrains the affine connection of the theory. On the other hand, the more general framework of the Palatini formulation of gravity \cite{Palatini:1919ffw,Ferraris:1982wci} has experienced a surge in popularity in recent years (see Refs. \cite{Bauer:2008zj, Bauer:2010bu, Tamanini:2010uq, Bauer:2010jg, Rasanen:2017ivk, Tenkanen:2017jih, Racioppi:2017spw, Markkanen:2017tun, Jarv:2017azx, Fu:2017iqg, Racioppi:2018zoy, Carrilho:2018ffi, Kozak:2018vlp, Rasanen:2018fom, Rasanen:2018ihz, Almeida:2018oid, Shimada:2018lnm, Takahashi:2018brt, Jinno:2018jei, Rubio:2019ypq, Bostan:2019uvv, Bostan:2019wsd, Tenkanen:2019xzn, Racioppi:2019jsp, Tenkanen:2020dge, Shaposhnikov:2020fdv, Borowiec:2020lfx, Jarv:2020qqm, Karam:2020rpa, McDonald:2020lpz, Langvik:2020nrs, Shaposhnikov:2020gts, Shaposhnikov:2020frq, Gialamas:2020vto, Mikura:2020qhc, Verner:2020gfa, Enckell:2020lvn, Reyimuaji:2020goi, Karam:2021wzz, Mikura:2021ldx, Kubota:2020ehu, Gomez:2021roj, Mikura:2021clt, Bombacigno:2018tyw, Enckell:2018hmo, Antoniadis:2018ywb, Antoniadis:2018yfq, Tenkanen:2019jiq, Edery:2019txq, Giovannini:2019mgk, Tenkanen:2019wsd, Gialamas:2019nly, Tenkanen:2020cvw, Lloyd-Stubbs:2020pvx, Antoniadis:2020dfq, Ghilencea:2020piz,Das:2020kff, Gialamas:2020snr, Ghilencea:2020rxc,TerenteDiaz:2023kgc,Bekov:2020dww, Dimopoulos:2020pas, Gomez:2020rnq, Karam:2021sno, Annala:2021zdt, Lykkas:2021vax, Gialamas:2021enw, AlHallak:2021hwb, Dioguardi:2021fmr, Dimopoulos:2022tvn,SanchezLopez:2023ixx,Kuralkar:2025hoz} for a non-comprehensive list). In the Palatini formalism, the metric and the connection are taken to be $\textit{a priori}$ independent fields, contrary to the metric formalism, where the connection takes the standard Levi-Civita (LC) form. In other words, in order to obtain the equations of motion, one needs to vary the action independently with respect to each of them. It is important to emphasise that this is not an additional ingredient introduced in the theory, but rather, a choice of the parametrisation of its degrees of freedom. For the Einstein-Hilbert action, both formalisms yield the same Einstein field equations, but this is no longer true when one considers deviations from GR, such as the aforementioned $f(R)$ gravity. Most strikingly, in metric $f(R)$ gravity, the conformal transformation necessary to express the action in the Einstein frame yields a kinetic term of a scalar field. Unlike this, in Palatini $f(R)$ gravity, the Ricci tensor is a function of the connection only and, therefore, the equivalent conformal transformation does not result in an additional degree of freedom.  Instead, the higher-order curvature terms act as additional effective matter sources in the Einstein equations. 

The considerations above apply, in particular, to the popular modified gravity model $f(R)=R +\alpha R^2$, the first model of inflation to be proposed \cite{Starobinsky:1980te} and one of the most successful to date \cite{Planck:2018jri,ACT:2025fju}. As is well known, in the metric formalism, the additional degree of freedom that appears in the Einstein frame, dubbed the scalaron, acts as the inflaton. Its flat potential is suitable for inflation, leading to values of the amplitude and tilt of the spectrum of the scalar perturbations, when taken in combination with the amplitude of the spectrum of primordial gravitational waves, in excellent agreement with the data, given that $\alpha \sim 10^8$. The Palatini formulation of this theory has also been widely successful \cite{Antoniadis:2019jnz,Antoniadis:2018yfq,Enckell:2018hmo,Dimopoulos:2022rdp,Dimopoulos:2020pas}. Now, the inclusion of the $\alpha R^2$ terms does not lead to a scalar field in the Einstein frame that can take the role of the inflaton, so it must be added by hand in the Jordan frame. With this, the Einstein frame action instead features a higher-order kinetic term, which can be neglected during slow-roll, and a re-scaled potential, which generally exhibits a plateau for large field values, making it ideal for inflationary model-building. In a nutshell, the Starobinsky-like term in the Palatini formalism leaves the amplitude and the tilt of the spectrum of scalar perturbations unchanged, but suppresses the tensor-to-scalar ratio \cite{Enckell:2018hmo}. This is the reason why models like quadratic or quartic chaotic inflation \cite{Linde:1983gd}, which have a correct value of $n_s$ but suffer from an overproduction of GWs, can be brought back into agreement with the data \cite{Antoniadis:2019jnz,Antoniadis:2018yfq,Dimopoulos:2020pas,Antoniadis:2018ywb}, as long as $\alpha\gtrsim 10^8$. Similarly, the addition of a non-minimal coupling between the scalar field and the curvature scalar $\xi \phi^2 R$ is useful for both inflationary \cite{Poisson:2023tja} and quintessence model-building. For example, the exponential potential \cite{Copeland:1997et}, for which it is difficult to satisfy the dark energy observational constraints without either giving up the idea of an attractor or including an amount of fine-tuning comparable to $\Lambda$CDM, becomes much more viable via the generation of a local minimum at large field values due to the $\xi$ term \cite{Antoniadis:2022cqh,Dimopoulos:2022rdp}. The setup is also successful in quintessential inflation, where the inflaton and quintessence fields are identified \cite{Dimopoulos:2020pas,Dimopoulos:2022tvn,Dimopoulos:2022rdp,Giovannini:2019mgk,Verner:2020gfa}.

Given that the seemingly innocuous choice of the Palatini formalism leads to a theory that is fundamentally different from its metric counterpart, it seems only natural to explore the ramifications of Palatini $R +\alpha R^2$ gravity in other cosmological scenarios. In this work, we consider the production of second-order tensor modes, sourced by first-order scalar perturbations, during a period of radiation domination.\footnote{See Ref.~\cite{Koivisto:2005yc} for an early study of linear cosmological perturbations in modified gravities formulated in Palatini gravity, where the focus is on the role of the scalar modes in the formation of the cosmic large-scale structure.} The effective new matter source terms, due to the $\alpha R^2$ term, depend on the trace of the background fluid. Since radiation, with $w=1/3$, has a null trace, we must consider the subdominant pressureless dust component, with $w=0$, as it is the only contribution to the trace. Due to its subdominant nature, it is natural to treat the corresponding modifications to GR perturbatively. We choose $\beta = \alpha\rho_{\rm m,0}/\m^4\sim 10^{-121}\alpha$, a combination that appears ubiquitously, as the perturbative parameter, allowing for large values of $\alpha$ without breaking the perturbative expansion. We proceed to obtain and solve the first-order scalar perturbation equations, to then use them as sources in the second-order transverse-traceless equations. We manage to obtain an accurate analytical expression for the final density spectrum under certain considerations, and solve it numerically otherwise. The effect of the higher-order curvature term leads to a spectrum with features different not only from GR, but from its metric counterpart. These features are testable across a wide range of frequencies by future gravitational wave experiments, highlighting the potential of SIGWs to probe the theory of gravity, as well as its degrees of freedom, at scales previously inaccessible.

The present work is structured as follows. We start Sec.~\ref{sec:2} with a brief overview of Palatini $f(R)$ gravity and lay out the background, first and second order perturbation equations, in the Newtonian gauge, in Secs.~\ref{sec:2.2}-\ref{sec:2.4}. We also give general expressions for the Riemann tensor, Ricci tensor and Ricci scalar to second order in the perturbations, without assuming any gauge, in App.~\ref{ap:second-order-calculations}. We then particularise the previously general treatment to $f(R)=R+\alpha R^2$ in the radiation-dominated era in Sec.~\ref{sec:3}. We treat the modifications to GR perturbatively and solve the first-order scalar perturbations both analytically and numerically in Secs.~\ref{sec:3.1}-\ref{sec:3.3}. We then consider some well-motivated approximations and solve for the kernel and the spectral density of SIGWs in Secs.~\ref{sec:3.4}-\ref{sec:3.4.2}. We finally obtain the spectrum, in two different regimes, in Sec.~\ref{sec:3.4.3}. Assuming a log-normal scalar power spectrum, we obtain that the effect of the $\alpha R^2$ is to amplify the signal with respect to GR in an scale-invariant way in one regime, while dampening the short-frequency tail in the other regime. We conclude and comment on future prospects in Sec.~\ref{sec:4}. 
 
Natural units for which $\m = (8\pi G)^{-1/2}$, where
\begin{equation}
    \m=2.44\cross 10^{18} \ \textrm{GeV}=3.81\cross 10^{56} \ \textrm{Mpc}^{-1}=3.70\cross 10^{42} \ \textrm{Hz}~,
\end{equation} 
is the reduced Planck mass, have been assumed in this work. $G$ is Newton's gravitational constant. The metric signature is the mostly positive one $(-+++)$. 

\section{$f(R)$ Gravity in the Palatini Formalism} \label{sec:2}
As indicated in the introduction, the present work focuses on $f(R)$ gravity. In this theory, the gravitational sector of the action features a function of the Ricci scalar $f(R)$, such that
\begin{equation}
    S = \int \textrm{d}^4x \sqrt{-g} \left[\frac{\m^2}{2} f(R)+\mathcal{L}_{\textrm{M}}(g_{\mu\nu},\Xi)\right],
\end{equation}
where $g_{\mu\nu}$ is the metric tensor and $\mathcal{L}_{\textrm{M}}$ is the Lagrangian of matter fields, these being collectively denoted by $\Xi$. The Ricci scalar $R$ is obtained by contracting the metric with the Ricci tensor $R_{\mu\nu}$, and the latter is the result of a contraction of the Riemann tensor $\tensor{R}{^{\alpha}_{\beta\mu\nu}}$. In the so-called `metric formalism', these tensors can be written in terms of derivatives of the metric given that the connection $\Gamma^{\alpha}_{\mu\nu}$ is uniquely determined by the latter   
\begin{equation}
    \label{eq:levi-civita-connection-eq}\Gamma^{\mu}_{\alpha\beta}=\frac{1}{2}g^{\mu\lambda}\left(\partial_{\alpha}g_{\lambda\beta}+\partial_{\beta}g_{\lambda \alpha}-\partial_{\lambda}g_{\alpha \beta}\right).
\end{equation}
This is called the `Levi-Civita' (LC) connection. However, \textit{a priori}, the metric and the connection need not be related. In GR one finds that, upon solving the connection field equations, the connection corresponds to the LC one plus an arbitrary vector field which can be removed via a projective transformation that leaves the action and the equations invariant \cite{Bernal:2016lhq,Dadhich:2012htv,Bejarano:2019zco}. In $f(R)$ gravity however, this is no longer the case given the presence of higher-order curvature terms, as it will be shown later. 

\subsection{Field Equations}
In this work, we assume that the metric and the connection are \emph{a priori} independent, meaning that they are both determined by their corresponding field equations. Therefore, we consider the `Palatini formalism' of gravity, where the action is varied independently with respect to both quantities. We denote geometric variables that rely on the arbitrary connection $\hat{\Gamma}^{\alpha}_{\mu\nu}$ with hats in order to distinguish them from those that are built upon the LC one, which shall not wear hats. The action in the Palatini formalism then reads\footnote{As can be seen, the matter action is required to not depend on the connection. This implies that free-falling particles follow geodesics of the metric \cite{Sotiriou:2008rp}.}
\begin{equation} \label{eq:mainaction}
    S = \int \textrm{d}^4x \sqrt{-g} \left[\frac{\m^2}{2} f(\hat{R})+\mathcal{L}_{\textrm{M}}(g_{\mu\nu},\Xi)\right],
\end{equation} 
where $\hat{R} \equiv g^{\mu\nu} \hat{R}_{\mu\nu}$ is the Ricci scalar in this formalism. The Riemann $\tensor{\hat{R}}{^{\alpha}_{\beta\mu\nu}}$ and Ricci $\hat{R}_{\mu\nu} \equiv \tensor{\hat{R}}{^{\alpha}_{\mu\alpha\nu}}$ tensors are generally determined by the independent connection $\hat{\Gamma}^{\alpha}_{\mu\nu}$ and its derivatives solely, given the definition
\begin{equation}
    \tensor{\hat{R}}{^{\alpha}_{\beta\mu\nu}} \equiv \partial_{\mu}\hat{\Gamma}^{\alpha}_{\nu\beta}-\partial_{\nu} \hat{\Gamma}^{\alpha}_{\mu\beta}+\hat{\Gamma}^{\alpha}_{\mu\lambda} \hat{\Gamma}^{\lambda}_{\nu\beta} - \hat{\Gamma}^{\alpha}_{\nu\lambda} \hat{\Gamma}^{\lambda}_{\mu\beta}~,
\end{equation}
while the Ricci scalar $\hat{R}$ depends on the inverse metric as well. 

The metric and connection field equations are obtained to be 
\begin{align}
\label{eq.metric.fR.Palatini}&F \hat{R}_{(\mu\nu)} -\frac{1}{2} g_{\mu\nu}f= \frac{1}{\m^2}T_{\mu\nu}~,\\
\nonumber&F \left[\tensor{Q}{_{\sigma}^{\mu\nu}}-\left(\tensor{Q}{_{\lambda}^{\mu\lambda}}-\frac{1}{2} \tensor{Q}{^{\mu\lambda}_{\lambda}}\right) \delta_{\sigma}^{\nu} -\frac{1}{2} \tensor{Q}{_{\sigma\lambda}^{\lambda}}g^{\mu\nu}+\tensor{T}{^{\nu}_{\sigma}^{\mu}}+\tensor{T}{^{\lambda}_{\lambda\sigma}}g^{\mu\nu} +\tensor{T}{_{\lambda}^{\mu\lambda}}\delta_{\sigma}^{\nu}\right]=\\
\label{eq.conn.fR.Palatini}&=g^{\mu\nu} \hat{\nabla}_{\sigma} F - g^{\mu\lambda} \hat{\nabla}_{\lambda} F \delta_{\sigma}^{\nu}~,
\end{align}
respectively, where we have defined $F(\hat{R})\equiv f_{,\hat{R}}(\hat{R})$. $\tensor{T}{^{\alpha}_{\mu\nu}} \equiv \hat{\Gamma}^{\alpha}_{\mu\nu}-\hat{\Gamma}^{\alpha}_{\nu\mu}$ and $Q_{\alpha\mu\nu}\equiv \hat{\nabla}_{\alpha} g_{\mu\nu}$ are the respective torsion and non-metricity tensors. Only the symmetric part of the Ricci tensor $\hat{R}_{\mu\nu}$ contributes to the metric field equations, such that $2\hat{R}_{(\mu\nu)}=\hat{R}_{\mu\nu}+\hat{R}_{\nu\mu}$. This tensor is not generally symmetric when written in terms of an arbitrary connection. Conversely, the energy-momentum tensor of the matter fields is symmetric and is defined as
\begin{equation}
    T_{\mu\nu} \equiv -\frac{2}{\sqrt{-g}}\frac{\delta(\sqrt{-g}\mathcal{L}_{\textrm{M}})}{\delta g^{\mu\nu}}~.
\end{equation}

The solution to the connection field equations \eqref{eq.conn.fR.Palatini} is
\begin{equation}
    \hat{\Gamma}^{\alpha}_{\mu\nu} = \Gamma^{\alpha}_{\mu\nu}+\frac{1}{2} \left[\delta^{\alpha}_{\nu}\partial_\mu \log(F)+\delta_{\mu}^{\alpha} \partial_{\nu} \log(F) - g_{\mu\nu} \partial^{\alpha} \log(F) \right],
\end{equation} 
up to a projective transformation that may remove the second addend. $\Gamma^{\alpha}_{\mu\nu}$ is the LC connection, which is symmetric and metric-compatible. $F$ should now be a function of the trace of the energy-momentum tensor $T\equiv g^{\mu\nu} T_{\mu\nu}$, obtained by solving the trace equation
\begin{equation}
    \label{trace-eq-fR-Palatini}F(\hat{R}) \hat{R} -2f(\hat{R}) = \frac{T}{m_{\textrm{P}}^2}~,
\end{equation}
of the metric field equations \eqref{eq.metric.fR.Palatini}. This is an algebraic equation for $\hat{R}$, which, if solvable, yields $F=F(T)$ and $f=f(T)$. If we split Eq.~\eqref{eq.metric.fR.Palatini} into the LC part and the non-Riemannian contributions, the Ricci tensor reads
\begin{equation}
\label{eq:Rmunu-F}\hat{R}_{\mu\nu} = R_{\mu\nu} -\frac{1}{F} \nabla_{\mu} \partial_{\nu} F +\frac{3}{2F^2} \partial_{\mu} F \partial_{\nu} F -\frac{1}{2F} g_{\mu\nu} \Box F~,
\end{equation}
and the Ricci scalar is 
\begin{equation}
\label{eq:Ricci_scalarmunu}\hat{R} = R -\frac{3}{F} \left(\Box F -\frac{1}{2F} \partial_{\sigma} F \partial^{\sigma} F \right),
\end{equation}
where $\Box \equiv g^{\mu\nu} \nabla_{\mu} \partial_{\nu}$ is the d'Alembert operator. We insist that $F$ on the right-hand side (RHS) of the equations is a function of the trace of the energy-momentum tensor. As we see, the Ricci tensor is fully symmetric, and therefore $\hat{R}_{\mu\nu} = \hat{R}_{(\mu\nu)}$. Inserting Eq.~\eqref{eq:Rmunu-F} in the metric field equations \eqref{eq.metric.fR.Palatini}, we arrive at
\begin{equation}
\label{eq:Palatini-form-fR}F \tensor{R}{_{\mu}^{\nu}} -\nabla_{\mu} \partial^{\nu} F +\frac{3}{2F} \partial_{\mu} F \partial^{\nu} F -\frac{1}{2} \delta_{\mu}^{\nu} \left(\Box F+f\right) = \frac{1}{m^2_{\textrm{P}}} \tensor{T}{_{\mu}^{\nu}}~,
\end{equation} 
written in terms of the mixed components (that were raised using the inverse metric $g^{\mu\nu}$). One may compare this equation to that in the metric formalism 
\begin{equation}
    \label{eq:Metric-form-fR-Rexpl}F\tensor{R}{_{\mu}^{\nu}}-\nabla_{\mu} \partial^{\nu}F -\frac{1}{2} \delta_{\mu}^{\nu}\left(-2\Box F+f\right) = \frac{1}{m^2_{\textrm{P}}}\tensor{T}{_{\mu}^{\nu}}~,
\end{equation}
and we see that they both differ in their explicit form but also in the different (implicit) dependence of the functions $F$ and $f$, which rely on $R$ in the metric case. 

It is important to mention that the energy-momentum tensor $\tensor{T}{_{\mu}^{\nu}}$ is covariantly conserved in both formalisms \cite{Koivisto:2005yk}. By taking the covariant divergence of Eq.~\eqref{eq:Palatini-form-fR} (and using the corresponding trace equation to replace $T$) we get the conservation equation 
\begin{equation}
    \label{eq:cons-eq-EM}\nabla_{\nu} \tensor{T}{_{\mu}^{\nu}} = 0~, 
\end{equation}
where the reader must notice that $\nabla_{\mu}f = F \nabla_{\mu}\hat{R}$, and $\nabla_{\nu} \tensor{R}{_{\mu}^{\nu}} =\frac{1}{2}\nabla_{\mu} R$ (remember that both Ricci scalars are related by Eq.~\eqref{eq:Ricci_scalarmunu}). The same conservation law holds for the metric formalism. 

Regarding the trace equation \eqref{trace-eq-fR-Palatini}, we may use Eq. \eqref{eq:Ricci_scalarmunu} to recast it as
\begin{equation}
    \label{eq:trace-equation-intermsof-R}FR-3\Box F +\frac{3}{2F}\partial_{\sigma} F \partial^{\sigma} F -2f = \frac{T}{m^2_{\textrm{P}}}~,
\end{equation}
and use it to rewrite the metric field equations \eqref{eq:Palatini-form-fR} as
\begin{equation}
    \label{eq:Palatini-form-fR-Einstein}F\tensor{G}{_{\mu}^{\nu}}-\nabla_{\mu} \partial^{\nu}F+\frac{3}{2F} \partial_{\mu} F \partial^{\nu} F +\frac{1}{2} \delta_{\mu}^{\nu} \left(2\Box F +f -\frac{3}{2F} \partial_{\sigma} F \partial^{\sigma} F \right) = \frac{1}{m^2_{\textrm{P}}}\left(\tensor{T}{_{\mu}^{\nu}}-\frac{1}{2} \delta_{\mu}^{\nu} T \right),
\end{equation}
where we have introduced the Einstein tensor
\begin{equation}
    \tensor{G}{_{\mu}^{\nu}}\equiv \tensor{R}{_{\mu}^{\nu}}-\frac{1}{2} \delta_{\mu}^{\nu} R~,
\end{equation}
which is the trace-reversed Ricci tensor in four spacetime dimensions. We shall consider this version of the metric field equations when extracting the homogeneous and perturbed equations.

Whether $f$ and $F$ depend on $R$ or $T$ however is not essential when it comes to disassembling their background and perturbation parts. In essence, we may write
\begin{align}
&f = f^{(0)}+f^{(1)} + \frac{1}{2} f^{(2)}~,\\
&F = F^{(0)} + F^{(1)} + \frac{1}{2} F^{(2)}~,
\end{align}
where the second-order contribution, denoted by `$(2)$', is retained given the presence of terms which are quadratic in the first-order perturbations. We denote background and linear orders by `$(0)$' and `$(1)$', respectively. In what follows, we will provide specific details about the perturbation order under consideration and the background dynamics. 

\subsection{Cosmological Perturbations} \label{sec:2.2}
We assume the following perturbed line element (see Eqs.~\eqref{eq:line-element-general}-\eqref{eq:C_tensor_decomposition} and App.~\ref{ap:second-order-calculations} for further details about the second-order cosmological perturbation theory): 
\begin{equation}
    \textrm{d}s^2 = -a^2(\eta)\left\{\left(1+2A\right) \textrm{d}\eta^2-\left[\left(1-2\psi\right)\delta_{ij} +\frac{1}{2}h_{ij} \right]\textrm{d}x^{i} \textrm{d}x^{j}\right\},
\end{equation} where the scalar perturbations $A(\eta,\mathbf{x})$ and $\psi(\eta,\mathbf{x})$ are first-order, and $h_{ij}(\eta,\mathbf{x})$ are second-order. $a(\eta)$ is the scale factor and $\eta$ the comoving or conformal time. The components of the metric are then
\begin{align}
&g_{00} = -a^2\left(1+2A\right),\\
&g_{0i} = g_{i0} = 0~,\\
&g_{ij} = a^2 \left(\delta_{ij} -2\psi \delta_{ij} +\frac{1}{2} h_{ij} \right).
\end{align}
$A$ and $\psi$ are the respective lapse function and curvature perturbations and $h_{ij}$ are traceless and divergenceless tensor perturbations; namely, $\delta^{ij} h_{ij} = \tensor{h}{_i^{i}}=0$ and $\partial_i \tensor{h}{^{i}_{j}} = 0$. The Newtonian or longitudinal gauge was assumed for simplicity, which amounts to shear-free slicings and the isotropic threading at linear order, so the scalar perturbation components of the metric are diagonal.\footnote{In sharp contrast to the linear counterparts, the tensor perturbations are \emph{not} gauge-invariant at second-order \cite{Domenech:2021ztg,Ali:2023moi,Comeau:2023mxi,Kugarajh:2025pjl}. This gauge issue, however, will not be discussed in the present work.} We dropped vector perturbations, first-order tensor perturbations and second-order scalar perturbations, simplifying the calculations greatly. We can ignore the second-order scalar and vector perturbations in particular because different types of perturbations do not couple at the same order, although lower-order perturbations of different type do act as source terms \cite{Malik:2008im}. In this work however, we will focus on the scalar-induced, cross terms, as pointed out in the introduction already. 

The inverse metric components, which are shown in Eqs.~\eqref{eq:comp-inverse-beginning}-\eqref{eq:comp-inverse-end} generally, reduce to 
\begin{align}
&g^{00} = -a^{-2} \left(1-2A+4A^2\right),\\
&g^{0i}=g^{i0} = 0~,\\
&g^{ij} = a^{-2} \left(\delta^{ij} +2\psi \delta^{ij}+4\psi^2 \delta^{ij} -\frac{1}{2} h^{ij}\right).
\end{align}
Using these and the metric components, we can calculate the Christoffel symbols (see Eqs.~\eqref{eq:Chris-beginning}-\eqref{eq:Chris-end})
\begin{align}
&\Gamma^{0}_{00} = \mathcal{H} + A^{'} -2AA^{'}~,\\
&\Gamma^{0}_{i0} = \Gamma^{0}_{0i} = \partial_i A -2A \partial_i A~,\\
&\Gamma^{i}_{00} = \partial^{i} A +2\psi \partial^{i} A~,\\
\nonumber&\Gamma^{0}_{ij} = \mathcal{H} \delta_{ij} -2\mathcal{H} A \delta_{ij} -\psi^{'} \delta_{ij} -2\mathcal{H} \psi \delta_{ij} +\frac{1}{4} h^{'}_{ij} +\frac{1}{2} \mathcal{H} h_{ij} +4\mathcal{H} A^2 \delta_{ij} +2A \psi^{'} \delta_{ij} +\\
&+4\mathcal{H} A \psi \delta_{ij}~,\\
&\Gamma^{i}_{0j} = \Gamma^{i}_{j0} = \mathcal{H} \delta_{j}^{i} -\psi^{'} \delta_j^{i} +\frac{1}{4} \tensor{h}{^{i}_{j}^{'}}-2\psi \psi^{'} \delta^{i}_{j}~,\\
\nonumber&\Gamma^{k}_{ij} =-\delta^{k}_{j} \partial_i \psi -\delta^{k}_{i} \partial_j \psi +\delta_{ij} \partial^{k} \psi +\frac{1}{4} \left(\partial_i \tensor{h}{^{k}_{j}}+\partial_j \tensor{h}{^{k}_{i}}-\partial^{k} h_{ij} \right) -2\psi \left(\delta^{k}_{j} \partial_i \psi + \partial^{k}_{i} \partial_j \psi -\right.\\
&\left.-\delta_{ij} \partial^{k} \psi \right),
\end{align}
$\mathcal{H} \equiv a^{'}/a$ being the comoving expansion rate (Hubble) parameter. Overprimes denote derivatives with respect to the conformal time. These expressions allow us to write the d'Alembertian of $F$, which appears in the field equations, up to second order
\begin{align}
\nonumber&\Box F = a^{-2} \left(-F^{(0)''} -2\mathcal{H} F^{(0)'}-F^{(1)''}-2\mathcal{H} F^{(1)'}+\partial_i \partial^{i} F^{(1)}+2AF^{(0)''}+A^{'} F^{(0)'}+4\mathcal{H} A F^{(0)'}+\right.\\
\nonumber&\left.+3\psi^{'}F^{(0)'}-\frac{1}{2} F^{(2)''}+\frac{1}{2} \partial_i \partial^{i} F^{(2)}-\mathcal{H} F^{(2)'}+2AF^{(1)''}+4\mathcal{H} A F^{(1)'}-4A^2 F^{(0)''}-8\mathcal{H}A^2 F^{(0)'}+\right.\\
\nonumber&\left.+3\psi^{'} F^{(1)'}+A^{'} F^{(1)'}-4AA^{'} F^{(0)'} -6A\psi^{'} F^{(0)'} +6\psi \psi^{'} F^{(0)'} +2\psi \partial_i \partial^{i} F^{(1)}+\partial_i A \partial^{i} F^{(1)}-\right.\\
&\left.-\partial_i \psi \partial^{i} F^{(1)}\right).
\end{align} 

The (mixed) Ricci components are (see Eqs.~\eqref{eq:R00-upt-second-order-full}-\eqref{eq:Rij-upt-second-order-full})
\begin{align}
\nonumber&\tensor{R}{_{0}^{0}} =a^{-2} \left(3\mathcal{H}^{'} -3\mathcal{H} A^{'} -3\mathcal{H} \psi^{'} -3\psi^{''} -6\mathcal{H}^{'} A-\partial_i \partial^{i} A+12\mathcal{H} A A^{'} -3\psi^{'2}+3\psi^{'} A^{'}+6\mathcal{H} A \psi^{'} -\right.\\
&\left.-6\mathcal{H} \psi \psi^{'}+6A \psi^{''} -6 \psi \psi^{''} +12\mathcal{H}^{'} A^2 +2A \partial_i \partial^{i} A-2\psi \partial_i \partial^{i} A+\partial_i A \partial^{i} A+\partial_i A \partial^{i} \psi\right), \\
&\tensor{R}{_{i}^{0}} = -a^{-2} \left(2\mathcal{H} \partial_i A +2\partial_i \psi^{'} -8\mathcal{H} A\partial_i A -2\psi^{'} \partial_i A -4A \partial_i \psi^{'} +4\psi \partial_i \psi^{'} +4\psi^{'} \partial_i \psi\right),\\
&\tensor{R}{_{0}^{i}} =a^{-2} \left(2\mathcal{H} \partial^{i} A +2\partial^{i} \psi^{'} -4\mathcal{H} A \partial^{i} A+4\mathcal{H} \psi \partial^{i} A-2\psi^{'} \partial^{i} A+8\psi \partial^{i} \psi^{'}+4\psi^{'} \partial^{i} \psi\right), \\
\nonumber&\tensor{R}{_{i}^{j}} = a^{-2} \left(2\mathcal{H}^2 \delta_i^{j} +\mathcal{H}^{'} \delta_i^{j}-4\mathcal{H}^2 A \delta_i^{j}-5\mathcal{H} \psi^{'} \delta_i^{j}-\mathcal{H} A^{'} \delta_i^{j}-2\mathcal{H}^{'} A\delta_i^{j} -\psi^{''} \delta_i^{j}+\delta_i^{j} \partial_k \partial^{k} \psi +\partial_i \partial^{j} \psi -\right.\\
\nonumber&\left.-\partial_i \partial^{j} A+\frac{1}{4} \tensor{h}{_{i}^{j''}}+\frac{1}{2} \mathcal{H} \tensor{h}{_{i}^{j'}} -\frac{1}{4} \partial_k \partial^{k} \tensor{h}{_{i}^{j}}+8\mathcal{H}^2 A^2 \delta_i^{j}+4\mathcal{H}^{'} A^2\delta_i^{j}+4\mathcal{H}AA^{'} \delta_i^{j}+A^{'} \psi^{'} \delta_i^{j}-10\mathcal{H} \psi \psi^{'} \delta_i^{j}+\right.\\
\nonumber&\left.+10 \mathcal{H} A \psi^{'} \delta_i^{j}+\psi^{'2} \delta_i^{j}+2A\psi^{''} \delta_i^{j}-2\psi \psi^{''} \delta_i^{j}+3\partial_{i} \psi \partial^{j}\psi+\partial_i A \partial^{j} A-\partial_i \psi \partial^{j} A -\partial_i A \partial^{j} \psi +2A \partial_i \partial^{j} A-\right.\\
&\left.-2\psi \partial_i \partial^{j} A+4\psi \partial_i \partial^{j} \psi +4\psi \delta_i^{j} \partial_k \partial^{k} \psi +\delta_i^{j} \partial_k \psi \partial^{k} \psi +\delta_i^{j} \partial_k \psi \partial^{k} A\right).
\end{align}
The Ricci scalar reads (see Eq.~\eqref{eq:Ricci-scalar-upt-second-order-full})
\begin{align}
\nonumber&R = a^{-2} \left(6\mathcal{H}^2+6\mathcal{H}^{'}-12\mathcal{H}^2 A-12\mathcal{H}^{'} A-6\mathcal{H} A^{'}-18\mathcal{H} \psi^{'}-6\psi^{''}-2\partial_i \partial^{i} A+4\partial_i \partial^{i} \psi+24\mathcal{H}^2 A^2+\right.\\
\nonumber&\left.+24\mathcal{H}^{'} A^2 +24\mathcal{H} AA^{'}+36\mathcal{H} A \psi^{'} -36\mathcal{H} \psi \psi^{'}+6A^{'} \psi^{'}+12A \psi^{''} -12\psi \psi^{''}+4A \partial_i \partial^{i} A -4\psi \partial_i \partial^{i} A+\right.\\
&\left.+2\partial_i A \partial^{i} A +2\partial_i A \partial^{i} \psi +6\partial_i \psi \partial^{i} \psi +16\psi \partial_i \partial^{i} \psi \right).
\end{align}

Regarding the energy-momentum tensor in Eq.~\eqref{eq:em-tensor-decomposition} and the four-velocity vector in Eq.~\eqref{eq:u-four-velocity-def}, we have that the components of the latter are
\begin{align}
&u^{0} = a^{-1} \left(1-A+\frac{3}{2} A^2+\frac{1}{2} \partial_i u \partial^{i} u \right),\\
&u^{i} = a^{-1} \partial^{i} u~,\\
&u_0 = -a\left(1+A -\frac{1}{2} A^2+\frac{1}{2} \partial_i u \partial^{i} u \right),\\
&u_i = a\left(\partial_i u -2\psi \partial_i u \right),
\end{align}
and we neglect the energy-flux contribution to the former and retain the second-order anisotropic stress, which is traceless despite Eq.~\eqref{eq:trace-anisotropic-stress}. The energy and isotropic pressure densities are split into homogeneous variables and first-order perturbations such that $\rho+\delta \rho$ and $P + \delta P$, respectively. Therefore, the mixed components of the energy-momentum tensor shall read (see Eqs.~\eqref{eq:T-mixed-00}-\eqref{eq:T-mixed-ij}) 
\begin{align}
&\tensor{T}{_0^{0}} =-\rho -\delta \rho -\left(\rho+P\right) \partial_i u \partial^{i} u~, \\
&\tensor{T}{_i^{0}} = \left(\rho+P\right) \partial_i u +\left(\delta \rho + \delta P \right) \partial_i u -A\left(\rho+P\right) \partial_i u -2\psi \left(\rho+P\right) \partial_i u~,\\
&\tensor{T}{_0^{i}} = -\left(\rho+P\right)\partial^{i} u -\left(\delta \rho + \delta P \right) \partial^{i} u-A\left(\rho+P\right) \partial^{i} u~, \\
&\tensor{T}{_{i}^{j}} = P \delta_i^{j} +\delta P \delta_i^{j} +\frac{1}{2} \tensor{\Pi}{_i^{j}}+\left(\rho + P \right) \partial_i u \partial^{j} u~,
\end{align}
where $\tensor{\Pi}{_i^{j}}\equiv \tensor{\Pi}{^{(2)}_i^{j}}$ is traceless, as pointed out earlier. The trace of the energy-momentum tensor is then
\begin{equation}
\label{eq:trace-energy-momentum-tensor-uptolinear}T = -\rho +3P -\delta \rho +3\delta P~,
\end{equation} 
which agrees with Eq.~\eqref{eq:trace-energy-momentum-tensor} if one neglects the second-order contributions. 

\subsection{The Equations of the Palatini Formalism} \label{sec:2.3}
In the Palatini formalism, $f=f(T)$ and $F=F(T)$, and hence
\begin{align}
\label{eq:first-f-order}&f^{(0)} = f(T^{(0)})~,\\
&f^{(1)} = f_{,T}(T^{(0)})T^{(1)}~,\\
&\frac{1}{2} f^{(2)} = \frac{1}{2} f_{,TT}(T^{(0)}) (T^{(1)})^2~,\\
&F^{(0)} = F(T^{(0)})~,\\
&F^{(1)} = F_{,T}(T^{(0)}) T^{(1)}~,\\
\label{eq:last-F-order}&\frac{1}{2}F^{(2)} = \frac{1}{2} F_{,TT}(T^{(0)})(T^{(1)})^2~,
\end{align}
where $T^{(n)}$ denotes the $n$-th order of the corresponding terms in Eq.~\eqref{eq:trace-energy-momentum-tensor-uptolinear}. Notice that, unlike the Ricci scalar, $T$ does not contain quadratic terms which would amount to second-order contributions $T^{(2)}$ to $f^{(2)}$ and $F^{(2)}$. 

The homogeneous equations are, firstly, the analogue of Friedmann's equation
\begin{equation}
    \label{eq:homogeneous-00}3\tilde{\mathcal{H}}^2 = \frac{a^2}{F^{(0)}} \left[\frac{\rho}{m^2_{\textrm{P}}}+\frac{1}{2} \left(f^{(0)}+\frac{T^{(0)}}{m^2_{\textrm{P}}}\right)\right],
\end{equation}
where we defined
\begin{equation}
    \label{eq:def-H-tilde}\tilde{\mathcal{H}} \equiv \mathcal{H}+\frac{F^{(0)'}}{2F^{(0)}}~,
\end{equation}
and, secondly, the diagonal part of the spatial components of the metric field equations
\begin{equation}
    \label{eq:diagonal-eq-homogeneous}-\tilde{\mathcal{H}}^2-2\tilde{\mathcal{H}}^{'} = \frac{a^2}{F^{(0)}}\left[\frac{P}{m^2_{\textrm{P}}}-\frac{1}{2} \left(f^{(0)}+\frac{T^{(0)}}{m^2_{\textrm{P}}}\right)\right].
\end{equation}
Additionally, combining the two, we arrive at an equation that does not include terms proportional to $\delta_{\mu}^{\nu}$ in the field equations
\begin{equation}
    \tilde{\mathcal{H}}^{'}-\tilde{\mathcal{H}}^2  = -\frac{a^2}{2m^2_{\textrm{P}}F^{(0)}}(\rho+P)~.
\end{equation}
On top of this, we have the continuity equation from the homogeneous contribution of Eq.~\eqref{eq:cons-eq-EM}
\begin{equation}
    \label{eq:continuity-equation-stress-energy}\rho^{'} = -3\mathcal{H}\left(\rho+P\right).
\end{equation}

The reader versed in certain modified gravity theories and in conformal frames may find the definition in Eq.~\eqref{eq:def-H-tilde} quite familiar. This is how the (comoving) Hubble parameter transforms under a conformal rescaling of the metric which recasts the action as that of the so-called Einstein frame \cite{Chiba:2008ia,Kuusk:2016rso,Racioppi:2021jai,Karciauskas:2022jzd}. The equations of the theory can indeed be simplified by rewriting the background and perturbation quantities in terms of the Einstein frame ones. This is just a redefinition of some variables however. We do this for the sake of simplicity but the analysis is done in the Jordan frame representation. Further details on the Einstein frame representation of the theory are left for the interested reader in App.~\ref{ap:Einstein-frame-discussion}. 

Regarding the linear perturbation equations, we can write the respective energy and momentum conservation equations
\begin{align}
    \label{eq:cons-eq-energy-density}&\delta \rho^{'} = -3\mathcal{H}\left(\delta \rho +\delta P \right) +3\left(\rho+P\right)\psi^{'} -\partial_i \partial^{i} \Psi~,\\
    \label{eq:cons-eq-momentum}&\Psi^{'} = -4\mathcal{H} \Psi-\delta P-\left(\rho+P\right)A~, 
\end{align}
where we defined the momentum density $\Psi \equiv \left(\rho+P\right)u$. From the metric field equations \eqref{eq:Palatini-form-fR-Einstein} we obtain the energy equation
\begin{align}
    \nonumber&6\mathcal{H}^2 A F^{(0)}-3\mathcal{H}^2 F^{(1)}+6\mathcal{H} \psi^{'} F^{(0)}-3\mathcal{H} F^{(1)'} +6\mathcal{H} AF^{(0)'} +3\psi^{'} F^{(0)'}-\frac{3}{2}F^{(1)'} \frac{F^{(0)'}}{F^{(0)}}+\\
    \label{eq:energy-eq-pert}&+\frac{3}{2}A\frac{F^{(0)'2}}{F^{(0)}}+\frac{3}{4}F^{(1)}\frac{F^{(0)'2}}{F^{(0)2}}+\partial_i \partial^{i} F^{(1)} -2F^{(0)} \partial_i \partial^{i} \psi =-a^2\left[\frac{\delta \rho}{m^2_{\textrm{P}}}+\frac{1}{2}\left(f^{(1)}+\frac{T^{(1)}}{m^2_{\textrm{P}}}\right)\right],
\end{align}
and the momentum equation
\begin{equation}
    \label{eq:momentum-equation}2F^{(0)}\left(\mathcal{H}A+\psi^{'}\right) -F^{(1)'}+F^{(0)'}A+\mathcal{H} F^{(1)}+\frac{3}{2}\frac{F^{(0)'}}{F^{(0)}}F^{(1)}=-\frac{a^2}{m^2_{\textrm{P}}}\Psi~.
\end{equation}
From the tracefree part of the spatial components of the metric field equations we derive the important equation
\begin{equation}
    \label{eq:tracefree-equation}\psi -A = \frac{F^{(1)}}{F^{(0)}}~,
\end{equation}
and from the corresponding trace
\begin{align}
    \nonumber&2\mathcal{H}^2 A F^{(0)}-\mathcal{H}^2 F^{(1)}+4\mathcal{H} \psi^{'} F^{(0)}-\mathcal{H} F^{(1)'}+2\mathcal{H} A F^{(0)'}+2\psi^{'} F^{(0)'}+\frac{3}{2} F^{(1)'}\frac{F^{(0)'}}{F^{(0)}}-\\
    \nonumber&-\frac{3}{2}A\frac{F^{(0)'2}}{F^{(0)}}-\frac{3}{4}F^{(1)}\frac{F^{(0)'2}}{F^{(0)2}}+\frac{2}{3} \partial_i \partial^{i} F^{(1)}-\frac{2}{3} F^{(0)}\partial_i \partial^{i} \psi +\frac{2}{3} F^{(0)}\partial_i \partial^{i} A -F^{(1)''}+\\
    \nonumber&+2\mathcal{H} A^{'} F^{(0)}-2\mathcal{H}^{'} F^{(1)}+A^{'} F^{(0)'}+2A F^{(0)''}+4\mathcal{H}^{'} A F^{(0)}+2\psi^{''} F^{(0)}=\\
    \label{eq:pressure-pert-eq}&=a^2 \left[\frac{\delta P}{m^2_{\textrm{P}}}-\frac{1}{2} \left(f^{(1)}+\frac{T^{(1)}}{m^2_{\textrm{P}}}\right)\right].
\end{align}

Using Eq.~\eqref{eq:tracefree-equation}, we can replace the $A$ perturbations in the energy equation \eqref{eq:energy-eq-pert} such that
\begin{equation}
    \label{eq:energy-pert-eq}3\tilde{\mathcal{H}}\left(\tilde{\psi}^{'}+\tilde{\mathcal{H}}\tilde{\psi}\right) -\partial_i \partial^{i} \tilde{\psi} -3\tilde{\mathcal{H}}^2\frac{F^{(1)}}{F^{(0)}}=-\frac{a^2}{2F^{(0)}}\left[\frac{\delta \rho}{m^2_{\textrm{P}}}+\frac{1}{2}\left(f^{(1)}+\frac{T^{(1)}}{m^2_{\textrm{P}}}\right)\right],
    \end{equation}
where we defined
\begin{equation} \label{Eq:fieldredefinition}
    \tilde{\psi} \equiv \psi -\frac{F^{(1)}}{2F^{(0)}}~.
\end{equation}
This perturbation quantity can be identified to be the conformal rescaling of the linear curvature perturbation $\psi$ and, in particular, $\tilde{\psi}$ is the curvature perturbation in the Einstein frame representation when the conformal factor is chosen to be $F$ \cite{Fakir:1992cg,Hwang:1990re,Hwang:1990jh,Hwang:1996xh,Hwang:1996np}. As explained below Eq.~\eqref{eq:continuity-equation-stress-energy}, the reason for introducing this perturbation variable here is that it allows to simplify the equations substantially and handle them in a much easier manner (see discussion in this respect below the aforementioned equation). As stressed already however, this is a simple field redefinition and the analysis of the dynamics of perturbations is done in the Jordan frame representation. 

Following a similar procedure, the equation of the pressure perturbation \eqref{eq:pressure-pert-eq} is turned into
\begin{equation}
    \label{eq:deltaP-tildes}\tilde{\psi}^{''} +3\tilde{\mathcal{H}}\tilde{\psi}^{'} +\left(\tilde{\mathcal{H}}^2+2\tilde{\mathcal{H}}^{'} \right)\left(\tilde{\psi}-\frac{F^{(1)}}{F^{(0)}}\right)=\frac{a^2}{2F^{(0)}}\left[\frac{\delta P}{m^2_{\textrm{P}}}-\frac{1}{2} \left(f^{(1)}+\frac{T^{(1)}}{m^2_{\textrm{P}}}\right)\right].
    \end{equation}
We define the adiabatic speed of sound \cite{Malik:2008im}, $c_s^2 \equiv P^{'}/\rho^{'}$, and the Equation-of-State (EoS) Parameter, $w\equiv P/\rho$, such that $c_s^2 = w$ when $w$ is constant. We may write the pressure perturbation in terms of the energy density perturbation as 
\begin{equation}
    \label{eq:relation-deltaP-and-deltarho}\delta P = c_s^2 \left[\delta \rho-3\rho(1+w)\mathcal{S}\right],
\end{equation}
where $\mathcal{S}$ denotes the total entropy perturbation \cite{Gordon:2000hv}
\begin{equation}
    \mathcal{S}\equiv \frac{\delta \rho - \delta P/c_s^2}{3\rho(1+w)}=\frac{\delta P}{P^{'}}-\frac{\delta \rho}{\rho^{'}}~.
\end{equation}
Using Eq.~\eqref{eq:relation-deltaP-and-deltarho} in Eq.~\eqref{eq:deltaP-tildes}, and the energy equation \eqref{eq:energy-pert-eq} again, we arrive at the result
\begin{align}
    \nonumber&\tilde{\psi}^{''} +3\tilde{\mathcal{H}}\left(1+c_s^2\right) \tilde{\psi}^{'} +\left[2\tilde{\mathcal{H}}^{'}+\tilde{\mathcal{H}}^2\left(1+3c_s^2\right)\right]\tilde{\psi} -c_s^2 \partial_i \partial^{i} \tilde{\psi}=a^2\left\{\left[\frac{c_s^2\rho-P}{m^2_{\textrm{P}}}+\frac{1+c_s^2}{2}\left(f^{(0)}+\right.\right.\right.\\
    \label{eq:the-equation-of-tilde-psi}&\left.\left.\left.+\frac{T^{(0)}}{m^2_{\textrm{P}}}\right)\right]\frac{F^{(1)}}{F^{(0)2}}-\frac{1+c_s^2}{4F^{(0)}}\left(f^{(1)}+\frac{T^{(1)}}{m^2_{\textrm{P}}}\right)-\frac{3\rho(1+w)c_s^2}{2m^2_{\textrm{P}}F^{(0)}}\mathcal{S}\right\},
\end{align}
where we used Eqs.~\eqref{eq:homogeneous-00} and \eqref{eq:diagonal-eq-homogeneous} on the RHS. In the remainder of the present work we set the entropy perturbations to zero $\mathcal{S}=0$.

\subsection{Scalar-Induced Gravitational Waves} \label{sec:2.4}
Finally, the scalar-induced tensor perturbations are governed by the following equation:
\begin{equation}
\label{eq:evolution-second-order-modes-tensor}    h_{ij}^{''}+2\tilde{\mathcal{H}}h_{ij}^{'}-\partial_k \partial^{k} h_{ij} = -4\tensor{\Delta}{_{ij}^{kl}}\mathcal{S}_{kl}+\frac{2a^2}{m^2_{\textrm{P}}F^{(0)}}\pi^{\textrm{TT}}_{ij}~,
\end{equation}
where $\pi^{\textrm{TT}}_{ij}$ is the traceless-transverse (TT) contribution to the anisotropic stress tensor at second order. 
$\tilde{\mathcal{H}}$ was defined in Eq.~\eqref{eq:def-H-tilde}, and $\mathcal{S}_{ij}$ is the source term, given by
\begin{align}
    \nonumber&\mathcal{S}_{ij} \equiv 3\partial_i \psi \partial_j \psi +\partial_i A \partial_j A -\partial_i \psi \partial_j A -\partial_i A \partial_j \psi +2(A-\psi) \partial_i \partial_j A+4\psi \partial_i \partial_j \psi +\\
    \nonumber&+\frac{F^{(1)}}{F^{(0)}}\partial_i \partial_j (\psi - A)+\frac{3\partial_i F^{(1)}\partial_j F^{(1)}}{2F^{(0)2}}-2\psi \frac{\partial_i \partial_j F^{(1)}}{F^{(0)}}-\frac{\partial_i F^{(1)}}{F^{(0)}}\partial_j \psi-\partial_i \psi \frac{\partial_j F^{(1)}}{F^{(0)}}-\\
    &-\frac{\partial_i\partial_j \Ftwo}{2F^{(0)}}-\frac{a^2\partial_i \Psi \partial_j \Psi}{m^2_{\textrm{P}}F^{(0)}(\rho+P)}~. 
\end{align}
$\tensor{\Delta}{_{ij}^{kl}}$ is the projection tensor that carries along the TT part of $\mathcal{S}_{ij}$ (see explicit form in Refs.~\cite{Baumann:2007zm,Zhou:2024doz}). In order to simplify this term, we use Eq.~\eqref{eq:tracefree-equation} and the definition \eqref{Eq:fieldredefinition}, such that
\begin{equation}
    \mathcal{S}_{ij} = -2\partial_i \tilde{\psi} \partial_j \tilde{\psi} +\partial_i \partial_j\left[\frac{1}{F^{(0)}} \left(\frac{F^{(1)2}}{F^{(0)}}-\frac{1}{2} F^{(2)}\right)+2\tilde{\psi}^2\right] -\frac{a^2 \partial_i \Psi \partial_j \Psi}{m^2_{\textrm{P}}F^{(0)}(\rho+P)}~,
\end{equation}
where we also used the chain rule for differentiation. We remind the reader that the projection tensor acts on this source term to extract the TT part, thereby eliminating all terms that go like $\sim \partial_i\partial_j$. Thus, the surviving source term reads 
\begin{equation}
    \label{eq:tilde-psi-Sij}\mathcal{S}_{ij}=-2\partial_i \tilde{\psi} \partial_j \tilde{\psi}-\frac{a^2\partial_i \Psi \partial_j \Psi}{m^2_{\textrm{P}}F^{(0)}(\rho+P)}~.
\end{equation}

In order to replace $\partial_i \Psi \partial_j \Psi$, we need to rewrite Eq.~\eqref{eq:momentum-equation} in terms of $\psi$ and $F^{(1)}$ on the left-hand side (LHS) only, using Eq.~\eqref{eq:tracefree-equation} again
\begin{equation}
    \psi^{'}-\frac{F^{(1)'}}{2F^{(0)}} + \mathcal{H}\left( \psi-\frac{F^{(1)}}{2F^{(0)}}\right)+\frac{F^{(0)'}}{2F^{(0)}}\left(\psi + \frac{F^{(1)}}{2F^{(0)}}\right)=-\frac{a^2}{2m^2_{\textrm{P}}F^{(0)}}\Psi~. 
\end{equation}
By virtue of Eqs.~\eqref{eq:def-H-tilde} and \eqref{Eq:fieldredefinition}, we arrive at
\begin{equation}
    \label{eq:new-eq-hatpsi}\tilde{\psi}^{'} +\tilde{\mathcal{H}}\tilde{\psi} = -\frac{a^2}{2m^2_{\textrm{P}}F^{(0)}}\Psi~.
\end{equation}
It then follows that
\begin{align}
    &-\frac{a^2\partial_i \Psi \partial_j \Psi}{m^2_{\textrm{P}}F^{(0)}(\rho+P)} =-\frac{4m^2_{\textrm{P}}F^{(0)}}{a^2(\rho+P)}\left[\partial_i \tilde{\psi}^{'} \partial_j \tilde{\psi}^{'} +\tilde{\mathcal{H}}\left(\partial_i \tilde{\psi} \partial_j \tilde{\psi}\right)^{'}+\tilde{\mathcal{H}}^2 \partial_i \tilde{\psi} \partial_j \tilde{\psi}\right].
\end{align}
By inserting this in the source term \eqref{eq:tilde-psi-Sij} we finally obtain
\begin{align} \label{eq:finalsourceterm}
    &\mathcal{S}_{ij} = -2\left[1+\frac{2\tilde{\mathcal{H}}^2m^2_{\textrm{P}}F^{(0)}}{a^2\rho(1+w)}\right]\partial_i \tilde{\psi} \partial_j \tilde{\psi}-\frac{4m^2_{\textrm{P}}F^{(0)}}{a^2\rho(1+w)}\left[\partial_i \tilde{\psi}^{'} \partial_j \tilde{\psi}^{'}+\tilde{\mathcal{H}}\left(\partial_i \tilde{\psi} \partial_j \tilde{\psi}\right)^{'} \right].
\end{align}
It is clear from this expression that we need to know the evolution of the rescaled perturbation $\tilde{\psi}$, and we should be able to solve Eq.~\eqref{eq:the-equation-of-tilde-psi} and obtain analytical solutions in some cases. This form of the source term resembles that of the GR one (albeit not the exact same one). The reason for this is the conformal invariance of the tensor perturbations at all perturbation orders. This is mentioned in App.~\ref{ap:Einstein-frame-discussion}, where one can find references in that respect. 

Regarding the anisotropic stress tensor of matter, $\pi^{\textrm{TT}}_{ij}$, its damping effect on the squared amplitude of linear tensor modes was studied in GR already \cite{Weinberg:2003ur,Dicus:2005rh,Watanabe:2006qe}, particularly for those modes reentering the horizon during the radiation domination era. The primary source of this anisotropic inertia is the freely streaming neutrinos, which decouple when all modes of cosmological scales of interest are still superhorizon, and one can calculate the exact contribution \cite{Bond:1983hb,Pritchard:2004qp} by solving a Boltzmann equation to first order in metric perturbations. At second order, the impact on the scalar-induced tensor modes was assessed in Refs.~\cite{Mangilli:2008bw,Saga:2014jca,Zhang:2022dgx} (see also Refs.~\cite{Bartolo:2010qu,Domenech:2025bvr}), where it was claimed that, in addition to an analogous damping effect, new source terms come about given the relevant contribution of the neutrinos to the total energy density during radiation domination, and their high velocity dispersion as well. Similar effects were shown in the case of photons \cite{Saga:2014jca,Saga:2018ont}. One may include these and add on the anisotropic stress induced by the modifications of gravity. As far as the authors know, no previous analysis was done in the metric formulation of the $f(R)$ modified gravity; namely, where both sources of anisotropic inertia were taken into account. We however notice that the prefactor in $\pi^{\textrm{TT}}_{ij}$ in Eq.~\eqref{eq:evolution-second-order-modes-tensor} is rescaled by $F^{(0)}$ compared to its counterpart in GR, and that the same rescaling is obtained in the metric formalism \cite{Zhou:2024doz,Kugarajh:2025rbt}. Although the two $F^{(0)}$ differ in the two formalisms, we shall treat perturbatively the gravity modifications, meaning that these matter contributions are not expected to manifest in distinctive features of the spectrum; that is, dependent on the formalism. Since the essential goal of this work is to analyse and exhibit the spectral features that may distinguish the two formalisms and GR, we hereon neglect the anisotropic stress contribution from the matter, in the vein of Refs.~\cite{Zhou:2024doz,Kugarajh:2025rbt}. Same reasoning applies to the anisotropic contribution to the linear equation \eqref{eq:tracefree-equation} which, along with the modified gravity contribution under examination, would make the two so-called `Bardeen potentials' \cite{Bardeen:1980kt,Baumann:2009ds}, $\psi$ and $A$, differ. This linear term is rescaled in the same way in the two formalisms and hence we do not anticipate substantial differences depending on the formulation of gravity.

\section{The Case of \texorpdfstring{$\hat{R}+\frac{\alpha}{2\m^2}\hat{R}^2$}{R\texttwosuperior} Gravity} \label{sec:3}
We may now proceed to make use of the equations derived in the preceding sections. As a well-known example, and very successful model of inflation also \cite{Starobinsky:1980te,Enckell:2018hmo,Dimopoulos:2022rdp,Antoniadis:2018ywb,Antoniadis:2018yfq,Dimopoulos:2020pas,Dimopoulos:2022tvn}, let us consider the analogue of Starobinsky gravity in the Palatini formalism, given by the following choice of the $f(\hat{R})$ function:
\begin{equation} \label{asdofbihwasfdbgqw}
    f(\hat{R})=\hat{R}+\frac{\alpha}{2\m^2}\hat{R}^2~,
\end{equation}
where $\alpha$ is a dimensionless constant. The trace equation \eqref{trace-eq-fR-Palatini} provides a simple GR-like relation 
\begin{equation}
    \hat{R}=-\frac{T}{\m^2}~.
\end{equation}
Plugging this back in Eq. \eqref{asdofbihwasfdbgqw}, we find
\begin{equation}
    f(T)=-\frac{T}{\m^2}\left(1-\frac{\alpha}{2\m^4}T\right),
\end{equation}
and (remember that $F=f_{,\hat{R}}=-m^2_{\textrm{P}}f_{,T}$)
\begin{equation}
    F(T)=1-\frac{\alpha}{\m^4}T~.
\end{equation}
As indicated below Eq.~\eqref{eq:the-equation-of-tilde-psi}, we shall assume adiabatic perturbations with constant EoS parameter, such that $\mathcal{S}=0$ and $w=c_s^2$. From Eq.~\eqref{eq:trace-energy-momentum-tensor-uptolinear}, we see that the trace of the energy-momentum tensor at background and first order read
\begin{align}
    \label{eq:T_zero_matterrad}&\Tzero=-\rho(1-3w)~,\\
    &\Tone=-\delta\rho(1-3c_s^2)~,
\end{align}
where $c_s^2=\delta P/\delta\rho$ given that $\mathcal{S}=0$. Thus, we obtain the following expressions for the different orders in the expansion of $f$ and $F$ (see Eqs.~\eqref{eq:first-f-order}-\eqref{eq:last-F-order}):
\begin{align}
    &\fzero = \frac{\rho(1-3w)}{\m^2}\left[1+\frac{\alpha\rho(1-3w)}{2\m^4}\right],\label{aiudsbfaiw}\\
    &\fone = \frac{\delta\rho(1-3c_s^2)}{\m^2}\left[1+\frac{\alpha\rho(1-3w)}{\m^4}\right],\\
    &\frac{1}{2}\ftwo = \frac{\alpha \delta\rho^2(1-3c_s^2)^2}{2\m^6}~,\\
    &F^{(0)} = 1+\frac{\alpha \rho(1-3w)}{\m^4}~,\\
    &\Fone = \frac{\alpha\delta\rho(1-3c_s^2)}{\m^4}~,\\
    &\frac{1}{2}\Ftwo = 0~.\label{aisudbfn}
\end{align}

\subsection{Radiation Domination} \label{sec:3.1}
If the content of the Universe is dominated by conformal matter (radiation), with $T=0$, then the trace equation \eqref{trace-eq-fR-Palatini} yields
\begin{equation}
    F(\hat{R})\hat{R}-2f(\hat{R})=0~,
\end{equation}
which implies (with the exception of the conformal case $f(\hat{R}) \propto \hat{R}^2$) that the Ricci scalar in the Palatini formulation (and therefore $F(\hat{R})$) is a constant $\hat{R} = \hat{R}_0$. Plugging this back in the field equations, we recover GR with a cosmological constant \cite{Sotiriou:2008rp,Olmo:2011uz}, given by
\begin{equation}
    \Lambda = \frac{\hat{R}_0 F(\hat{R}_0)-f(\hat{R}_0)}{2F(\hat{R}_0)}~.
\end{equation}
For $f(\hat{R})$ given by Eq.~\eqref{asdofbihwasfdbgqw}, $\hat{R}_0=0$ and the cosmological constant is zero. 

Since the present work seeks to analyse the radiation dominated era, one might naively conclude that the corresponding dynamics of the SIGWs is trivially that of GR. We emphasise, however, that one must take into account the subdominant pressureless dust component, present during the aforementioned era. Not only will it modify the dynamics with respect to GR, but its contribution to the energy density will be small in comparison to that of the radiation. Consequently, we shall be able to treat the modifications perturbatively and obtain analytical expressions. To this end, we first notice that $\rho w = \rho_{\textrm{r}}/3$, where $w_{\textrm{r}} =1/3$ and $w_{\textrm{m}} = 0$. Also, $\delta \rho c_s^2 = \delta \rho_{\textrm{r}}/3$ as $c_s^2 = w$. Therefore, Eqs. \eqref{aiudsbfaiw}-\eqref{aisudbfn} read
\begin{align}
    &\fzero = \frac{\rhom}{\m^2}\left(1+\frac{\alpha\rhom}{2\m^4}\right),\label{aoisdbfuibwaeubrwqer}\\
    &\fone = \frac{\delta\rhom}{\m^2}\left(1+\frac{\alpha\rhom}{\m^4}\right),\\
    &\frac{1}{2}\ftwo = \frac{\alpha \delta\rhom^2}{2\m^6}~,\\
    \label{eq:F0_matterrad}&F^{(0)} = 1+\frac{\alpha \rhom}{\m^4}~,\\
    &\Fone = \frac{\alpha\delta\rhom}{\m^4}~,\\
    &\frac{1}{2}\Ftwo = 0~. \label{asodfbwaefbwer}
\end{align}
With these, the analogue of Friedmann equation \eqref{eq:homogeneous-00} reads
\begin{equation}
    \label{eq:friedmann-R2-Palatini}3\mathcal{H}^2 \left(1-\frac{\alpha\rhom}{2\m^4}\right)^2 = a^2 \left(1+\frac{\alpha\rhom}{\m^4}\right)\left[\frac{\rho_{\textrm{r}}}{\m^2}+\frac{\rhom}{\m^2}\left(1+\frac{\alpha\rhom}{4\m^4}\right)\right],
\end{equation}
where the continuity equation \eqref{eq:continuity-equation-stress-energy} was used to find $\rhom' = -3\mathcal{H} \rhom$. Note that the product $a^3 \rhom$ is constant and in particular it can be evaluated at the present time, such that
\begin{equation}
    \label{eq:cont_a3rhom}a^3\rho_{\textrm{m}} = a_0^3 \rho_{\textrm{m},0} = \rho_{\textrm{m},0}~,
\end{equation}
where we take $a_0 = 1$ and $\rho_{\textrm{m},0} \simeq 3.28\cross 10^{-121} \m^4$ \cite{Planck:2018vyg}. The same applies to the energy density of radiation, whose continuity equation is $\rho_{\textrm{r}}'=-4\mathcal{H}\rho_{\textrm{r}}$, and then 
\begin{equation}
    \label{eq:cont_a4rhor}a^4 \rho_{\textrm{r}} = a_0^4\rho_{\textrm{r},0} = \rho_{\textrm{r},0}~,
\end{equation}
where $\rho_{\textrm{r},0} \simeq 9.64\cross 10^{-125} \m^4$.

We may also write the perturbation equation \eqref{eq:the-equation-of-tilde-psi} in Fourier space as (remember that $c_s^2=w$ and $\mathcal{S}=0$)
\begin{equation}
\label{eq:tilde-psi-source-term}\tilde{\psi}^{''}_k +3\tilde{\mathcal{H}}(1+w)\tilde{\psi}^{'}_k +\left[wk^2+2\tilde{\mathcal{H}}^{'}+\tilde{\mathcal{H}}^2(1+3w)\right]\tilde{\psi}_k = -a^2\frac{\alpha(1+w)\rhom\delta \rho^{\textrm{m}}_{k}}{4\m^6}\left(1+\frac{\alpha\rhom}{\m^4}\right)^{-2},
\end{equation}
where 
\begin{equation}
    w = \frac{\rho_{\textrm{r}}}{3(\rho_r+\rhom)}~.
\end{equation}
As can be seen, the density perturbation of (pressureless) matter acts as a source in the equation of $\tilde{\psi}_k$. We may define alternatively the density contrast $\delta_{\textrm{m}} \equiv \delta \rhom/\rhom$ \cite{Baumann:2022mni}. Note that no approximations were made to obtain Eqs.~\eqref{eq:friedmann-R2-Palatini} and \eqref{eq:tilde-psi-source-term}. In order to simplify these, one may leverage the smallness of the dimensionless parameter $\beta$ (as long as $\alpha < 10^{120}$, so that $\beta \ll 1$), defined as
\begin{equation} \label{aosdfbiawner}
    \beta \equiv \frac{\alpha \rho_{\textrm{m},0}}{\m^4} \simeq 3.28\times 10^{-121}\alpha~,
\end{equation}
which is present in Eq.~\eqref{eq:friedmann-R2-Palatini} such that
\begin{equation}
    3\mathcal{H}^2 \left(1-\frac{\beta}{2} a^{-3} \right)^2 = a^2 \left(1+\beta a^{-3} \right)\left[\frac{\rho_{\textrm{r},0}}{\m^2}a^{-4}+\frac{\rho_{\textrm{m},0}}{\m^2} a^{-3} \left(1+\frac{\beta}{4}a^{-3}\right)\right],
\end{equation}
where we remind the reader that the continuity equation has the same form as that of GR \cite{Koivisto:2005yk} and we use it in such a way that $\rhom = \rho_{\textrm{m},0} a^{-3}$ and $\rho_{\textrm{r}} = \rho_{\textrm{r},0}a^{-4}$ with $a_0 =1$ (see Eqs.~\eqref{eq:cont_a3rhom} and \eqref{eq:cont_a4rhor}, respectively). We then expand to linear order in $\beta$ and arrive at the equation
\begin{equation}
    \label{eq:Friedmann-like-first-appr}3\mathcal{H}^2\simeq \frac{\rho_{\textrm{r},0}}{\m^2}a^{-2}+\frac{\rho_{\textrm{m},0}}{\m^2}a^{-1}+2\beta\left(\frac{\rho_{\textrm{r},0}}{\m^2}a^{-5}+\frac{9\rho_{\textrm{m},0}}{8\m^2}a^{-4}\right).
\end{equation}
As can be seen, this equation splits into a GR-like part and an additional contribution which relies on the small (but non-zero) value of $\beta$. We may expand the scale factor and (comoving) Hubble parameter similarly then
\begin{align}
    &a = a_{\textrm{GR}}+\beta a_{\textrm{MG}}~,\\
    &\mathcal{H} = \mathcal{H}_{\textrm{GR}}+\beta \mathcal{H}_{\textrm{MG}}~.
\end{align}
This approach to modified gravity corrections was developed within the metric formalism in Ref.~\cite{Bertacca:2011wu} and later applied to the study of SIGWs in Ref.~\cite{Kugarajh:2025rbt} in the same formalism. It is worth noting that, while $\alpha$ is their choice of expansion parameter, here we opt for $\beta$ which accommodates large values of $\alpha$ due to how small the order of magnitude of $\rho_{\textrm{m},0}$ is in Planck units. Also, the authors of the aforementioned paper are taking $\alpha$ with dimensions of inverse mass squared, while our $\beta$ is dimensionless.

Following this procedure, Eq.~\eqref{eq:Friedmann-like-first-appr} leads to the GR (standard) equation
\begin{equation}
    \label{eq:GR-eq-Friedmann-matterandrad}3\mathcal{H}_{\textrm{GR}}^2 = \frac{\rho_{\textrm{r},0}}{\m^2} a_{\textrm{GR}}^{-2} +\frac{\rho_{\textrm{m},0}}{\m^2}a^{-1}_{\textrm{GR}}~,
\end{equation}
and the modified gravity one
\begin{equation}
    \label{eq:mod-grav-approx-eq}\mathcal{H}_{\textrm{GR}}\mathcal{H}_{\textrm{MG}} = \frac{\rho_{\textrm{r},0}}{3\m^2}a_{\textrm{GR}}^{-3}\left(a_{\textrm{GR}}^{-2}-a_{\textrm{MG}}\right)+\frac{3\rho_{\textrm{m},0}}{8\m^2} a_{\textrm{GR}}^{-2} \left(a^{-2}_{\textrm{GR}}-\frac{4}{9} a_{\textrm{MG}}\right).
\end{equation}
To relate the GR and modified gravity parts of $\mathcal{H}$ to those of $a$, we simply resort to the definition $\mathcal{H} = a^{'}/a$. Then $\mathcal{H}_{\textrm{GR}} = a^{'}_{\textrm{GR}}/a_{\textrm{GR}}$ and $\mathcal{H}_{\textrm{MG}} = a_{\textrm{MG}}^{'}/a_{\textrm{GR}}-a^{'}_{\textrm{GR}}a_{\textrm{MG}}/a^2_{\textrm{GR}}$. Assuming $\rho_{\textrm{m},0} \ll \rho_{\textrm{r},0} a^{-1}_{\textrm{GR}}$, which amounts to radiation domination, we neglect the second term in Eq.~\eqref{eq:GR-eq-Friedmann-matterandrad} such that
\begin{equation}
    \label{eq:standard_GR_raddom}a'_{\textrm{GR}} = \sqrt{\frac{\rho_{\textrm{r},0}}{3\m^2}}~,
\end{equation}
and obtain from solving this equation that 
\begin{equation}
    a_{\textrm{GR}}(\eta) = \mathcal{A}\eta + \mathcal{B}~,
\end{equation}
where $\mathcal{B}$ is a constant of integration and $\mathcal{A}$ is defined as
\begin{equation} \label{eq:Adefinition}
    \mathcal{A} \equiv \sqrt{\frac{\rho_{\textrm{r},0}}{3\m^2}}\simeq 5.67\cross 10^{-63} \m = 2.16\cross 10^{-6}\text{Mpc}^{-1}=2.10\cross 10^{-20}\text{Hz}~.
\end{equation}
While $\mathcal{B}$ is a constant that relies on the initial condition of $a_{\textrm{GR}}$, $\mathcal{A}$ is simply the RHS of Eq. \eqref{eq:standard_GR_raddom}, and its value is determined, as we see, by the ratio between the energy density of radiation at the present time, $\rho_{\textrm{r},0}$ (see the comment in that respect below Eq.~\eqref{eq:friedmann-R2-Palatini}), and the Planck mass squared $\m^2$. Consequently, $\mathcal{H}_{\textrm{GR}} = \mathcal{A}/a_{\textrm{GR}}$, and by inserting this in Eq.~\eqref{eq:mod-grav-approx-eq} we arrive at
\begin{equation}
    \mathcal{H}_{\textrm{MG}} \simeq \mathcal{A} a_{\textrm{GR}}^{-2} \left(a^{-2}_{\textrm{GR}}-a_{\textrm{MG}}\right),
\end{equation}
where we neglected the pressureless matter contribution (second term on the RHS) in Eq.~\eqref{eq:mod-grav-approx-eq} accordingly. Using $\mathcal{H}_{\textrm{MG}}$ defined below Eq.~\eqref{eq:mod-grav-approx-eq} and $\mathcal{H}_{\textrm{GR}} = \mathcal{A}/a_{\textrm{GR}}$, the equation above becomes
\begin{equation}
    a_{\textrm{MG}}' = \frac{\mathcal{A}}{(\mathcal{A}\eta+\mathcal{B})^3}~.
\end{equation}
Upon solving this equation one obtains
\begin{equation}
    a_{\textrm{MG}}(\eta) = \mathcal{C}-\frac{1}{2(\mathcal{A}\eta+\mathcal{B})^2}~,
\end{equation}
$\mathcal{C}$ being another constant of integration, which is set by the initial condition on $a_{\textrm{MG}}$. One may want $a_{\textrm{MG}}(\eta)$ to be non-negative, and hence
\begin{equation}
    \mathcal{C} > \frac{1}{2\left(\mathcal{A}\eta_{\textrm{ini}}+\mathcal{B}\right)^2}~,
\end{equation}
meaning that $a_{\textrm{GR}}(\eta_{\textrm{ini}}) >1/\sqrt{2\mathcal{C}}$. Keeping in mind the condition under which the matter contribution is neglected when compared to the dominant component of radiation, we arrive at
\begin{equation} \label{eq:conditionCfirst}
    \frac{\rho_{\textrm{m},0}}{\rho_{\textrm{r},0}} \simeq 3.40 \cross 10^{3} \ll \sqrt{2\mathcal{C}} \Rightarrow \mathcal{C} \gg 5.78\cross 10^6~.
\end{equation}
Either possibility however (large or small $\mathcal{C}$ compared to the $\eta^{-2}$ term) would not be problematic because the modified gravity part is smaller than the GR one and the total scale factor remains positive. 

Given the approximations above, we may write the scale factor and comoving Hubble parameter as
\begin{align}
    \label{eq:scale-factor-total-GRMG}&a(\eta) = \mathcal{A}\eta + \beta \left(\mathcal{C}-\frac{1}{2\mathcal{A}^2\eta^2}\right),\\
    &\mathcal{H}(\eta) = \frac{1}{\eta}+\frac{\beta}{\mathcal{A}\eta^2}\left(\frac{3}{2\mathcal{A}^2\eta^2}-\mathcal{C}\right)\label{oaisdbfiabsdfawe},
\end{align}
respectively, where a shift in conformal time was employed conveniently to remove the constant of integration $\mathcal{B}$. A similar treatment based upon the smallness of $\beta$ can be given to the perturbations, which shall be explored in the next section in order to extract analytical expressions for the transfer function of $\tilde{\psi}$ and for more relevant perturbation quantities. 

\subsection{The Transfer Function} \label{sec:3.2}
Using Eqs.~\eqref{eq:homogeneous-00} and \eqref{eq:diagonal-eq-homogeneous} we have
\begin{equation}
    \label{eq:new-eq-explanatory}2\tilde{\mathcal{H}}^{'}+\tilde{\mathcal{H}}^2(1+3w) = a^2\frac{1+w}{2F^{(0)}}\left(f^{(0)}+\frac{T^{(0)}}{\m^2}\right)\simeq a^2(1+w)\frac{\alpha\rhom^2}{4\m^6}~,
\end{equation}
where we employed Eqs.~\eqref{eq:T_zero_matterrad}, \eqref{aoisdbfuibwaeubrwqer} and \eqref{eq:F0_matterrad} too, and neglected the term that is high-order in $\beta$, which was defined in Eq.~\eqref{aosdfbiawner}.
The perturbation equation \eqref{eq:tilde-psi-source-term} can then be written as 
\begin{equation}
    \label{eq:preliminary_version_before}\tilde{\psi}^{''}_k +4\mathcal{H}\left(1-\frac{3\alpha\rho_{\textrm{m}}}{2\m^4}\right) \tilde{\psi}_k^{'} +\frac{1}{3} \left(k^2+a^2\frac{\alpha\rhom^2}{\m^6}\right) \tilde{\psi}_k = -a^2 \frac{\alpha\rhom^2}{3\m^6}\delta^{\textrm{m}}_k~,
\end{equation}
where we set $w=1/3$ and used Eqs.~\eqref{eq:def-H-tilde} and \eqref{eq:F0_matterrad} such that (by means of the continuity equation \eqref{eq:continuity-equation-stress-energy} and Eq.~\eqref{eq:cont_a3rhom} as well)
\begin{equation}
    3\tilde{\mathcal{H}}(1+w) \simeq 4\left(\mathcal{H}+\frac{\alpha\rhom'}{2\m^4}\right)=\mathcal{H}\left(1-\frac{3}{2}\beta a^{-3}\right).
\end{equation}
The high-order term in $\beta$ on the RHS of Eq.~\eqref{eq:tilde-psi-source-term} was neglected accordingly. Assuming (for large $\mathcal{C}$; see Eqs.~\eqref{eq:scale-factor-total-GRMG} and \eqref{oaisdbfiabsdfawe} respectively) 
\begin{align}
    &a(\eta) = \mathcal{A} \eta +\beta \mathcal{C}~,\\
    &\mathcal{H}(\eta) = \frac{1}{\eta}-\frac{\beta\mathcal{C}}{\mathcal{A}\eta^2}~,
\end{align}
Eq.~\eqref{eq:preliminary_version_before} becomes
\begin{equation}
    \label{eq:psitilde-second-order}\tilde{\psi}^{''}_k +\frac{4}{\eta}\left(1-\frac{\beta\mathcal{C}}{\mathcal{A}\eta}\right)\tilde{\psi}^{'}_k+\frac{1}{3}\left(k^2+\frac{\beta \mathcal{D}}{\eta^4}\right)\tilde{\psi}_k =-\frac{\beta \mathcal{D}}{3\eta^4}\delta_k^{\textrm{m}}~,
\end{equation}
where we neglected $-3\beta /(2\mathcal{A}^3\eta^3)$ within the first parenthesis as well; namely
\begin{align}
    \nonumber&4\mathcal{H}\left(1-\frac{3\alpha \rhom}{2\m^4}\right) \simeq \frac{4}{\eta}\left(1-\frac{\beta\mathcal{C}}{\mathcal{A}\eta}\right)\left(1-\frac{3\beta}{2\mathcal{A}^3\eta^3}\right)\simeq \frac{4}{\eta}\left[1-\frac{\beta}{\mathcal{A}\eta}\left(\mathcal{C}+\frac{3}{2\mathcal{A}^2\eta^2}\right)\right]\simeq\\
    &\simeq \frac{4}{\eta} \left(1-\frac{\beta \mathcal{C}}{\mathcal{A}\eta}\right). 
\end{align}
$\mathcal{D}$ is defined as (see Eq.~\eqref{eq:cont_a3rhom})
\begin{equation} \label{asdofjubaiwerawer}
    \mathcal{D} \equiv \frac{\rho_{\textrm{m},0}}{\m^2 \mathcal{A}^4} \simeq 3.17\cross 10^{128}\m^{-2}=2.18\cross 10^{15}\text{Mpc}^2=2.25\cross 10^{43}\text{Hz}^{-2}~,
\end{equation}
and, as can be seen, depends on $\mathcal{A}$ which was introduced in Eq.~\eqref{eq:Adefinition} and the energy density of presureless matter at the present time (see discussion below Eq.~\eqref{eq:friedmann-R2-Palatini}). Unlike $\mathcal{C}$ and the $\mathcal{B}$ that was absorbed by a shift in the conformal time, this $\mathcal{D}$ is a prefactor in the source term (RHS of Eq.~\eqref{eq:preliminary_version_before}) brought about by the density contrast of pressureless matter $\delta_{\textrm{m}}$. Both values of $\mathcal{A}$ and $\mathcal{D}$ depend on current values of the energy densities of radiation and dust, and of the fundamental constant $\m^2$, meaning that they are fixed.

In the same vein as with the scale factor and the comoving Hubble parameter, we differentiate the GR solution from the small contribution from modified gravity, which is determined by the smallness of $\beta$
\begin{equation} \label{eq:psiseparation}
    \tilde{\psi}_k = \psi^{\textrm{GR}}_k + \beta \tilde{\psi}^{\textrm{MG}}_k~,
\end{equation}
where we have dropped the tilde on the GR solution because, when $\beta=0$, $\tilde{\psi}_k$ becomes $\psi_k=\psi^{\textrm{GR}}_k$ given that we have no $\hat{R}^2$ term in the action, and the theory is fully equivalent to GR. Same procedure applies to the density contrast $\delta^{\textrm{m}}_k$.  Working perturbatively to first order in $\beta$, Eq.~\eqref{eq:psitilde-second-order} can be rewritten as a differential equation for $\tilde{\psi}^{\textrm{MG}}_k$ such that 
\begin{equation} \label{eq:def-eq-psiMG}
    \tilde{\psi}^{\textrm{MG}''}_k +\frac{4}{\eta} \tilde{\psi}^{\textrm{MG}'}_k +\frac{1}{3} k^2 \tilde{\psi}^{\textrm{MG}}_k = \frac{4\mathcal{C}}{\mathcal{A}\eta^2}\psi^{\textrm{GR}'}_k-\frac{\mathcal{D}}{3\eta^4}\left(\delta^{\textrm{m},\textrm{GR}}_k + \psi^{\textrm{GR}}_k\right).
\end{equation}
In the case of small $\mathcal{C}$ we would have the same equation without the term proportional to $\psi^{\textrm{GR}'}_k$ on the RHS. $\psi^{\textrm{GR}}_k$ is solved by the equation
\begin{equation}
    \label{eq:psi-GR-rad}\psi^{\textrm{GR}''}_k +\frac{4}{\eta}\psi^{\textrm{GR}'}_k +\frac{1}{3} k^2 \psi^{\textrm{GR}}_k = 0~,
\end{equation}
whereas the equation for $\delta^{\textrm{m},\textrm{GR}}_k$ can be obtained from the energy and momentum conservation equations, Eqs.~\eqref{eq:cons-eq-energy-density} and \eqref{eq:cons-eq-momentum}, respectively (the stress-energy tensors of the radiation and the matter are assumed to be separately conserved as was considered at background level already)
\begin{align}
    \label{eq:energy-conservation-matter}&\delta_{\textrm{m}}^{'} = 3\psi^{'} -\partial_i \partial^{i} \left(\frac{\Psi_{\textrm{m}}}{\rho_{\textrm{m}}}\right),\\
    &\left(a\frac{\Psi_{\textrm{m}}}{\rho_{\textrm{m}}}\right)^{'} = -a A~,
\end{align}
where the continuity equation \eqref{eq:continuity-equation-stress-energy} was employed. Deriving the first equation with respect to the conformal time $\eta$ and using the second one and Eq.~\eqref{eq:tracefree-equation} to get rid of $A$, we arrive at
\begin{equation}
    \delta_{\textrm{m}}^{''}+\mathcal{H} \delta_{\textrm{m}}^{'} = 3\left(\psi^{''} + \mathcal{H} \psi^{'}\right) +\partial_i \partial^{i}\left(\psi -\frac{F^{(1)}}{F^{(0)}}\right).
\end{equation}
Writing the equation in terms of $\tilde{\psi}$
\begin{equation}
    \left(\delta_{\textrm{m}}-\frac{3F^{(1)}}{2F^{(0)}}\right)^{''} +\mathcal{H}\left(\delta_{\textrm{m}}-\frac{3F^{(1)}}{2F^{(0)}}\right)^{'} +\partial_i \partial^{i}\left(\frac{F^{(1)}}{2F^{(0)}}\right)= 3\left(\tilde{\psi}^{''}+\mathcal{H} \tilde{\psi}^{'}\right)+\partial_i \partial^{i} \tilde{\psi}~.
\end{equation}
In the case of $\hat{R}+\frac{\alpha}{2\m^2}\hat{R}^2$ we have $F^{(1)} = -\delta_{\textrm{m}}(1-F^{(0)})$ and
\begin{equation}
    \delta_{\textrm{m}} -\frac{3F^{(1)}}{2F^{(0)}}=\delta_{\textrm{m}}\frac{3-F^{(0)}}{2F^{(0)}}\simeq \delta_{\textrm{m}}\left(1-\frac{3}{2}\frac{\beta\mathcal{A}^3}{\eta^3}\right)~,
\end{equation}
such that
\begin{equation}
    \label{eq:deltam-GR-eq}\delta_k^{\textrm{m},\textrm{GR}''}+\frac{1}{\eta}\delta_k^{\textrm{m},\textrm{GR}'}=-\frac{9}{\eta}\psi^{\textrm{GR}'}_k-2k^2 \psi^{\textrm{GR}}_k~,
\end{equation}
where we used Eq.~\eqref{eq:psi-GR-rad} as well.

Solving Eq.~\eqref{eq:psi-GR-rad}, we obtain
\begin{equation}
    \psi^{\textrm{GR}}_k(\eta) = \frac{1}{\sqrt{k}\eta^2}\left\{c_1\left[\frac{\sqrt{3}}{k\eta}\sin\left(\frac{k\eta}{\sqrt{3}}\right)-\cos\left(\frac{k\eta}{\sqrt{3}}\right)\right]+c_2\left[\sin\left(\frac{k\eta}{\sqrt{3}}\right)+\frac{\sqrt{3}}{k\eta}\cos\left(\frac{k\eta}{\sqrt{3}}\right)\right]\right\},
\end{equation}
where $c_1$ and $c_2$ are constants of integration. It is known that $\psi_k^{\textrm{GR}}$ becomes constant on superhorizon scales (when $k|\eta| \rightarrow 0$) \cite{Baumann:2022mni}. Therefore, we must set $c_2 = 0$ and $c_1 \propto k^{-3/2}$, such that
\begin{equation}
    \label{eq:sol-after-superh-cond}\psi_k^{\textrm{GR}}(\eta) \propto \frac{1}{k^2\eta^2}\left[\frac{\sqrt{3}}{k\eta}\sin\left(\frac{k\eta}{\sqrt{3}}\right)-\cos\left(\frac{k\eta}{\sqrt{3}}\right)\right].
\end{equation}
To find the constant of proportionality it is convenient that we relate $\psi_{k}^{\textrm{GR}}$ to the initial condition after inflation, given by the constant value of the comoving curvature perturbation on superhorizon scales, $\mathcal{R}_k$ (this is the curvature perturbation on comoving slicings) \cite{Baumann:2022mni,Weinberg:2003sw,Lyth:1984gv}. Moreover, in order to determine the deviation from the GR solution due to the modified gravity correction, we shall assume the same initial condition for both. Consequently, we can write (see Ref.~\cite{Kugarajh:2025rbt} for a similar separation)
\begin{equation}
    \label{eq:separation-transfer-functions}\tilde{\psi}_k \equiv \frac{3(1+w)}{5+3w} \tilde{T}_{\tilde{\psi}}(k\eta) \mathcal{R}_k = \frac{3(1+w)}{5+3w}\left[T^{\textrm{GR}}_{\psi}(k\eta)+\beta T^{\textrm{MG}}_{\tilde{\psi}}(k\eta)\right]\mathcal{R}_k~.
\end{equation}
The $T(k\eta)$ functions are the so-called `transfer functions' of each solution, such that $T^{\textrm{GR}}_{\psi}(0) = T^{\textrm{MG}}_{\tilde{\psi}}(0) = 1$, and hence\footnote{Even if the initial condition for the modified gravity correction is not the assumed one, we can always redefined this initial value by changing the constant parameter $\alpha$ in $\beta$ accordingly.} $\psi^{\textrm{GR}}_{k,i} = \tilde{\psi}^{\textrm{MG}}_{k,i} = \frac{2}{3} \mathcal{R}_k$. This means that there is a slight difference between the initial value of the rescaled curvature perturbation $\tilde{\psi}$ and the GR one, weighed by the parameter $\beta$. We can make notation less cluttered by means of the following definitions: $T_{\psi}(k\eta)\equiv T_{\psi}^{\rm GR}(k\eta)$ and $T_{\tilde{\psi}}(k\eta)\equiv T_{\tilde{\psi}}^{\rm MG}(k\eta)$.

The constant of proportionality in Eq.~\eqref{eq:sol-after-superh-cond} is $6\mathcal{R}_k$ then, yielding
\begin{equation}
    \label{eq:sol-psiGR-ini-cond}\psi^{\textrm{GR}}_k (\eta) = \frac{6\mathcal{R}_k}{k^2\eta^2}\left[\frac{\sqrt{3}}{k\eta}\sin\left(\frac{k\eta}{\sqrt{3}}\right)-\cos\left(\frac{k\eta}{\sqrt{3}}\right)\right].
\end{equation}
Using Eq.~\eqref{eq:sol-psiGR-ini-cond} we may solve Eq.~\eqref{eq:deltam-GR-eq} which reads
\begin{equation}
    \delta_k^{\textrm{m},\textrm{GR}''}+\frac{1}{\eta}\delta_k^{\textrm{m},\textrm{GR}'}=-\frac{12\mathcal{R}_k}{k^2\eta^4}\left[\frac{\sqrt{3}}{2}\left(5k\eta-\frac{27}{k\eta}\right)\sin\left(\frac{k\eta}{\sqrt{3}}\right)+\left(\frac{27}{2}-k^2\eta^2\right)\cos\left(\frac{k\eta}{\sqrt{3}}\right)\right].
\end{equation}
The solution is 
\begin{equation}
    \delta_k^{\textrm{m},\textrm{GR}}(\eta) = c_1+c_2\ln\left(\frac{k\eta}{\sqrt{3}}\right) +6\mathcal{R}_k \left[\textrm{Ci}\left(\frac{k\eta}{\sqrt{3}}\right)-\frac{\sqrt{3}}{k\eta}\left(1-\frac{3}{k^2\eta^2}\right)\sin \left(\frac{k\eta}{\sqrt{3}}\right)-\frac{3}{k^2\eta^2}\cos\left(\frac{k\eta}{\sqrt{3}}\right)\right],
\end{equation}
$c_1$ and $c_2$ being constants of integration and $\textrm{Ci}(x)$ the cosine integral. Bearing in mind the constancy of $\delta_k^{\textrm{m},\textrm{GR}}$ on superhorizon scales (see Eq.~\eqref{eq:energy-conservation-matter} and Ref.~\cite{Baumann:2022mni}), we expand the expression above on these scales and obtain
\begin{equation}
    \delta_k^{\textrm{m},\textrm{GR}} \simeq c_1+6\mathcal{R}_k\left(\gamma-\frac{2}{3}\right)+\left(c_2+6\mathcal{R}_k\right)\ln\left(\frac{k\eta}{\sqrt{3}}\right), 
\end{equation}
$\gamma$ being the Euler-Mascheroni constant. Setting $c_2 = -6\mathcal{R}_k$ we remove the time dependence on scales beyond the horizon. In order to determine $c_1$, it can be shown that the constant value of $\delta_{k}^{\textrm{m},\textrm{GR}}$ is $-\mathcal{R}_k$ given the adiabaticity of matter perturbations and the well-known fact that the curvature perturbation on uniform-density and comoving slicings coincide on superhorizon scales in GR \cite{Baumann:2022mni,Weinberg:2003sw}. Hence $c_1 = -3(2\gamma-1) \mathcal{R}_k$. We then write
\begin{equation}
    \label{eq:sol-deltamGR-ini-cond}\delta_k^{\textrm{m},\textrm{GR}} (\eta) =-6\mathcal{R}_k\left[\textrm{Cin}\left(\frac{k\eta}{\sqrt{3}}\right)-\frac{1}{2}+\frac{\sqrt{3}}{k\eta}\left(1-\frac{3}{k^2\eta^2}\right)\sin\left(\frac{k\eta}{\sqrt{3}}\right)+\frac{3}{k^2\eta^2}\cos\left(\frac{k\eta}{\sqrt{3}}\right)\right],
\end{equation}
where
\begin{equation}
    \textrm{Cin}(x) \equiv \int^{x}_0 \frac{1-\cos(t)}{t} \textrm{d}t=\gamma+\ln(x) -\textrm{Ci}(x)~.
\end{equation}
In the vein of Eq.~\eqref{eq:separation-transfer-functions}, we introduce the transfer function of the GR density contrast as follows
\begin{equation}
    \delta_k^{\textrm{m},\textrm{GR}}\equiv -T_{\delta}(k\eta) \mathcal{R}_k~,
\end{equation}
such that
\begin{equation}
    T_{\delta}(k\eta) \equiv -\left[\textrm{Cin}\left(\frac{k\eta}{\sqrt{3}}\right)-\frac{1}{2}+\frac{\sqrt{3}}{k\eta}\left(1-\frac{3}{k^2\eta^2}\right)\sin\left(\frac{k\eta}{\sqrt{3}}\right)+\frac{3}{k^2\eta^2}\cos\left(\frac{k\eta}{\sqrt{3}}\right)\right],
\end{equation}
and thus the equation of $\tilde{\psi}^{\textrm{MG}}_k$ \eqref{eq:def-eq-psiMG} can be rewritten as one of transfer functions
\begin{equation}
    \label{eq:for-transfer-functions}T_{\tilde{\psi}}'' +\frac{4}{x}T_{\tilde{\psi}}' +\frac{1}{3}T_{\tilde{\psi}}=\frac{4\mathcal{C}k}{\mathcal{A}x^2}T_{\psi}'+\frac{\mathcal{D}k^2}{3x^4}\left(\frac{3}{2}T_{\delta}-T_{\psi}\right),
\end{equation}
where primes now denote derivatives with respect to $x\equiv k\eta$. The solution to this equation is
\begin{equation} \label{adsofubewarawer}
    T_{\tilde{\psi}}(x)=T_{\tilde{\psi}}^{{\rm p}}(x)+\int_{x_{\rm reh}}^{x\textbf{}}\text{d}u\,G(x,u)\frac{1}{u^2}\left[\frac{4\mathcal{C}k}{\mathcal{A}}T_{\psi}'(u)+\frac{\mathcal{D}k^2}{3u^2}\left(\frac{3}{2}T_{\delta}(u)-T_{\psi}(u)\right)\right],
\end{equation}
where $T_{\tilde{\psi}}^{{\rm p}}(x)$ is the solution to the homogeneous part of Eq. \eqref{eq:for-transfer-functions}. We also define $x_{\rm reh}\equiv k\eta_{\rm reh}$. $\eta_{\rm reh}$ (the start of the radiation dominated era) is taken to be the initial time, and we assume that all the corresponding modes are superhorizon. Since the homogeneous problem is the same as its GR counterpart, we obtain the same solution
\begin{equation} \label{asdfiuabwfewaeraew}
    T_{\tilde{\psi}}^{{\rm p}}(x) = \frac{9}{x^2}\left[\frac{\sqrt{3}}{x}\sin\left(\frac{x}{\sqrt{3}}\right)-\cos\left(\frac{x}{\sqrt{3}}\right)\right].
\end{equation}
Green's function $G(x,\tau)$ for this problem is given by 
\begin{equation}
    \mathcal{L}G(x,\tau)\equiv \left(\frac{\textrm{d}^2}{\textrm{d}x^2} +\frac{4}{x} \frac{\textrm{d}}{\textrm{d}x}+\frac{1}{3}\right)G(x,\tau)=\delta^{(1)}(x-\tau)~.
\end{equation}
When $x\neq \tau$, it satisfies the homogeneous problem $\mathcal{L}G=0$. This means it is given by a linear combination of two linearly independent solutions, $\{\tilde{\psi}_1,\tilde{\psi}_2\}$, which are given by
\begin{align}
    &\tilde{\psi}_1(x)=\frac{1}{x^2}\left[\frac{\sqrt{3}}{x}\cos\left(\frac{x}{\sqrt{3}}\right)+\sin\left(\frac{x}{\sqrt{3}}\right)\right],\nonumber\\
    &\tilde{\psi}_2(x)=\frac{1}{x^2}\left[\frac{\sqrt{3}}{x}\sin\left(\frac{x}{\sqrt{3}}\right)-\cos\left(\frac{x}{\sqrt{3}}\right)\right]. \label{asodfubauybwerwqer}
\end{align}
The Wronskian of these two solutions reads
\begin{equation}
    W[\tilde{\psi}_1,\tilde{\psi}_2]\equiv\tilde{\psi}_1\tilde{\psi}_2'-\tilde{\psi}_1'\tilde{\psi}_2=\frac{1}{\sqrt{3}x^4}~.
\end{equation}
Using Green's method, we arrive at
\begin{equation}
    G(x,\tau)=\sqrt{3}\left[\tau^4\tilde{\psi}_1(\tau)\tilde{\psi}_2(x)-\tau^4\tilde{\psi}_2(\tau)\tilde{\psi}_1(x)\right]\Theta(x-\tau)~,
\end{equation}
where we introduced the Heaviside step function since $G=0$ for $x<\tau$. The solution is obtained to be
\begin{align} 
    &T_{\tilde{\psi}}(x)=\frac{9}{x^2}\left[\frac{\sqrt{3}}{x}\sin\left(\frac{x}{\sqrt{3}}\right)-\cos\left(\frac{x}{\sqrt{3}}\right)\right]+\nonumber\\
    &+\frac{\tilde{\psi}_2(x)\mathcal{D}k^2}{\sqrt{3}}\int_{x_{\rm reh}}^{x}\text{d}\tau\,\tilde{\psi}_1(\tau)\left(\frac{3}{2}T_{\delta}(\tau)-T_{\psi}(\tau)\right)-\frac{\tilde{\psi}_1(x)\mathcal{D}k^2}{\sqrt{3}}\int_{x_{\rm reh}}^{x}\text{d}\tau\,\tilde{\psi}_2(\tau)\left(\frac{3}{2}T_{\delta}(\tau)-T_{\psi}(\tau)\right)+\nonumber \\
    &+\frac{\tilde{\psi}_2(x)4\sqrt{3}\mathcal{C}k}{\mathcal{A}}\int_{x_{\rm reh}}^x \text{d}\tau \,\tau^2 \tilde{\psi}_1(\tau)T^{'}_{\psi}(\tau)-\frac{\tilde{\psi}_1(x)4\sqrt{3}\mathcal{C}k}{\mathcal{A}}\int_{x_{\rm reh}}^x \text{d}\tau \,\tau^2 \tilde{\psi}_2(\tau)T^{'}_{\psi}(\tau)~,  \label{aisdbfiuwaerawer}
\end{align}
where the integration range was simplified using the Heaviside function. 

\begin{figure}[h]
     \centering
     \begin{subfigure}[b]{0.49\textwidth}
         \centering
         \includegraphics[width=\textwidth]{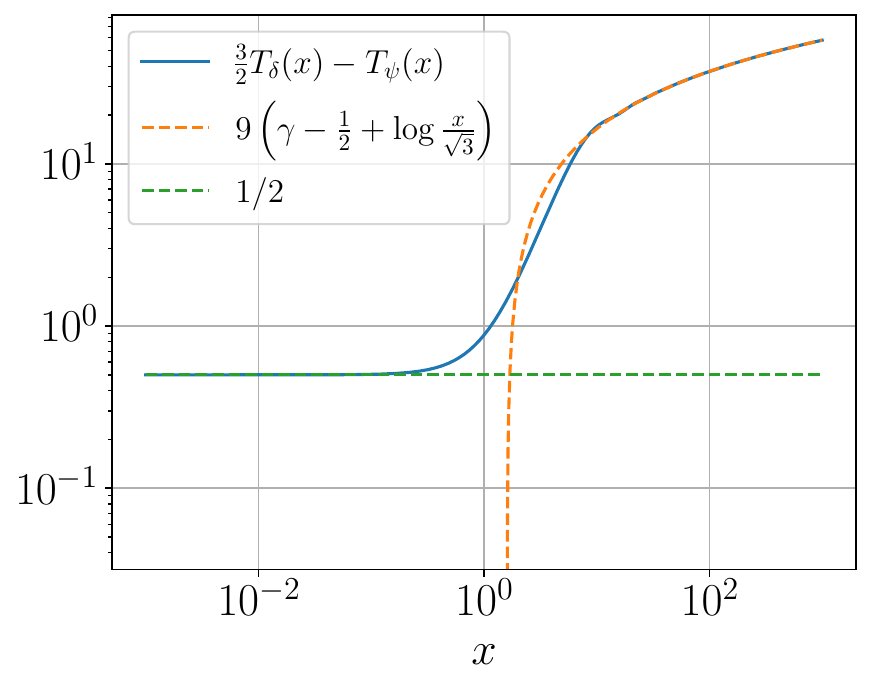}
     \end{subfigure}
     \begin{subfigure}[b]{0.49\textwidth}
         \centering
         \includegraphics[width=\textwidth]{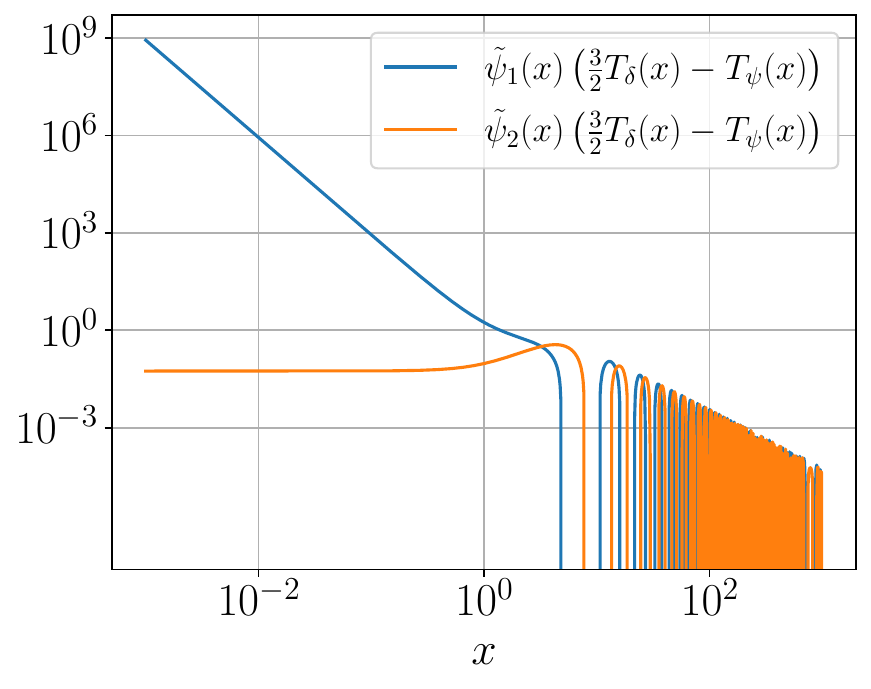}
     \end{subfigure}
     \caption{Left: Source term (blue) with its subhorizon (dashed green) and superhorizon (dashed orange) limits. Right: Integrands of the first (blue) and second (orange) terms in the solution for the transfer function of $\tilde{\psi}^{\textrm{MG}}_k$.}
     \label{fig:limits_and_integrands}
\end{figure}

The terms proportional to $\mathcal{D}$ lead to different behaviour than those proportional to $\mathcal{C}$. In what follows, we shall study their effect separately, identifying two different regimes. Let us start by assuming the $\mathcal{D}$ terms dominate over the $\mathcal{C}$ terms; that is, we consider the solution 
\begin{align} 
    &T_{\tilde{\psi}}(x)=\frac{9}{x^2}\left[\frac{\sqrt{3}}{x}\sin\left(\frac{x}{\sqrt{3}}\right)-\cos\left(\frac{x}{\sqrt{3}}\right)\right]+\nonumber\\
    &+\frac{\tilde{\psi}_2(x)\mathcal{D}k^2}{\sqrt{3}}\int_{x_{\rm reh}}^{x}\text{d}\tau\,\tilde{\psi}_1(\tau)\left(\frac{3}{2}T_{\delta}(\tau)-T_{\psi}(\tau)\right)-\frac{\tilde{\psi}_1(x)\mathcal{D}k^2}{\sqrt{3}}\int_{x_{\rm reh}}^{x}\text{d}\tau\,\tilde{\psi}_2(\tau)\left(\frac{3}{2}T_{\delta}(\tau)-T_{\psi}(\tau)\right). \label{asidfbawawerawer}
\end{align}
These are complicated integrals that cannot be solved with full generality. However, we may look into the limits of the integrands more closely. We begin with the superhorizon limit, which gives
\begin{equation}
    \lim_{x\to 0}\left(\frac{3}{2}T_{\delta}(x)-T_{\psi}(x)\right)=\frac{1}{2}~,
\end{equation}
and, from Eq. \eqref{asodfubauybwerwqer}, 
\begin{align}
    &\tilde{\psi}_1(x)\left(\frac{3}{2}T_{\delta}(x)-T_{\psi}(x)\right)\xrightarrow[ x\to 0 ]{}\frac{\sqrt{3}}{2x^3}~,\nonumber\\
    &\tilde{\psi}_2(x)\left(\frac{3}{2}T_{\delta}(x)-T_{\psi}(x)\right)\xrightarrow[ x\to 0 ]{}\frac{1}{18}~.
\end{align}
Regarding the subhorizon limit, we find
\begin{equation}
   \frac{3}{2}T_{\delta}(x)-T_{\psi}(x)\xrightarrow[ x\to \infty ]{}9\left[\gamma -\frac{1}{2}+\log\left(\frac{x}{\sqrt{3}}\right)\right].
\end{equation}
Although this has logarithmic growth, $\tilde{\psi}_1(x)$ and $\tilde{\psi}_2(x)$ are oscillatory functions which decay as $\sim 1/x^2$ for large $x$, yielding 
\begin{align}
    &\tilde{\psi}_1(x)\left(\frac{3}{2}T_{\delta}(x)-T_{\psi}(x)\right)\xrightarrow[ x\to \infty ]{}\frac{9}{x^2}\sin{\left(\frac{x}{\sqrt{3}}\right)}\left[\gamma -\frac{1}{2}+\log\left(\frac{x}{\sqrt{3}}\right)\right],\nonumber\\
    &\tilde{\psi}_2(x)\left(\frac{3}{2}T_{\delta}(x)-T_{\psi}(x)\right)\xrightarrow[ x\to \infty ]{}-\frac{9}{x^2}\cos{\left(\frac{x}{\sqrt{3}}\right)}\left[\gamma -\frac{1}{2}+\log\left(\frac{x}{\sqrt{3}}\right)\right],
\end{align}
which means that their contribution to the integral becomes negligible very quick. We show the integrands, without any approximation, in the right panel of Fig.~\ref{fig:limits_and_integrands} and the different limits of $\frac{3}{2}T_{\delta}(x)-T_{\psi}(x)$ in the left panel of Fig.~\ref{fig:limits_and_integrands}.

It can be seen that the superhorizon limit $\frac{3}{2}T_{\delta}(x)-T_{\psi}(x)=1/2$ is an excellent approximation in the integrals of Eq. \eqref{asidfbawawerawer}. This allows to obtain analytical expressions, which read
\begin{equation} \label{asodufbinawerfwae}
    J_1(x)\equiv\int_{x_{\rm reh}}^x\text{d}\tau\frac{\tilde{\psi}_1(\tau)}{2}=\frac{1}{12} \left[ \sqrt{3} \, \mathrm{Ci}\left(\frac{x}{\sqrt{3}}\right) -3x\tilde{\psi}_1(x) \right]-\frac{1}{12} \left[ \sqrt{3} \, \mathrm{Ci}\left(\frac{x_{\rm reh}}{\sqrt{3}}\right) -3x\tilde{\psi}_1(x_{\rm reh})\right],
\end{equation}
\begin{equation}
    J_2(x)\equiv\int_{x_{\rm reh}}^x\text{d}\tau\frac{\tilde{\psi}_2(\tau)}{2}=\frac{1}{12} \left[ \sqrt{3}\,\mathrm{Si}\left(\frac{x}{\sqrt{3}}\right)- 3x\tilde{\psi}_2(x)\right]-\frac{1}{12} \left[\sqrt{3}\,\mathrm{Si}\left(\frac{x_{\rm reh}}{\sqrt{3}}\right) - 3x\tilde{\psi}_2(x_{\rm reh})\right],
\end{equation}
where $\mathrm{Si}(x)$ is the sine integral. Consequently, the analytical solution for the transfer function of $\tilde{\psi}_k^{\rm MG}$ reads
\begin{equation} \label{fasidbfaidsfasdf}
    T_{\tilde{\psi}}(k,x)=\tilde{\psi}_2(x)\left(9+\frac{\mathcal{D}k^2}{\sqrt{3}}J_1(x)\right)-\tilde{\psi}_1(x)\frac{\mathcal{D}k^2}{\sqrt{3}}J_2(x)~.
\end{equation}
Note that the term $9\tilde{\psi}_2(x)$ corresponds to the homogeneous solution. Therefore, when $\mathcal{D}k^2J_2(x)/\sqrt{3}\gg 9$, the solution will deviate from the standard GR behaviour. 
\begin{figure}[h]
     \centering
     \begin{subfigure}[b]{0.49\textwidth}
         \centering
         \includegraphics[width=\textwidth]{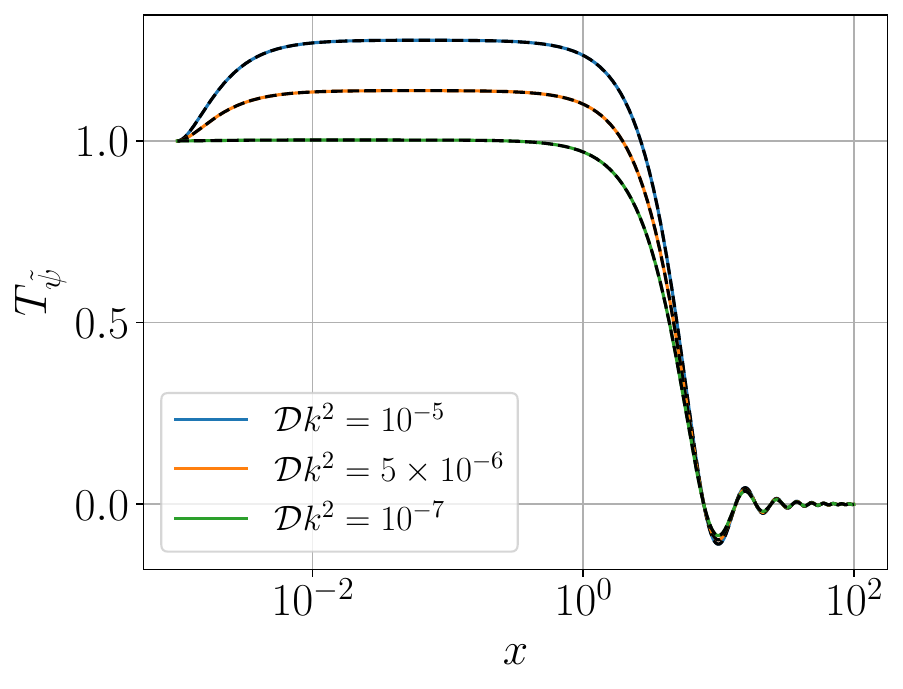}
     \end{subfigure}
     \begin{subfigure}[b]{0.49\textwidth}
         \centering
         \includegraphics[width=\textwidth]{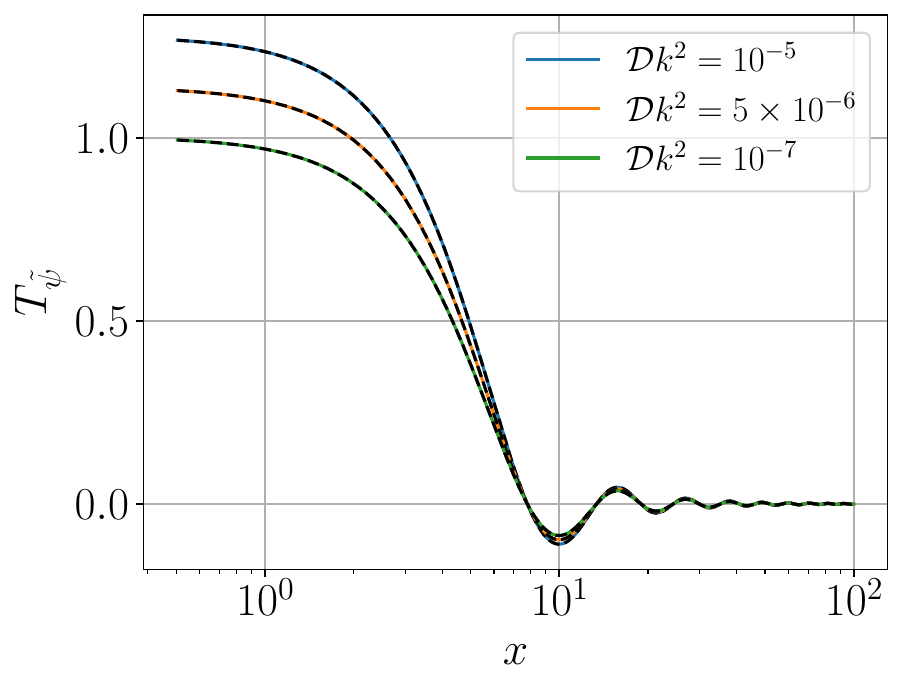}
     \end{subfigure}
     \caption{Left: Transfer function of $\tilde{\psi}_k^{\rm MG}$ for different values of $\mathcal{D}k^2$ in the parameter range close to the particular solution being negligible. The initial integration time is set to $x_{\rm reh}=10^{-3}$. We show the full numerical solution, without any approximations, in solid blue, orange, and green. For the green curve, the particular solution is negligible, meaning that the full solution is given by the homogeneous solution. In dashed black we show the analytical solutions corresponding to each numerical solution.  Right: A zoom-in of the left panel after horizon crossing.}
     \label{fig:transfer1}
\end{figure}

\begin{figure}[h]
     \centering
     \begin{subfigure}[b]{0.49\textwidth}
         \centering
         \includegraphics[width=\textwidth]{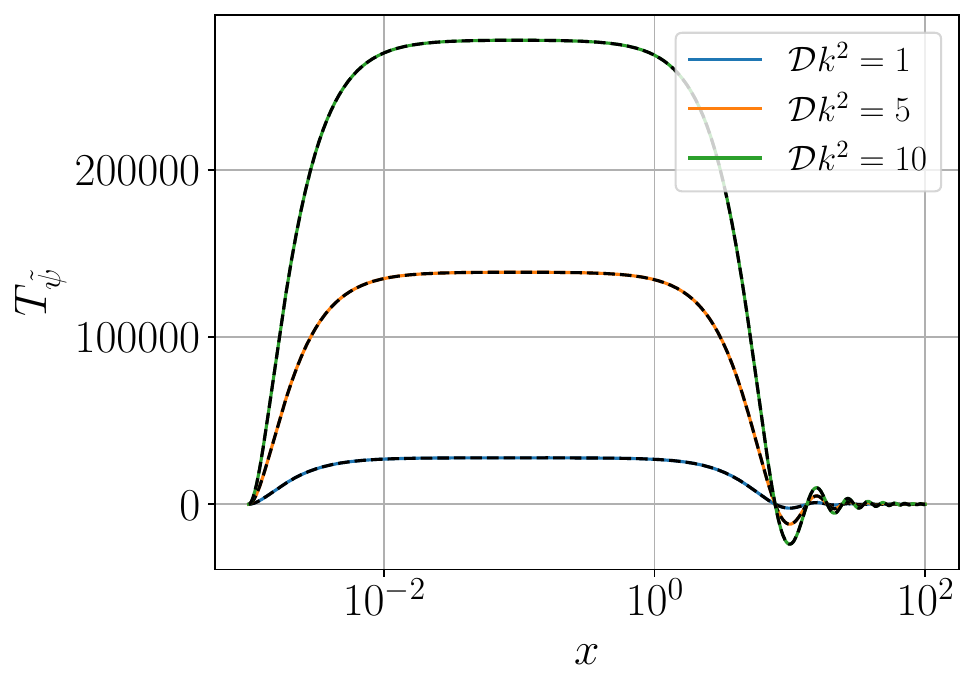}
     \end{subfigure}
     \begin{subfigure}[b]{0.49\textwidth}
         \centering
         \includegraphics[width=\textwidth]{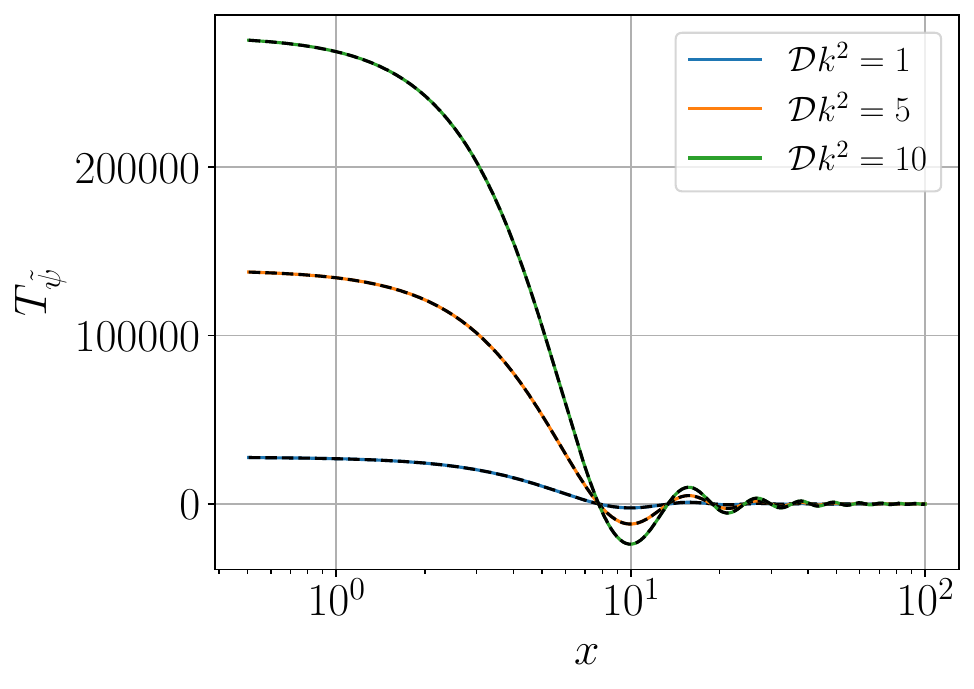}
     \end{subfigure}
     \caption{Left: Transfer function of $\tilde{\psi}_k^{\rm MG}$ for different values of $\mathcal{D}k^2$ in the parameter range where the homogeneous solution is negligible. The initial integration time is set to $x_{\rm reh}=10^{-3}$. We show the full numerical solution, without any approximations, in solid blue, orange, and green. In dashed black we show the analytical solutions corresponding to each numerical solution.  Right: A zoom-in of the left panel after horizon crossing.}
     \label{fig:transfer2}
\end{figure}

In Figs.~\ref{fig:transfer1}-\ref{fig:transfer2} we plot the numerical solution to Eq.~\eqref{adsofubewarawer} in addition to its analytical counterpart \eqref{fasidbfaidsfasdf} for different values of $\mathcal{D}k^2$. The agreement between both is excellent. In Fig. \ref{fig:transfer1} we show the behaviour of the transfer function in the regime where $\mathcal{D}k^2J_2(x)/\sqrt{3}\sim 9$; namely, when the magnitude of the homogeneous and particular solutions is comparable. For the green curve in that figure, we show the evolution when the particular solution is negligible. In Fig.~\ref{fig:transfer2} we find that for larger values of $\mathcal{D}k^2$ the particular solution fully dominates and the transfer function is largely amplified.

One might think that the perturbative expansion defined in Eq. \eqref{eq:psiseparation} does not hold for such a large amplification in $\tilde{\psi}^{\rm MG}$. However, the reader should bear in mind that it is $\beta\tilde{\psi}^{\rm MG}$ the quantity that should be small when compared to $\psi^{\textrm{GR}}$. Using Eq. \eqref{aosdfbiawner}, we find that as long as 
\begin{equation} \label{asdoifbahwefrwerwer}
    \alpha T_{\tilde{\psi}}<3.05\cross 10^{120}~,
\end{equation}
our perturbative expansion shall hold. We may translate this into a concrete bound on $\alpha$ for realistic temperatures of reheating, such as the one around the GUT scale. Using Eqs.~\eqref{asdofjubaiwerawer}, and \eqref{aosdfbaiwenwe} and \eqref{adoufbanwefrwer}, we may replace $T_{\tilde{\psi}}$ with $T_{\psi}$ and $\eta_{\textrm{reh}}$ with $T_{\textrm{reh}}$ in the corresponding prefactor in Eq.~\eqref{asdoifbahwefrwerwer}, such that
\begin{equation}
    \alpha\left(\frac{T_{\textrm{reh}}}{\m}\right)^2T_{\psi} < 1.22 \cross 10^{55}~.
\end{equation}
This condition must be satisfied by all modes, particularly by those that are superhorizon, where $T_{\psi} \rightarrow 1$ (see Eq.~\eqref{aoisdbfhafwae}). Setting the temperature of reheating to be the one at GUT ($T_{\textrm{reh}} \sim 10^{15}\textrm{GeV}$), we obtain
\begin{equation}
    \label{eq:bound_on_alpha_GUT}\alpha < 6.09 \times 10^{61}~.
\end{equation}
For larger values of $\alpha$ one needs to include higher-order terms which may translate into backreaction of the modified gravity terms on the GR evolution equations. We notice however that, at first order, the modified gravity corrections to the (conformal) Hubble parameter in Eq.~\eqref{oaisdbfiabsdfawe} go like $1/\eta^2$ at most compared to the GR prescription, which is $1/\eta$, meaning that they become weaker at background level as the radiation dominated era progresses (see Ref.~\cite{Kugarajh:2025rbt} for similar assertions in the metric formalism). We then expect the backreacting terms altering the GR evolution to decay at similar or even faster rate.

As with previous perturbation quantities, we may split the source term in Eq.~\eqref{eq:finalsourceterm}
\begin{equation}
    \mathcal{S}_{ij} =-2\partial_i\tilde{\psi}\partial_j\tilde{\psi}-\frac{4\m^2F^{(0)}\tilde{\mathcal{H}}^2}{a^2\rho(1+w)}\partial_i\left(\tilde{\psi}+\frac{\tilde{\psi}'}{\tilde{\mathcal{H}}}\right)\partial_j\left(\tilde{\psi}+\frac{\tilde{\psi}'}{\tilde{\mathcal{H}}}\right),
\end{equation}
as
\begin{equation}
    \mathcal{S}_{ij}=\mathcal{S}^{\rm GR}_{ij}+\beta\mathcal{S}^{\rm MG}_{ij}.
\end{equation}
Expanding to linear order in $\beta$, we see that, from Eq.~\eqref{eq:homogeneous-00}, the radiation energy density dominates the RHS of the analogue of Friedmann's equation
\begin{equation}
    3\tilde{\mathcal{H}}^2F^{(0)}=\frac{a^2\rho}{\m^2}+\frac{\alpha a^2\rho_{\rm m}^2}{4\m^6}\simeq \frac{a^2\rho_{\textrm{r}}}{\m^2}~,
\end{equation}
and hence
\begin{equation}
    \frac{4\m^2F^{(0)}\tilde{\mathcal{H}}^2}{a^2\rho(1+w)}\simeq \frac{4}{3(1+w)}~.
\end{equation}
Also, $\tilde{\mathcal{H}} \simeq \mathcal{H}_{\textrm{GR}}(1-3\beta w_{\rm m}/(2a^3))=\mathcal{H}_{\textrm{GR}}$, since $w_{\rm m}=0$. Consequently, $\mathcal{S}^{\rm MG}_{ij}$ shall be given by
\begin{align}
    &\mathcal{S}^{\rm MG}_{ij}=-2\partial_i\psi_{\textrm{GR}}\partial_j\tilde{\psi}_{\textrm{MG}}-2\partial_i\tilde{\psi}_{\textrm{MG}}\partial_j\psi_{\textrm{GR}}-\frac{4}{3(1+w)}\partial_i\left(\tilde{\psi}_{\textrm{MG}}+\frac{\tilde{\psi}'_{\textrm{MG}}}{\mathcal{H}_{\rm GR}}\right)\partial_j\left(\psi_{\textrm{GR}}+\frac{\psi'_{\textrm{GR}}}{\mathcal{H}_{\rm GR}}\right)-\nonumber\\
    \label{eq:S-mod-grav}&-\frac{4}{3(1+w)}\partial_i\left(\psi_{\textrm{GR}}+\frac{\psi'_{\textrm{GR}}}{\mathcal{H}_{\rm GR}}\right)\partial_j\left(\tilde{\psi}_{\textrm{MG}}+\frac{\tilde{\psi}'_{\textrm{MG}}}{\mathcal{H}_{\rm GR}}\right).
\end{align}
It is worth noting that this expression for the source is valid regardless of whether $\mathcal{C}$ dominates, which shall be considered in the next section. 

\subsection{The $\mathcal{C}$-Dominated Regime} \label{sec:3.3}
In this subsection we analyse the regime in which the terms proportional to $\mathcal{C}$ dominate the solution in Eq. \eqref{aisdbfiuwaerawer}; namely, we consider the solution
\begin{align} 
    &T_{\tilde{\psi}}(x)=\frac{9}{x^2}\left[\frac{\sqrt{3}}{x}\sin\left(\frac{x}{\sqrt{3}}\right)-\cos\left(\frac{x}{\sqrt{3}}\right)\right]+\nonumber\\
    &+\frac{\tilde{\psi}_2(x)4\sqrt{3}\mathcal{C}k}{\mathcal{A}}\int_{x_{\rm reh}}^x \text{d}\tau \,\tau^2 \tilde{\psi}_1(\tau)T^{'}_{\psi}(\tau)-\frac{\tilde{\psi}_1(x)4\sqrt{3}\mathcal{C}k}{\mathcal{A}}\int_{x_{\rm reh}}^x \text{d}\tau \,\tau^2 \tilde{\psi}_2(\tau)T^{'}_{\psi}(\tau)~.
\end{align}
These integrals have simple analytical solutions. They read
\begin{equation}
    K_1(x)\equiv \int\text{d}x \,x^2 \tilde{\psi}_1(x)T^{'}_{\psi}(x)=\frac{9}{4}\left[-\frac{2}{\sqrt{3}x}-\frac{3\sqrt{3}\cos{\left(\frac{2x}{\sqrt{3}}\right)}}{x^3}+\frac{9\sin{\left(\frac{2x}{\sqrt{3}}\right)}}{2x^4}-\frac{\sin{\left(\frac{2x}{\sqrt{3}}\right)}}{x^2}\right],
\end{equation}
\begin{equation}
    K_2(x)\equiv\int\text{d}x \,x^2 \tilde{\psi}_2(x)T^{'}_{\psi}(x)=\frac{9}{4}\left[\frac{9}{2x^4}+\frac{2}{x^2}-\frac{9\cos{\left(\frac{2x}{\sqrt{3}}\right)}}{2x^4}+\frac{\cos{\left(\frac{2x}{\sqrt{3}}\right)}}{x^2}-\frac{3\sqrt{3}\sin{\left(\frac{2x}{\sqrt{3}}\right)}}{x^3}\right].
\end{equation}
The superhorizon limit of these two integrals reads
\begin{equation}
    \lim_{x\to 0}K_1(x)=-\frac{x}{5\sqrt{3}}+\mathcal{O}(x^2)=0~,
\end{equation}
\begin{equation}
    \lim_{x\to 0}K_2(x)=3/4+\mathcal{O}(x^2)~.
\end{equation}
Following our previous assumption that all modes are superhorizon at the beginning of the radiation-dominated era, we use $x_{\rm reh}\ll 1$. Therefore 
\begin{align}
    &T_{\tilde{\psi}}(x) = T_{\tilde{\psi}}^{{\rm p}}(x)+\frac{4\sqrt{3}\mathcal{C}k}{\mathcal{A}}\left[\tilde{\psi}_2(x)K_1(x) - \tilde{\psi}_1(x)K_2(x) + \frac{3}{4}\tilde{\psi}_1(x)\right]=\nonumber \\
    &=T_{\tilde{\psi}}^{{\rm p}}(x)+ \frac{4\sqrt{3}\mathcal{C}k}{\mathcal{A}}\left[\frac{3}{4}\tilde{\psi}_1(x)-\frac{9}{2\sqrt{3}x}\tilde{\psi}_2(x)-\frac{9}{4x^4}\sin{\left(\frac{x}{\sqrt{3}}\right)}\right] = T_{\tilde{\psi}}^{{\rm p}}(x)+\frac{\mathcal{C}k}{\mathcal{A}}T_{\psi}'(x)~,\label{asjdfbafasdf}
\end{align}
where we have identified that the term in brackets is $T_{\psi}'(x)/(4\sqrt{3})$. If the $\mathcal{C}$ terms dominate over the $\mathcal{D}$ ones, they also take over the homogeneous solution (see below), yielding
\begin{equation}
    T_{\tilde{\psi}}(k\eta) = \frac{\mathcal{C}k}{\mathcal{A}}T_{\psi}'(k\eta)~.
\end{equation}
\begin{figure}[h]
     \centering
     \begin{subfigure}[b]{0.49\textwidth}
         \centering
         \includegraphics[width=\textwidth]{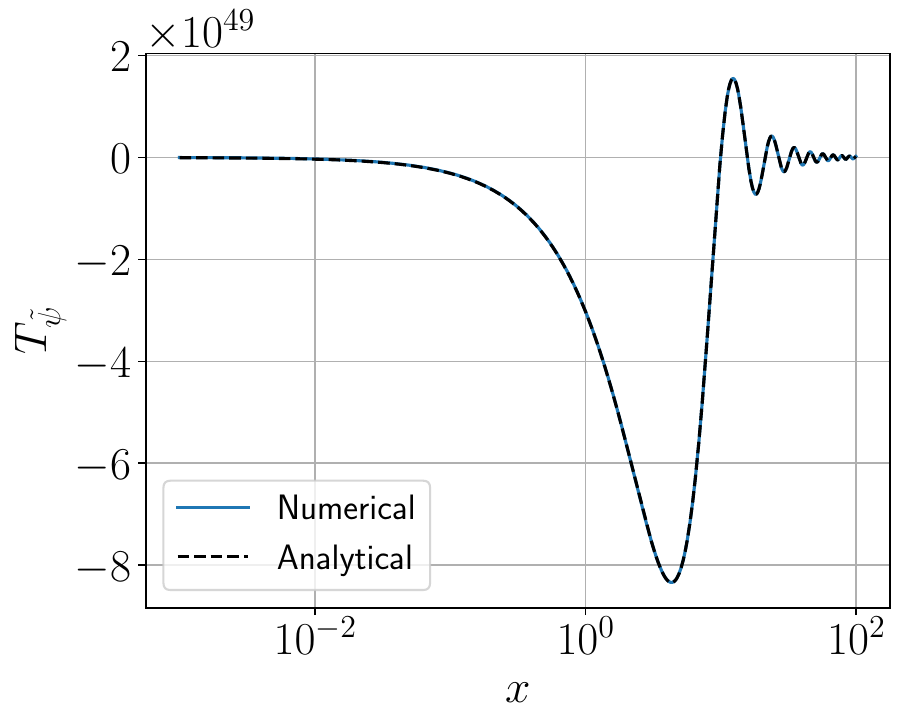}
     \end{subfigure}
     \begin{subfigure}[b]{0.49\textwidth}
         \centering
         \includegraphics[width=\textwidth]{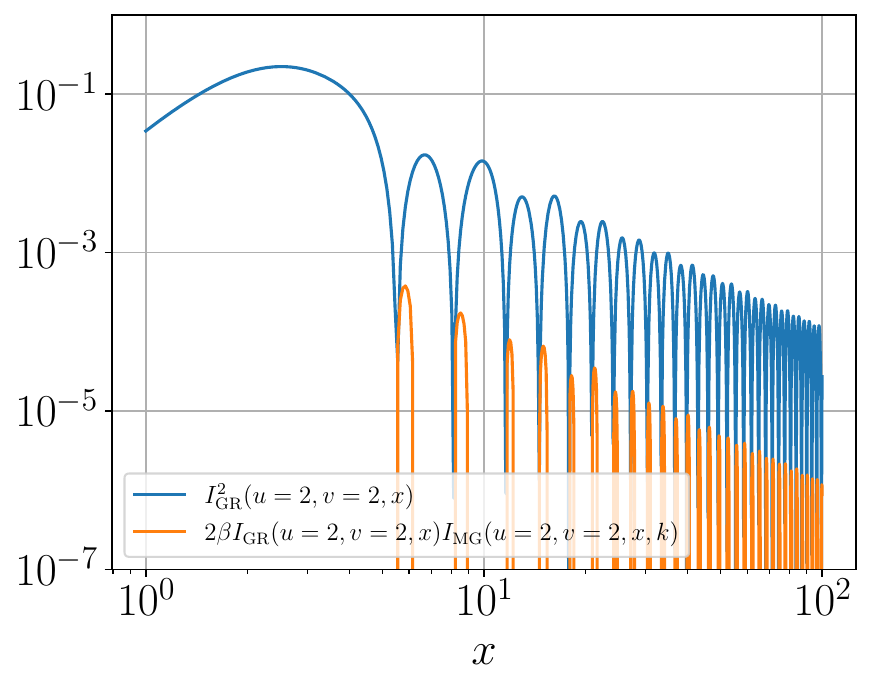}
     \end{subfigure}
     \caption{Left: Numerical (full blue) and analytical (dashed black) solutions for the modified gravity transfer function. Right: GR kernel (blue) and the modified gravity contribution (dashed orange), for $u=v=2$ and $\alpha = 10^{67}$. In both panels we use $k=10^{17}\text{Mpc}^{-1}$ and $\mathcal{C}=10^{30}$, so that the terms proportional to $\mathcal{C}$ dominate in the modified gravity transfer function.}
     \label{aisdfbaijndfawef}
\end{figure}
Due to the additional scale dependence, we must be careful to not violate the perturbativity condition \eqref{asdoifbahwefrwerwer}, which, assuming $T_{\psi}'(k\eta)\sim\mathcal{O}(1)$, gives
\begin{equation}
    \mathcal{C}\lesssim \frac{\mathcal{A}}{\beta k_{\rm max}}\sim \frac{10^{90}\m}{\alpha T_{\rm reh}}~,
\end{equation}
where we have chosen a maximum cutoff wavenumber $k_{\rm max}$ corresponding to the end of inflation and used Eqs. \eqref{aosdfbiawner}, \eqref{eq:Adefinition}, \eqref{adoufbanwefrwer}, assuming instantaneous reheating. The required value of $\mathcal{C}$ is roughly 
\begin{equation}
    \mathcal{C}>\mathcal{D}\mathcal{A}k\sim 10^{25}~,
\end{equation}
resorting to a typical $k=10^{17}\text{Mpc}^{-1}$ of interest. For example, for $\alpha =10^{60}$, $T_{\rm reh}=10^{-4}\m$ and $\mathcal{C}=10^{30}$, the terms proportional to $\mathcal{C}$ dominate while also remaining in the perturbativity regime. The condition in Eq. \eqref{eq:conditionCfirst} is also satisfied. In Fig. \ref{aisdfbaijndfawef} we show the solution \eqref{asjdfbafasdf} and the numerical solution to the corresponding differential equation. 

\subsection{Density Spectrum of Scalar-Induced Gravitational Waves} \label{sec:3.4}
In this section, the most observationally relevant quantities in the study of the SIGWs will be considered. We turn our attention to the spectral energy density parameter per logarithmic wavelength interval, which is defined as 
\begin{equation} \label{asd9fubnqerwer}
    \Omega_{\rm GW}(k,\eta)\equiv \frac{1}{\rho_{\rm c}(\eta)}\frac{\text{d}\rho_{\rm GW}}{\text{d}\log(k)}~,
\end{equation}
where $\rho_{\rm c}$ is the critical energy density. Note that, in view of Eq.~\eqref{eq:homogeneous-00}, one needs to account for the modified gravity corrections to determine this density. However, these shall be negligible given that we are essentially interested in evaluating Eq.~\eqref{asd9fubnqerwer} at the present time. In what follows, we therefore use $\rho_{\rm c}=3\m^2 a^{-2}\mathcal{H}^2$. 

The energy density on subhorizon scales can be written as\footnote{One may notice that the homogeneous part of the evolution equation of the second-order tensor modes, Eq.~\eqref{eq:evolution-second-order-modes-tensor}, is identical to that of GR, except for the modified friction term. In order to calculate the energy-momentum tensor of these modes, we ignore this modified gravity correction for the same reason as in the case of the critical energy density.} \cite{Maggiore:1999vm}
\begin{equation} \label{aoisdbfhqwe}
    \rho_{\rm GW}=\frac{\m^2}{16 a^2(\eta)}\langle\overline{\partial_k h_{ij}\partial^k h^{ij}}\rangle~,
\end{equation}
where the overbar denotes time averaging and the spatial indices are contracted with Kronecker's delta. We then write the Fourier decomposition of the tensor modes as
\begin{equation} \label{aosdbfbiqwer}
    h_{ij}(\eta,\textbf{x})=\sum_{s=+,\cross}\int\frac{\text{d}^3\textbf{k}}{(2\pi)^3}e_{ij}^s(\textbf{k}) h^s_{\textbf{k}}(\eta)e^{i \textbf{k}\cdot\textbf{x}}~,
\end{equation}
where $s=+,\cross$ denotes the polarisations and $e_{ij}^s(\textbf{k})$ are the usual polarisation tensors, defined in terms of an orthonormal basis $\{e_i,\bar{e}_j\}$ of vectors orthogonal to $\textbf{k}$ as $e_{ij}^{+}(\textbf{k})=(e_i(\textbf{k})e_j(\textbf{k})-\bar{e}_i(\textbf{k})\bar{e}_j(\textbf{k}))/\sqrt{2}$ and $e_{ij}^{\cross}(\textbf{k})=(e_i(\textbf{k})\bar{e}_j(\textbf{k})+\bar{e}_i(\textbf{k})e_j(\textbf{k}))/\sqrt{2}$. These are transverse ($k^{i}e^s_{ij}(\textbf{k})=0$), traceless ($\sum_i e^s_{ii}=0$) and satisfy the normalisation condition 
\begin{equation} \label{oajsdbfinawer}
    \sum_{i,j}e^s_{ij}(\textbf{k})e^{s'}_{ij}(\textbf{k})=\delta^{s,s'}~.
\end{equation}
Lastly, imposing that the tensor modes $h_{ij}(\eta,\textbf{x})$ are real implies that $e_{ij}^s(-\textbf{k})=e_{ij}^s(\textbf{k})$ and $(h^s_{\textbf{k}}){}^{*}=h^s_{-\textbf{k}}$.

Plugging Eq.~\eqref{aosdbfbiqwer} in Eq.~\eqref{aoisdbfhqwe} and using the properties of the polarisation tensors gives
\begin{equation}
    \rho_{\rm GW}=-\frac{\m^2}{16 a^2(\eta)}\sum_{s,s'=+,\cross}\int \frac{\text{d}^3\textbf{k}\text{d}^3\textbf{q}}{(2\pi)^6}e_{ij}^s(\textbf{k})e_{ij}^{s'}(\textbf{q})\langle\overline{h_\textbf{k}^s(\eta)h_{\textbf{q}}^{s'}(\eta)}\rangle k_m q_m e^{i(\textbf{k}+\textbf{q})\cdot\textbf{x}}~.
\end{equation}
By introducing the GW dimensionless power spectrum
\begin{equation}
    \label{eq:relation-dimensionless-powerspectrum}\langle h_{\textbf{k}}^s(\eta)h_{\textbf{k}'}^{s'}(\eta)\rangle=\delta^{ss'}(2\pi)^3\delta^{(3)}(\textbf{k}+\textbf{k}')\frac{2\pi^2}{k^3}\Delta_{h}^2(\eta,k)~,
\end{equation}
plugging everything back in Eq.~\eqref{asd9fubnqerwer} and keeping Eq.~\eqref{oajsdbfinawer} in mind, we find
\begin{equation} \label{asdoifbwqerqwewer}
    \Omega_{\rm GW}(k,\eta)=\frac{1}{24}\left(\frac{k}{\mathcal{H}(\eta)}\right)^2\overline{\Delta_h^2(\eta,k)}~,
\end{equation}
where we have summed over both polarisations, since they contribute equally. The subsequent calculation reduces to obtaining the GW dimensionless power spectrum, and hence we must determine the correlator 
\begin{equation}
    \sum_{s=+,\cross}\langle h_{\textbf{k}}^s(\eta)h_{\textbf{q}}^s(\eta)\rangle~,
\end{equation}
where it can be noted that, in order to obtain the power spectrum, we need to sum over all possible polarisations. All the statistical information is encoded in the initial values $\psi_{\textbf{k},i}$, related to the inflationary comoving curvature perturbation. This means that we will be able to express the power spectrum of the SIGWs as a function of the inflationary scalar power spectrum.

\subsubsection{Power Spectrum of Scalar-Induced Gravitational Waves} \label{sec:3.4.1}
We first write Eq.~\eqref{eq:evolution-second-order-modes-tensor} in Fourier space
\begin{equation} 
    \label{eq:second-order-tensor-inhomogeneous}(h_{\textbf{k}}^s){}''+2\mathcal{H}(h_{\textbf{k}}^s){}'+k^2h_{\textbf{k}}^s=4\int\frac{\text{d}^3\textbf{q}}{(2\pi)^3}e^{s}_{jk}(\textbf{k})q^j q^k\tilde{\psi}_{\textbf{q},i}\tilde{\psi}_{\textbf{k}-\textbf{q},i}f(\eta,q,\abs{\textbf{k}-\textbf{q}})~,
\end{equation}
where we neglect the anisotropic stress tensor (as indicated below the aforementioned equation) as well as the modified gravity correction to the Hubble friction term (in line with previous considerations). $\tilde{\psi}_{\mathbf{k},i}$ denotes the initial value of the perturbation $\tilde{\psi}_{\mathbf{k}}$. The minus sign is absorbed into the source term. Also  
\begin{equation}
    f(\eta,q,\abs{\textbf{k}-\textbf{q}})\equiv f_{\rm GR}(\eta,q,\abs{\textbf{k}-\textbf{q}})+\beta f_{\rm MG}(\eta,q,\abs{\textbf{k}-\textbf{q}})~,
\end{equation}
$f$ being the source function which we split into the GR one and the modified gravity corrections. The part associated with GR is the standard one  
\begin{align}
    &f_{\rm GR}(\eta,q,\abs{\textbf{k}-\textbf{q}})=2T_{\psi}(q\eta)T_{\psi}(\abs{\textbf{k}-\textbf{q}}\eta)+\nonumber\\
    &+\frac{4}{3(1+w)}\left(T_{\psi}(q\eta)+\frac{T'_{\psi}(q\eta)}{\mathcal{H}_{\textrm{GR}}}\right)\left(T_{\psi}(\abs{\textbf{k}-\textbf{q}}\eta)+\frac{T'_{\psi}(\abs{\textbf{k}-\textbf{q}}\eta)}{\mathcal{H}_{\textrm{GR}}}\right),
\end{align}
while the modified gravity part reads (see Eq.~\eqref{eq:S-mod-grav})
\begin{align}
    &f_{\rm MG}(\eta,q,\abs{\textbf{k}-\textbf{q}})=2T_{\psi}(q\eta)T_{\tilde{\psi}}(\abs{\textbf{k}-\textbf{q}}\eta)+2T_{\psi}(\abs{\textbf{k}-\textbf{q}}\eta)T_{\tilde{\psi}}(q\eta)+\nonumber\\
    &+\frac{4}{3(1+w)}\left(T_{\psi}(q\eta)+\frac{T'_{\psi}(q\eta)}{\mathcal{H}_{\rm GR}}\right)\left(T_{\tilde{\psi}}(\abs{\textbf{k}-\textbf{q}}\eta)+\frac{T'_{\tilde{\psi}}(\abs{\textbf{k}-\textbf{q}}\eta)}{\mathcal{H}_{\rm GR}}\right)+\nonumber \\
    &+\frac{4}{3(1+w)}\left(T_{\psi}(\abs{\textbf{k}-\textbf{q}}\eta)+\frac{T'_{\psi}(\abs{\textbf{k}-\textbf{q}}\eta)}{\mathcal{H}_{\rm GR}}\right)\left(T_{\tilde{\psi}}(q\eta)+\frac{T'_{\tilde{\psi}}(q\eta)}{\mathcal{H}_{\rm GR}}\right).
\end{align}

We now take a closer look at the solution for the transfer function of $\tilde{\psi}_k^{\rm MG}$ in Eq. \eqref{fasidbfaidsfasdf}. Given the value of $\mathcal{D}$, for the range of wavenumbers that is observationally interesting; namely, $k\sim 10^{7}\text{Mpc}^{-1}$~-~$10^{17}\text{Mpc}^{-1}$, the particular solution dominates over the homogeneous one yielding
\begin{equation} \label{aosdfnaingfawre}
        T_{\tilde{\psi}}(k,x)=\frac{\mathcal{D}k^2}{\sqrt{3}}\left(\tilde{\psi}_2(x)J_1(x)-\tilde{\psi}_1(x)J_2(x)\right).
\end{equation}
\begin{figure}[h]
     \centering
     \begin{subfigure}[b]{0.49\textwidth}
         \centering
         \includegraphics[width=\textwidth]{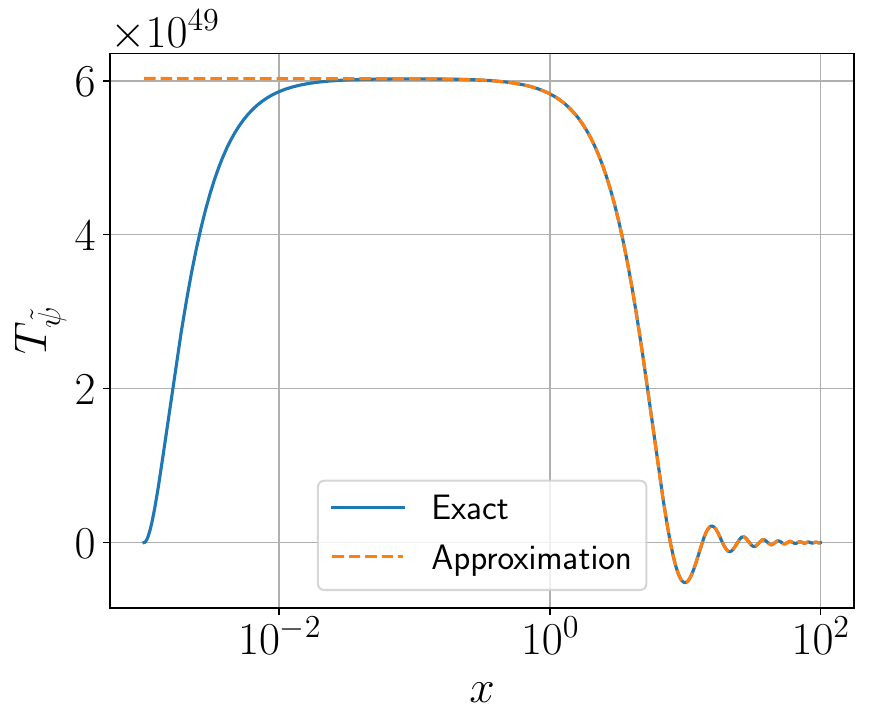}
     \end{subfigure}
     \begin{subfigure}[b]{0.49\textwidth}
         \centering
         \includegraphics[width=\textwidth]{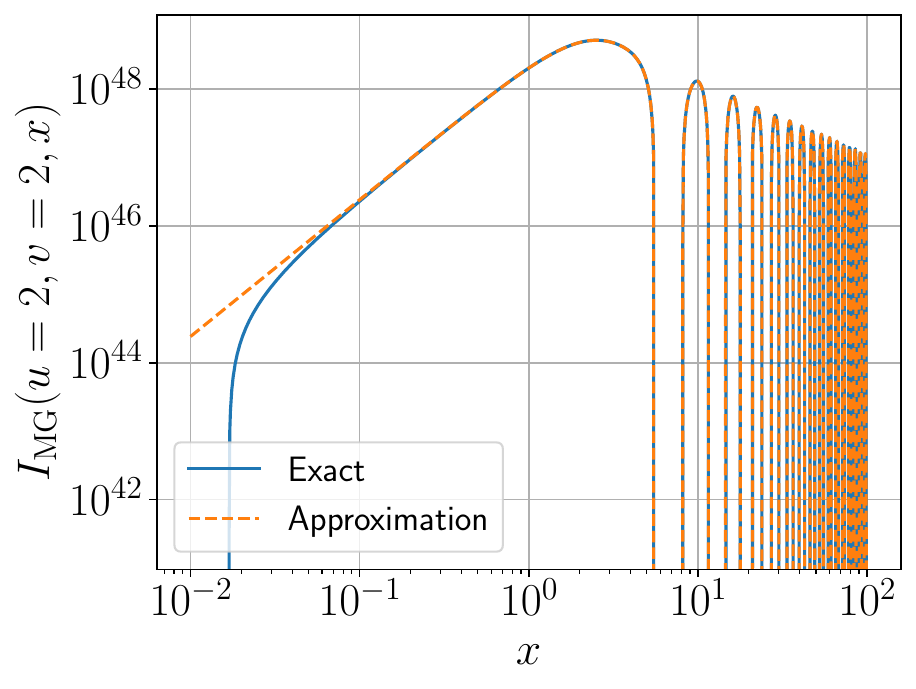}
     \end{subfigure}
     \caption{Left: Comparison between the exact solution for the transfer function of $\tilde{\psi}_k^{\rm MG}$ (solid blue) and its approximation (dashed orange), for $k=10^{15}\text{Mpc}^{-1}$. Right: Comparison between the exact solution for the modified gravity kernel $I_{\rm MG}$, evaluated at $u=v=2$ (solid blue) and its approximation (dashed orange), for $k=10^{15}\text{Mpc}^{-1}$.}
     \label{fig:approximation}
\end{figure}
Focusing on $\tilde{\psi}_1(x)J_2(x)$, we note that in the region $x\in [0,\infty)$ the sine integral $\text{Si}(x)$ is bounded between $[0,\pi/2)$, while the function $\tilde{\psi}_2(x)$ lies within the range $(0,1]$. $\tilde{\psi}_1(x)$ diverges as $\sim 1/x^3$ for $x\to 0$ but quickly approaches 0 as $x$ grows. Since, in order to obtain the spectrum of SIGWs, we are interested in the late time behaviour, $x\to \infty$, this means that we can safely neglect the second term in Eq. \eqref{aosdfnaingfawre}. As for $\tilde{\psi}_2(x)J_1(x)$, both $\text{Ci}(x)$ and $\tilde{\psi}_1(x)$ tend to zero for large $x$, so we can ignore the first term in Eq. \eqref{asodufbinawerfwae}. For small $x$, we have that $\text{Ci}(x)\xrightarrow[ x\to 0 ]{}\gamma + \log(x)$, whereas $\tilde{\psi}_1(x)\xrightarrow[ x\to 0 ]{}\sqrt{3}/x^3$, which is much faster. Therefore, we can write the approximate expression
\begin{equation}  \label{aosdfbaiwenwe}
        T_{\tilde{\psi}}(k,\eta)=\frac{\mathcal{D}k^2}{4\sqrt{3}}x_{\rm reh}\tilde{\psi}_1(x_{\rm reh})\tilde{\psi}_2(k\eta)=\frac{\mathcal{D}k^2}{36x_{\rm reh}^2}T_{\psi}(k\eta)=\frac{\mathcal{D}}{36\eta_{\rm reh}^2}T_{\psi}(k\eta)~,
\end{equation}
where the transfer function of $\psi_k^{\rm GR}$ is given by (see Eqs.~\eqref{eq:separation-transfer-functions} and \eqref{eq:sol-psiGR-ini-cond})
\begin{equation} \label{aoisdbfhafwae}
    T_{\psi}(k\eta) = \frac{9}{(k\eta)^2}\left[\frac{\sqrt{3}}{k\eta}\sin\left(\frac{k\eta}{\sqrt{3}}\right)-\cos\left(\frac{k\eta}{\sqrt{3}}\right)\right].
\end{equation}
In the left panel of Fig.~\ref{fig:approximation} we compare Eq.~\eqref{aosdfbaiwenwe} to the exact solution for $k=10^{15}$Mpc$^{-1}$. In order to quantify the validity of the approximation in a systematic way, we compute the relative error $\mathcal{E}_{\rm rel}\equiv (T^{\rm exact}-T^{\rm approx})/T^{\rm exact}$, where $T^{\rm exact}$ is given by Eq. \eqref{aisdbfiuwaerawer} and $T^{\rm approx}$ by Eq. \eqref{aosdfbaiwenwe}, for a range of wavenumbers spanning all observationally relevant scales, from $k=10^7$Mpc$^{-1}$ (PTA scales) to $k=10^{17}$Mpc$^{-1}$ (scales accessible to ET and LVK). We show the result as a heat map in Fig. \ref{fig:heatmap}. As expected, the approximation is excellent for large $x$.

\begin{figure}[h]
     \centering
     \begin{subfigure}[b]{0.49\textwidth}
         \centering
         \includegraphics[width=\textwidth]{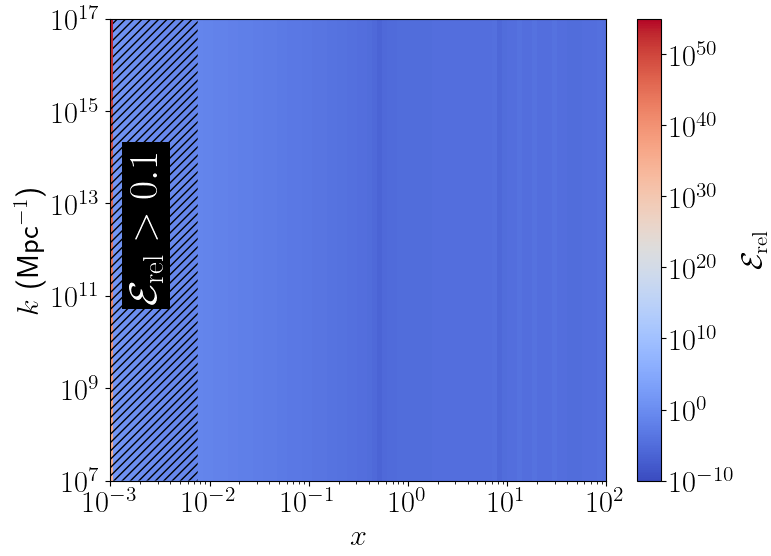}
     \end{subfigure}
     \begin{subfigure}[b]{0.49\textwidth}
         \centering
         \includegraphics[width=\textwidth]{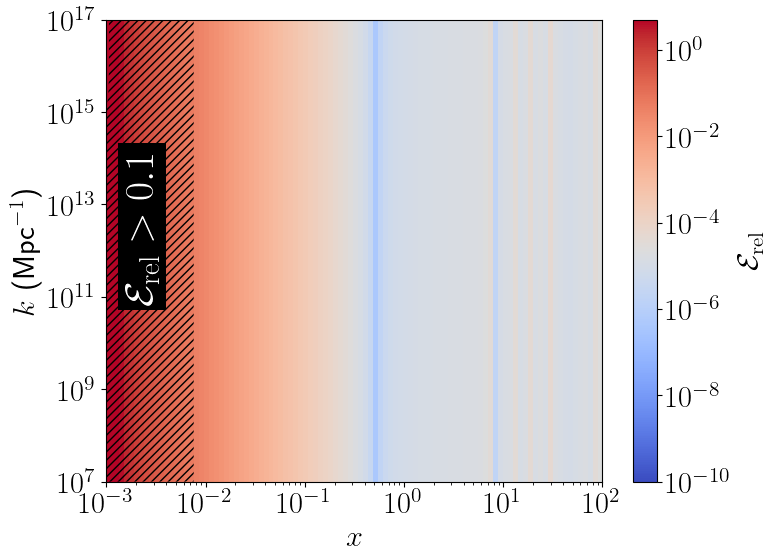}
     \end{subfigure}
     \caption{Left: Relative error $\mathcal{E}_{\rm rel}$ between the exact and approximated versions of $T_{\tilde{\psi}}$. The shaded region is the locus of points $(k,x)$ satisfying $\mathcal{E}_{\rm rel}>0.1$. As expected, the relative error is huge only near $x=10^{-3}$ (notice the thin red band), but the approximation quickly becomes excellent. Right: Zoom-in of the left panel, where we only show $\mathcal{E}_{\rm rel}<5$, for improved visibility.}
     \label{fig:heatmap}
\end{figure}

With Eqs.~\eqref{aosdfnaingfawre}-\eqref{aoisdbfhafwae}, we can easily simplify the modified gravity contribution to the source function, which becomes
\begin{equation}
    f_{\rm MG}(\eta,q,\abs{\textbf{k}-\textbf{q}}) = \frac{\mathcal{D}}{18\eta_{\rm reh}^2}f_{\rm GR}(x,u,v)~, \label{adsifbuiwaeawer}
\end{equation}
where we introduce the variables
\begin{equation}
    u=\frac{\abs{\mathbf{k}-\mathbf{q}}}{k}~,\quad v=\frac{q}{k}~.
\end{equation}
In the $\mathcal{C}$-dominated regime, the source function is given by
\begin{align}
    f_{\text{MG}}(u, v, x, k) &= \frac{Ck}{A} \Bigg\{ 
    2 v\, T_{\psi}(ux)\, T_{\tilde{\psi}}'(vx) + 2 u\, T_{\psi}(vx)\, T_{\tilde{\psi}}'(ux) +\nonumber\\
    &+ u\, \left[ T_{\psi}(vx) + vx\, T_{\psi}'(vx) \right] \left[ T_{\tilde{\psi}}'(ux) + ux\, T_{\tilde{\psi}}''(ux) \right] +\nonumber\\
    &+v\, \left[ T_{\psi}(ux) + ux\, T_{\psi}'(ux) \right] \left[ T_{\tilde{\psi}}'(vx) + vx\, T_{\tilde{\psi}}''(vx) \right]\Bigg\}, \label{eq:sourcefuncitonC}
\end{align}
instead. This expression does not allow for a straightforward analytical expression for the kernel. In the right panel of Fig. \ref{aisdfbaijndfawef}, we show the modified gravity contribution to the GR kernel, for some example parameter values, obtained numerically.

Solving Eq.~\eqref{eq:second-order-tensor-inhomogeneous} using Green's function method, we have 
\begin{equation} \label{asdifubauewwqer}
    \langle h_{\textbf{k}}^s(\eta)h_{\textbf{q}}^s(\eta)\rangle=\frac{1}{a^2(\eta)}\int_{\eta_i}^{\eta}\text{d}\tau_1\int_{\eta_i}^{\eta}\text{d}\tau_2 G(\eta,\tau_1)G(\eta,\tau_2)a(\tau_1)a(\tau_2)\langle S^s(\tau_1,\textbf{k})S^s(\tau_2,\textbf{q})\rangle~,
\end{equation}
where
\begin{equation} \label{asdoifbwaerwer}
    S^s(\eta,\textbf{k})=4\int\frac{\text{d}^3\textbf{q}}{(2\pi)^3}e^{s}_{jk}(\textbf{k})q^j q^k\tilde{\psi}_{\textbf{q},i}\tilde{\psi}_{\textbf{k}-\textbf{q},i}f(\eta,q,\abs{\textbf{k}-\textbf{q}})~.
\end{equation}
This two-point function of the source term can be shown to be 
\begin{align}
    &\sum_{s=+,\cross}\langle S^s(\tau_1,\textbf{k})S^s(\tau_2,\textbf{q})\rangle=\nonumber\\
    &=8(2\pi)^3\frac{2\pi^2}{k^3}\delta^{(3)}(\textbf{k}+\textbf{q})\int_0^{\infty}\text{d}v\int_{\abs{1-v}}^{1+v}\text{d}u\,k^4\left[\frac{4v^2-(1+v^2-u^2))}{4vu}\right]^2f(\tau_1,kv,ku)f(\tau_2,kv,ku)\times\nonumber\\
    &\times\Delta_{\tilde{\psi}}^2(ku)\Delta_{\tilde{\psi}}^2(kv)~,
\end{align}
where we sum over the two polarisation states, and we made use of Wick's theorem on the four-point functions of $\tilde{\psi}_{i,\mathbf{k}}$ (along with other calculations which are similar to those carried out in the case of GR). 

We define the kernel
\begin{equation}
    I(x,u,v)\equiv \frac{1}{a(x/k)}\int_{x_i}^{x}\text{d}y\, k\, G(x/k,y/k)a(y/k)f(y,u,v)~,
\end{equation}
and then
\begin{align}\label{adsoufiubnwer}
    &\sum_{s=+,\cross}\langle h_{\textbf{k}}^s(\eta)h_{\textbf{q}}^s(\eta)\rangle=\nonumber\\
    &=8(2\pi)^3\frac{2\pi^2}{k^3}\delta^{(3)}(\mathbf{k}+\mathbf{q})\int_0^{\infty}\text{d}v\int_{\abs{1-v}}^{1+v}\text{d}u\,\left[\frac{4v^2-(1+v^2-u^2))}{4vu}\right]^2I^2(x,u,v)\Delta_{\tilde{\psi}}^2(ku)\Delta_{\tilde{\psi}}^2(kv)~.
\end{align}
Comparing this with the definition of the power spectrum, Eq.~\eqref{eq:relation-dimensionless-powerspectrum}, that of the SIGWs can be written as (noting that $\Delta_h^2(k,\eta)=\sum_s \Delta_{h,s}^2(k,\eta)$ and that we have already summed over the polarisations in Eq. \eqref{adsoufiubnwer})
\begin{equation}
    \Delta_h^2(k,\eta)=4\int_0^{\infty}\text{d}v\int_{\abs{1-v}}^{1+v}\text{d}u\,\left[\frac{4v^2-(1+v^2-u^2))}{4vu}\right]^2I^2(x,u,v)\Delta_{\tilde{\psi}}^2(ku)\Delta_{\tilde{\psi}}^2(kv)~.
\end{equation}
We now separate the GR part of the kernel and the modified gravity contribution
\begin{align}
    \Delta_h^2(k,\eta)=4\int_0^{\infty}\text{d}v\int_{\abs{1-v}}^{1+v}&\text{d}u\,\left[\frac{4v^2-(1+v^2-u^2))}{4vu}\right]^2\times\nonumber\\
    &\times\left(I^2_{\rm GR}(x,u,v)+2\beta I_{\rm GR}(x,u,v)I_{\rm MG}(x,u,v)\right)\Delta_{\tilde\psi}^2(ku)\Delta_{\tilde\psi}^2(kv)~,
\end{align}
where
\begin{equation}
    I_{\rm GR}(x,u,v)\equiv \frac{1}{a(\eta)}\int_{x_i}^{x}\text{d}y\, k\, G(\eta,y/k)a(y/k)f_{\rm GR}(y,u,v)~,
\end{equation}
and
\begin{equation}
    I_{\rm MG}(x,u,v)\equiv \frac{1}{a(\eta)}\int_{x_i}^{x}\text{d}y\, k\, G(\eta,y/k)a(y/k)f_{\rm MG}(y,u,v)~.
\end{equation}
For the density spectrum, we need the time-averaged power spectrum, namely
\begin{align}
    \label{asdjfaijsdfawer}
    \overline{\Delta_h^2(k,\eta)}=4\int_0^{\infty}\text{d}v\int_{\abs{1-v}}^{1+v}&\text{d}u\,\left[\frac{4v^2-(1+v^2-u^2))}{4vu}\right]^2\times\nonumber\\
    &\times\left(\overline{I^2_{\rm GR}(x,u,v)}+2\beta\overline{I_{\rm GR}(x,u,v)I_{\rm MG}(x,u,v)}\right)\Delta_{\tilde\psi}^2(ku)\Delta_{\tilde\psi}^2(kv)~.
\end{align}
Since all the time dependence is in the kernel, one needs to perform the time-averaging over $I(x,u,v)$ solely.

\subsubsection{Time-Averaged Kernel} \label{sec:3.4.2}
From the source functions \eqref{adsifbuiwaeawer}, it immediately follows that
\begin{equation} \label{asdfobaewnfawe}
    I_{\rm MG}(u,v,k,\eta)=\frac{\mathcal{D}}{18\eta_{\rm reh}^2}I_{\rm GR}(k\eta,u,v)~.
\end{equation}
We can easily estimate $\eta_{\rm reh}$, relating it to the reheating temperature. Indeed, using the fact that during radiation domination $a\propto \eta$, we get
\begin{equation}
    \frac{a_{\rm eq}}{a_{\rm reh}}=\frac{\eta_{\rm eq}}{\eta_{\rm reh}}\quad \Rightarrow\quad \eta_{\rm reh}=\frac{a_{\rm reh}}{a_{\rm eq}}\eta_{\rm eq}=\left(\frac{\rho_{\rm eq}}{\rho_{\rm reh}}\right)^{1/4}\eta_{\rm eq}=\frac{1}{T_{\rm reh}}\left(\frac{30\rho_{\rm eq}}{\pi^2g_{*}(T_{\rm reh})}\right)^{1/4}\eta_{\rm eq}~,
\end{equation}
where we have used that $\rho\propto a^{-4}$ during radiation domination and $g_{*}(T_{\rm reh})=106.75$ is the effective number of relativistic degrees of freedom at reheating. Using $\rho_{\rm eq}=1.27\cross 10^{-110}\m^4$ and $\eta_{\rm eq}=2H_0^{-1}(\sqrt{2a_{\rm eq}}-\sqrt{a_{\rm eq}})/\sqrt{\Omega_{{\rm m},0}}=4.30\cross 10^{58}\m^{-1}$ \cite{Planck:2018vyg}, we obtain
\begin{equation} \label{adoufbanwefrwer}
    \eta_{\rm reh}=\frac{5.94\cross 10^{30}}{T_{\rm reh}}~,
\end{equation}
where we consider the simplifying assumption of instant reheating.

In the right panel of Fig. \ref{fig:approximation} we show the modified gravity kernel, obtained numerically from the full solution \eqref{fasidbfaidsfasdf} and from the approximation \eqref{aosdfbaiwenwe}. Except for very small values of $x$, the approximate curve matches that from the numerical simulations, as we may expect (see also Fig. \ref{fig:heatmap}). Plugging Eq. \eqref{asdfobaewnfawe} in Eq. \eqref{asdjfaijsdfawer} gives
\begin{align}
    &\overline{\Delta_h^2(\eta,k)}=4\int_0^{\infty}\text{d}v\int_{\abs{1-v}}^{1+v}\text{d}u\,\left[\frac{4v^2-(1+v^2-u^2))}{4vu}\right]^2\left[1+\beta\frac{\mathcal{D}}{18\eta_{\rm reh}^2}\right]\times\nonumber\\
    &\times\overline{I^2_{\rm GR}(x,u,v)}\Delta_{\mathcal{R}}^2(ku)\Delta_{\mathcal{R}}^2(kv)~,
\end{align}
where Eq.~\eqref{eq:separation-transfer-functions} was used to relate the power spectrum of $\tilde{\psi}_{\textrm{MG}}$ to that of $\mathcal{R}$. An analytical expression of the GR kernel, for radiation domination in the subhorizon ($x\gg 1$) limit, was obtained in Ref. \cite{Kohri:2018awv}, where they find
\begin{align}
    &I_{\rm RD}(x\to \infty,u,v) = \frac{3(u^2+v^2-3)}{4u^3v^3 x}\Bigg[\sin{(x)}\left(-4uv+(u^2+v^2-3)\log\abs{\frac{3-(u+v)^2}{3-(u-v)^2}}\right)- \nonumber\\
    &- \pi\cos{(x)}(u^2+v^2-3)\Theta(v+u-\sqrt{3})\Bigg].
\end{align}
Squaring, taking the oscillation average, and remembering that $\overline{\cos^2(x)}=\overline{\sin^2(x)}=1/2$ and $\overline{\cos(x)\sin(x)}=0$, gives
\begin{align} 
    &\overline{I^2_{\rm RD}(x\to \infty,u,v)}=\frac{1}{2}\left(\frac{3(u^2+v^2-3)}{4u^3v^3x}\right)^2\Bigg[\left(-4uv+(u^2+v^2-3)\log\abs{\frac{3-(u+v)^2}{3-(u-v)^2}}\right)^2+\nonumber\\
    &+\pi^2(u^2+v^2-3)^2\Theta(u+v-\sqrt{3})\Bigg], \label{asdiofubaiwefrawerf}
\end{align}
where we remind the reader that $\Theta(u+v-\sqrt{3})$ is the Heaviside theta function with $\Theta(x)=1$ if $x>0$ and $\Theta(x)=0$ if $x\leq 0$. Therefore, the density spectrum of SIGWs reads
\begin{equation}
    \Omega_{\rm SIGW}(k,\eta)=\frac{1}{24}\left(\frac{k}{a(\eta)H(\eta)}\right)^2\overline{\Delta_h^2(\eta,k)}=\frac{1}{24}x^2\overline{\Delta_h^2(\eta,k)}=\left(1+\Delta(\alpha,T_{\rm reh})\right)\Omega_{\rm SIGW}^{\rm GR}(k)~.
\end{equation}
We obtain a scale-independent amplification of the density spectrum of SIGWs with respect to GR. Additionally, an amplification parameter has been defined as
\begin{equation}
    \Delta(\alpha,T_{\rm reh})\equiv\frac{\beta\mathcal{D}}{18\eta_{\rm reh}^2}=1.63\alpha\left(\frac{T_{\rm reh}}{\m}\right)^2\cross 10^{-55}=2.76\alpha\left(\frac{T_{\rm reh}}{\text{GeV}}\right)^2\cross 10^{-92}~,
\end{equation}
where we have used Eqs. \eqref{asdofjubaiwerawer}, \eqref{aosdfbiawner} and \eqref{adoufbanwefrwer}.

As mentioned above, instant reheating is assumed. As an example, taking the end of inflation to occur around the GUT scale, the reheating temperature is $T_{\rm GUT}\sim 10^{15}\text{GeV}$. We then have
\begin{equation}
    \label{asdifbaiwefawer}
    \Delta(\alpha, T_{\rm GUT})\sim \alpha \cross 10^{-62}~.
\end{equation}
In order to obtain a $10\%$ amplification with respect to GR, one needs $\alpha \sim 10^{61}$. For smaller values of $T_{\rm reh}$, $\alpha$ needs to be larger in order to keep $\Delta$ fixed. Note that $\Delta<1$ is precisely the perturbativity condition of the scalar perturbations given by Eq. \eqref{asdoifbahwefrwerwer}. Therefore, it is satisfied even for $\alpha\sim 10^{61}$. For these values of $\alpha$ and $T_{\rm reh}$, the perturbativity condition of the background, $\beta / a_{\rm reh}^3<1$, is violated at early times. However, we remind the reader that the background quickly asymptotes to GR (see the $1/\eta^4$ dependence in Eq.~\eqref{oaisdbfiabsdfawe}, for example). We therefore neglect this effect, as is done in Ref. \cite{Kugarajh:2025rbt}.

Regarding the $\mathcal{C}$-dominated regime, we do not have analytical expressions below the source function \eqref{eq:sourcefuncitonC}. We show the density spectrum, obtained numerically, in Fig. \ref{fig:spectrumC}, for one representative value $\alpha=10^{67}$. The opposite effect compared to the $\mathcal{D}$-dominated regime can be found. Not only is the spectrum not amplified, but becomes suppressed in the long-wavelength tail. This makes sense as $T_{\tilde{\psi}}$ is now negative. Furthermore, because $T_{\tilde{\psi}}$ in the $\mathcal{C}$-dominated regime does not depend on the initial integration time $\eta_{\rm reh}$, the background does not violate the perturbativity condition, for all relevant values of $\alpha$.

\begin{figure}[h]
     \centering
    \includegraphics[width=0.9\textwidth]{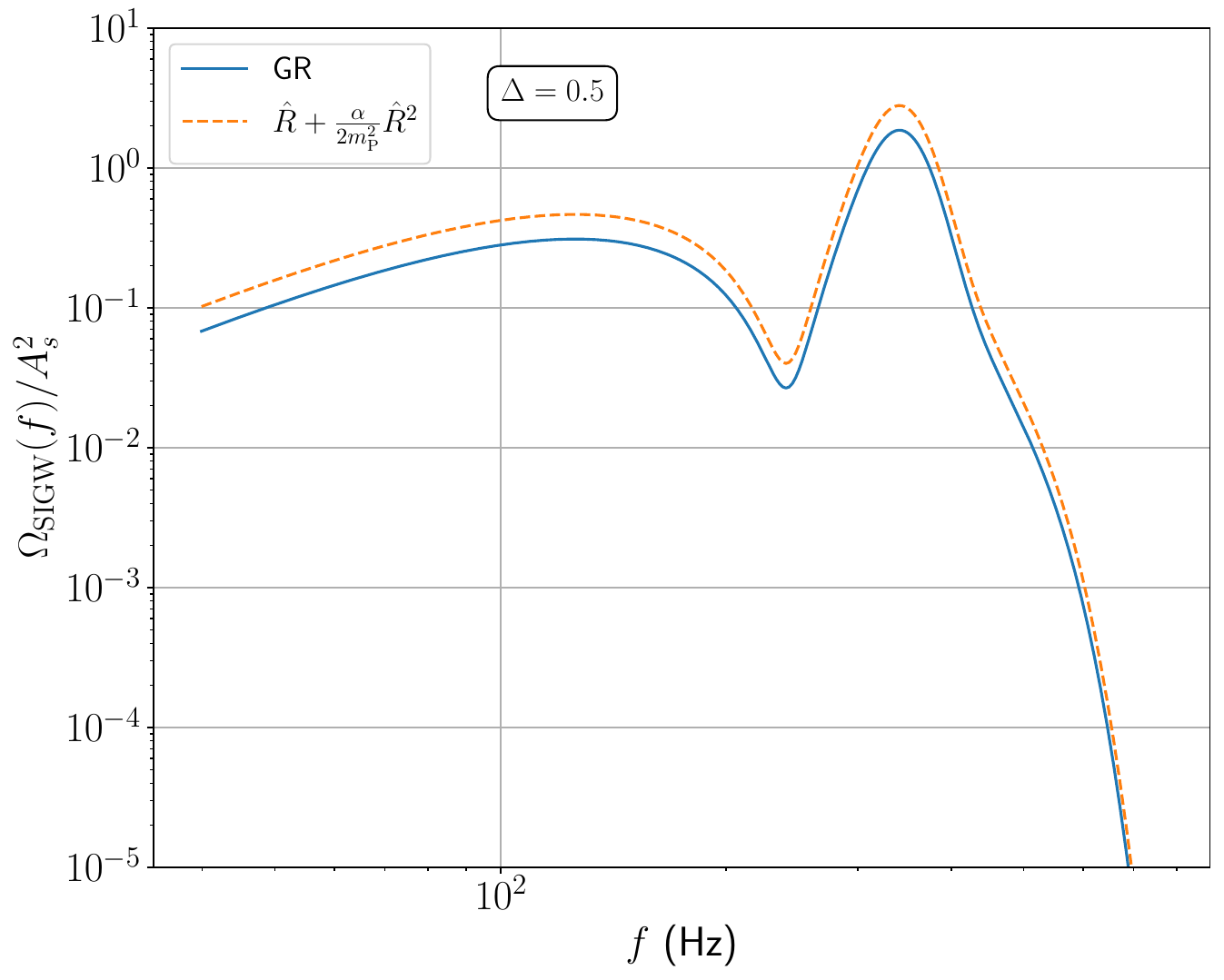}
     \caption{Spectral energy density of the scalar-induced gravitational waves $\Omega_{\rm SIGW}$, normalised by $A_s^2$, as a function of frequency, for the regime where terms proportional to $\mathcal{D}$ dominate in the modified gravity transfer function. We use a log-normal scalar power spectrum with width $\sigma=0.1$ and centered at $k_{*}=1.96\cross 10^{17}\text{Mpc}^{-1}$. In solid blue we show the spectrum corresponding to GR, while in dashed orange we show the spectrum corresponding to $\hat{R}+\frac{\alpha}{2\m^2} \hat{R}^2$ gravity in the Palatini formalism. We take $\Delta=0.5$.}
     \label{fig:spectrum}
\end{figure}

\begin{figure}[h]
     \centering
    \includegraphics[width=0.9\textwidth]{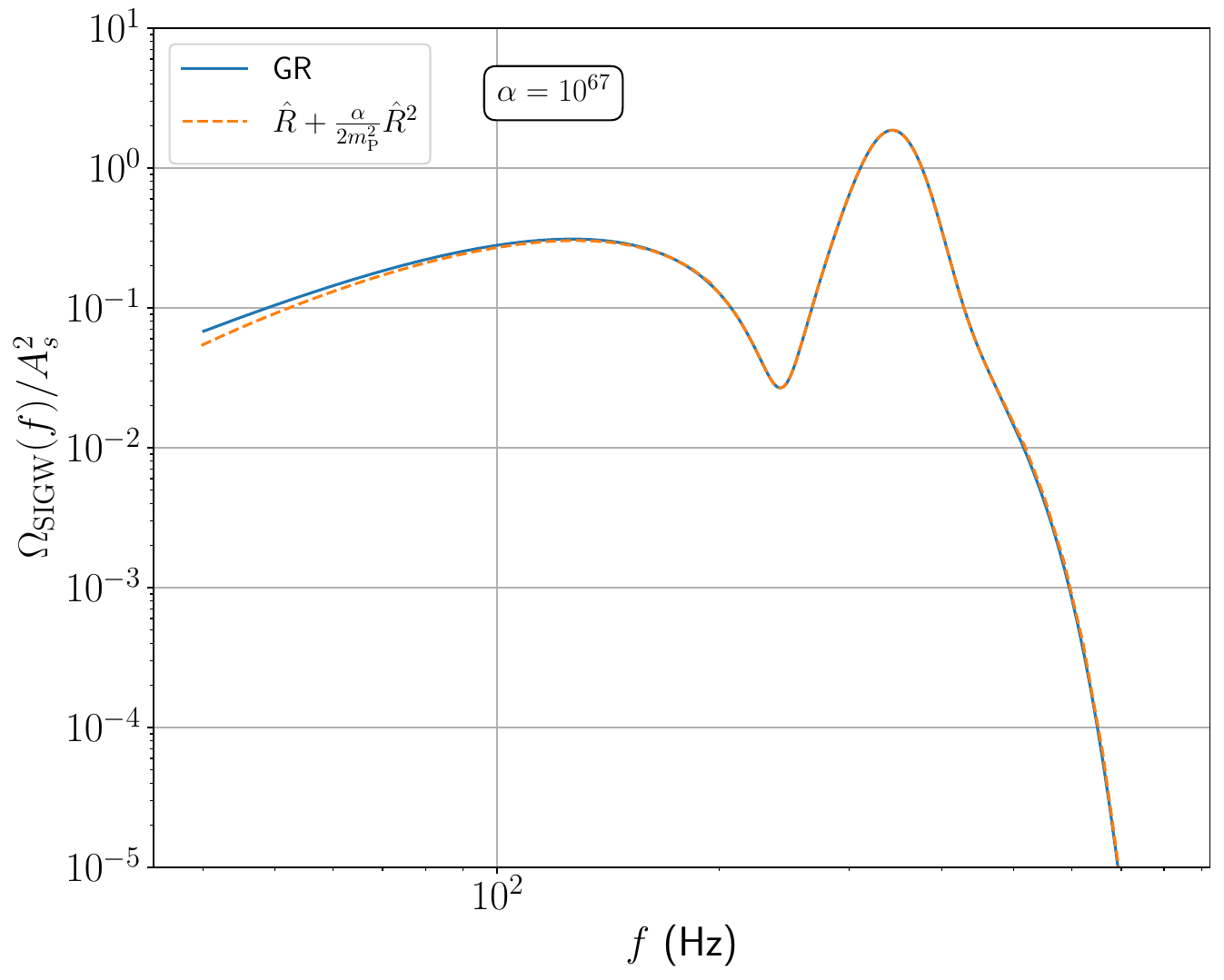}
     \caption{Spectral energy density of the scalar-induced gravitational waves $\Omega_{\rm SIGW}$, normalised by $A_s^2$, as a function of frequency, for the regime where terms proportional to $\mathcal{C}$ dominate in the modified gravity transfer function. We use a log-normal scalar power spectrum with width $\sigma=0.1$ and centered at $k_{*}=1.96\cross 10^{17}\text{Mpc}^{-1}$. In solid blue we show the spectrum corresponding to GR, while in dashed orange we show the spectrum corresponding to $\hat{R}+\frac{\alpha}{2\m^2} \hat{R}^2$ gravity in the Palatini formalism. We take $\mathcal{C}=10^{30}$ and $\alpha = 10^{67}$.}
     \label{fig:spectrumC}
\end{figure}

\begin{figure}[h]
     \centering
    \includegraphics[width=0.9\textwidth]{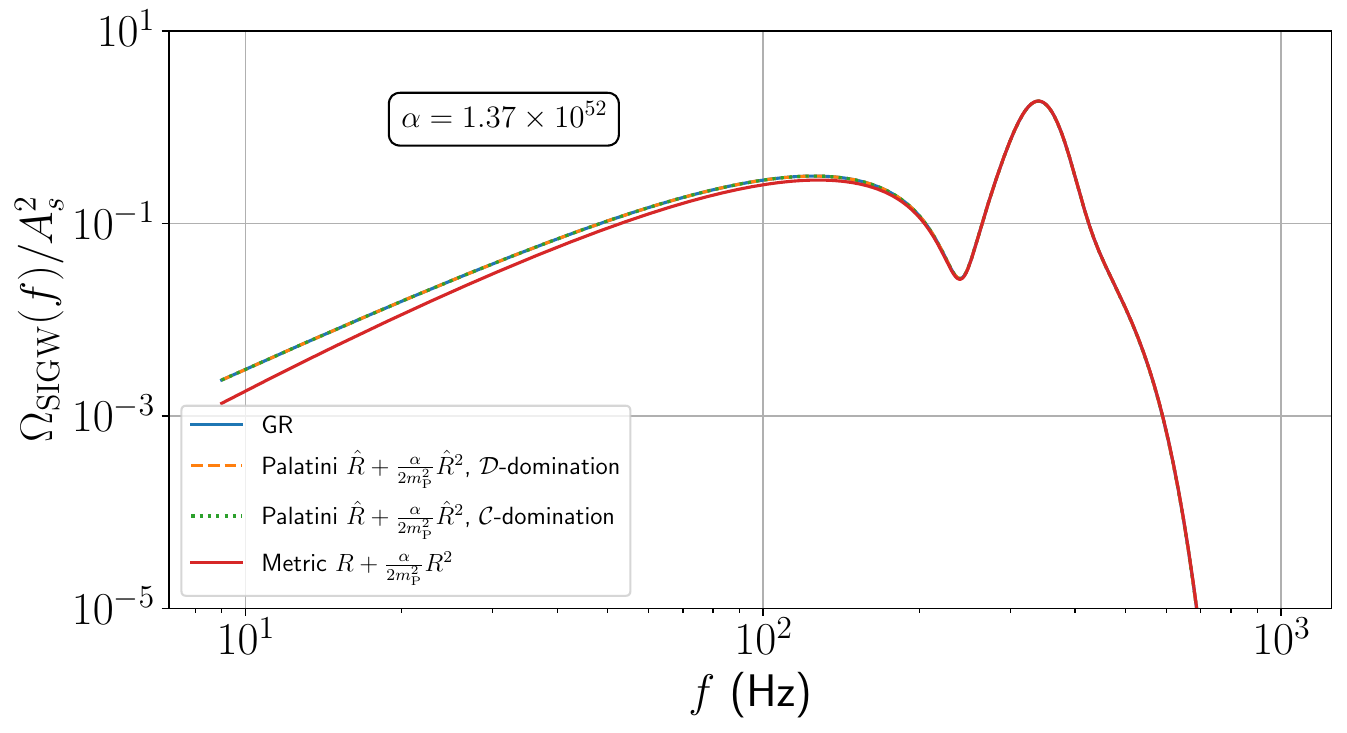}
     \caption{Comparison between GR (full blue), Palatini $\hat{R}+\frac{\alpha}{2\m^2} \hat{R}^2$ gravity, in the $\mathcal{D}$-dominated (dashed orange) and $\mathcal{C}$-dominated (dotted green) regimes, and metric $R+\frac{\alpha}{2\m^2} R^2$ gravity (full red), for $\alpha = 10^{52}$. We use a log-normal scalar power spectrum with width $\sigma=0.1$ and centered at $k_{*}=1.96\cross 10^{17}\text{Mpc}^{-1}$. For this value of $\alpha$, the maximum amplification possible in the $\mathcal{D}$-dominated regime is $\Delta \sim 10^{-10}$ (see Eq. \eqref{asdifbaiwefawer}), making it effectively indistinguishable from GR. A similar effect happens in the $\mathcal{C}$-dominated regime. A larger value of $\alpha$, around $\alpha\sim 10^{60}$, is required for the Palatini formalism to display its distinctive features. We thank the authors of Ref. \cite{Kugarajh:2025rbt} for generously sharing with us the necessary data to plot the result in the metric formalism.}
     \label{fig:comparisonmetric}
\end{figure}

It is interesting to relate the value of $\alpha$ to inflation in Palatini $\hat{R}+\alpha \hat{R}^2/(2\m^2)$ gravity. Recent studies in this context have shown that models previously discarded by the Planck data, such as quadratic and quartic chaotic inflation, or exponential inflation, can be brought back within observational bounds as long as $\alpha \gtrsim 10^{10}$ \cite{Antoniadis:2018ywb,Antoniadis:2018yfq,Dimopoulos:2020pas,Dimopoulos:2022tvn,Dimopoulos:2022rdp}. There is no upper bound on the value $\alpha$ can take, although it is well known that the tensor-to-scalar ratio $r$ scales as $1/\alpha$ \cite{Enckell:2018hmo}. In this way, the large value of $\alpha \sim 10^{61}$ discussed above, would lead to an unobservable $r$. If the secondary GWs signal is boosted, the primary one is strongly suppressed.

\subsubsection{The Spectrum} \label{sec:3.4.3}
\begin{figure}[h]
     \centering
    \includegraphics[width=0.8\textwidth]{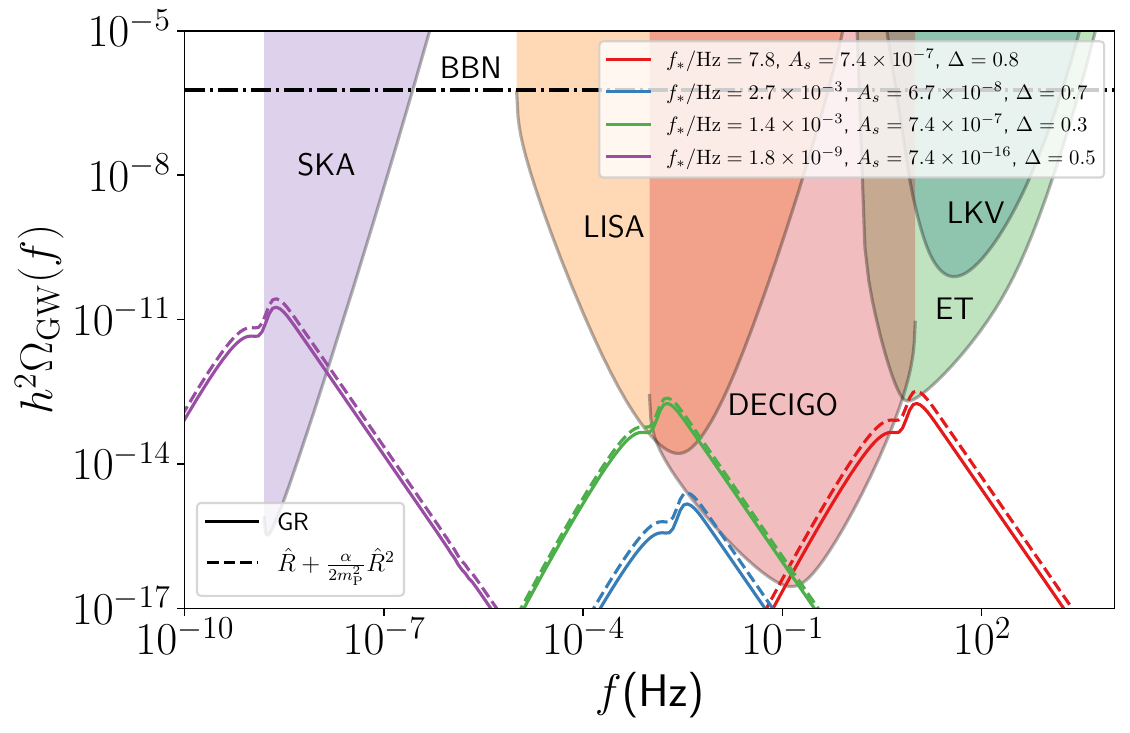}
     \caption{Spectral energy density of the scalar-induced gravitational waves $\Omega_{\rm SIGW}$ as a function of frequency induced by a broken power law power spectrum, superimposed with the PLIC for different upcoming GW experiments. In full are the spectra obtained in GR while in dashed are the spectra obtained in Palatini $\hat{R}+\alpha \hat{R}^2/(2\m^2)$ gravity, in the $\mathcal{D}$-dominated regime. We choose a few relevant pivot frequencies $f_{*}$, amplitudes of the power spectrum $A_s$, and amplifications $\Delta$.}
     \label{fig:PLICS}
\end{figure}

We now proceed to compute some example spectra. For the sake of comparison, we take the recent study of Starobinsky-like gravity in the metric formalism; namely, Ref.~\cite{Kugarajh:2025rbt}, and consider a log-normal scalar power spectrum given by \cite{Pi:2020otn}
\begin{equation}
    \Delta_{\mathcal{R}}^2(k)=\frac{A_{s}}{\sqrt{2\pi}\sigma}\text{exp}\left[-\frac{\log^2\left(k/k_{*}\right)}{2\sigma^2}\right],
\end{equation}
where the amplitude is normalised such that $\int_0^{\infty}\text{d}\log k\,\Delta_{\mathcal{R}}^2(k)=A_s$. $k_{*}$ is the wavenumber at which the peak, with width $\sigma$, is centered. We take $\sigma = 0.1$ and $k_{*}=1.96\cross 10^{17}\text{Mpc}^{-1}$, a wavenumber relevant for the observability range of LVK \cite{Harry:2010zz,VIRGO:2014yos,LIGOScientific:2014pky,LIGOScientific:2019lzm,KAGRA:2020tym} and ET \cite{Punturo:2010zz,Hild:2010id}. We do not fix $A_s$ since we plot the density parameter normalised by $A_s^2$. Note that LVK has not detected a gravitational wave signal associated to this spectrum, which means that $A_s\lesssim 10^{-5}$ (see the PLIC of LVK in Fig. \ref{fig:PLICS}). However, this bound is much larger than the one obtained from CMB measurements $A_s(k_0)= 2.099\times 10^{-9}$ (TT,TE,EE+lowE+lensing) \cite{Planck:2018vyg}, with the pivot scale given by $k_0=0.05$Mpc$^{-1}$. This is possible since we are working at scales much smaller than the CMB, $k_{*}\gg k_0$. In fact, the amplification of the power spectrum at small scales \cite{Byrnes:2018txb,Inomata:2018epa} has been the subject of intense study in the literature, since it may lead to the formation of primordial black holes (PBHs) \cite{Sasaki:2018dmp,Green2016} and, if so, it would inevitably lead to the production of SIGWs \cite{Yuan:2021qgz}. This is especially interesting since there are no observational constraints on PBHs in the mass range between $\sim 10^{19}$g and $\sim 10^{22}$g \cite{Carr:2020gox}, which means that they could constitute the totality of dark matter \cite{Carr:2025kdk}.

We show the result for the $\mathcal{D}$-dominated regime in Fig.~\ref{fig:spectrum}, for $\Delta =0.5$. As expected, we obtain a scale-independent amplification with respect to GR, accessible to experiments such as BBO \cite{Harry:2006fi}, DECIGO \cite{Kawamura:2006up,Kawamura:2011zz,Kawamura:2020pcg}, ET \cite{Punturo:2010zz,Hild:2010id}, LISA \cite{Bartolo:2016ami,Caprini:2019pxz,LISACosmologyWorkingGroup:2022jok} (see Ref.~\cite{LISACosmologyWorkingGroup:2025vdz} for detection prospects for the SIGWs), and SKA \cite{Janssen:2014dka}. This result is contrary to what the metric formalism prescribes, where the signal is actually suppressed (see Figs. 5.1-5.3 in Ref.~\cite{Kugarajh:2025rbt}) in a scale-dependent way, with a stronger suppression for smaller values of frequency.\footnote{We put emphasis on the fact that, in $f(R)$ gravity and in the case of $R^2$ in particular, the only limit where both the metric and Palatini formalisms agree is when $\alpha$ vanishes and one is left with the Einstein-Hilbert action. As a matter of fact, in GR, by varying the Palatini action with respect to $\hat{\Gamma}^{\alpha}_{\mu\nu}$, one obtains that the connection is the LC one (see Eq.~\eqref{eq:levi-civita-connection-eq}) plus a term that can be removed by virtue of a projective transformation of the connection. The LC connection is the one assumed in the metric formalism, as mentioned in the Introduction. Additionally, the two formalisms agree when there are no matter fields (vacuum) or when the trace of those is zero (pure radiation or conformally invariant fields), these two not being the cases here as remarked throughout the present paper.} We show the result for the $\mathcal{C}$-dominated regime in Fig.~\ref{fig:spectrumC}. In this case, the effect of the $\alpha\hat{R}^2$ term seems to give a similar effect to its metric counterpart; that is, a suppression at the low frequency tail. When both the $\mathcal{C}$ and $\mathcal{D}$ terms are comparable in magnitude, we expect an addition of both effects, boosting the spectrum in a scale-invariant way, but with an extra suppression at the long-wavelength tail.

A direct comparison between the Palatini and metric formalisms is shown in Fig. \ref{fig:comparisonmetric}. In Ref. \cite{Kugarajh:2025rbt}, the authors take $\alpha = 5\cross 10^{-34}\text{Hz}^{-2}$, where $\alpha$ is the coupling constant of their $f(R)=R+\alpha R^2$. With our definition $f(\hat{R})=\hat{R}+\alpha \hat{R}^2/(2\m^2)$, it corresponds to $\alpha = 1.37\cross 10^{52}$. From Eq. \eqref{asdifbaiwefawer}, this means that $\Delta \simeq 10^{-10}$, leading to an amplification with respect to GR that is not visible in the plot. We also show the spectrum in the $\mathcal{C}-$dominated regime, with a similar result. We need $\alpha\lesssim 10^{62}$ for the spectrum in the Palatini formalism to be significantly amplified with respect to GR.

Next, we consider a broken power law scalar power spectrum \cite{Byrnes:2018txb,Ozsoy:2019lyy,Carrilho:2019oqg,Atal:2018neu} given by
\begin{equation}
    \Delta^2_{\mathcal{R}}(k)=A_s\begin{cases}
                                \left(\frac{k}{k_{*}}\right)^{n_{\rm IR}} & k \leq k_{*} \\
                                \left(\frac{k}{k_{*}}\right)^{-n_{\rm UV}} & k > k_{*}.
                                \end{cases}
\end{equation}
This spectrum was studied analytically in Refs. \cite{Liu:2020oqe,Atal:2021jyo,Xu:2019bdp,Clesse:2018ogk,Riccardi:2021rlf}. We fix $n_{\rm IR}=4$, $n_{\rm UV}=1$ and solve it numerically for GR and Palatini $\hat{R}+\alpha \hat{R}^2/(2\m^2)$ gravity in the $\mathcal{D}$-dominated regime. We also compute the power-law integrated curves (PLIC) \cite{Thrane:2013oya} for LVK \cite{LIGOScientific:2019lzm}, DECIGO \cite{Kawamura:2006up,Kawamura:2011zz,Kawamura:2020pcg}, ET \cite{Punturo:2010zz,Hild:2010id}, LISA \cite{Bartolo:2016ami,Caprini:2019pxz,LISACosmologyWorkingGroup:2022jok}, and SKA \cite{Janssen:2014dka}, and present them in superposition with a few example spectra in Fig. \ref{fig:PLICS}. As expected, the modified gravity contribution provides a scale-invariant amplification of the density spectrum. For some combinations of parameters, like $f_{*}=2.7\times 10^{-3}$, $A_s=6.7\times 10^{-8}$ (in blue) the signal from GR is not detectable for any experiment, while for $\Delta > 0.5$ the signal becomes detectable by DECIGO. As another example, for $f_{*}=7.8$, $A_s=7.4\times 10^{-7}$ (in red), the GR signal is only detectable by DECIGO, while for  $\Delta > 0.4$ it becomes detectable also by ET. We expect analogous modifications for other example spectra (\textit{e.g.}, for an inflaton potential with a small bump \cite{Yang:2024ntt}).

From our results, it seems possible that future GW experiments may be able to distinguish not only GR predictions from those of modified gravity, but also between the different formalisms of the theory of gravity. Since the GR spectrum is well-known, a detection of a signal with the same profile as GR, but with boosted amplitude, would be supporting evidence for a quadratic correction to the Einstein-Hilbert action, as well as for the Palatini formalism. Additionally, it would also provide information regarding the value of $\alpha$, if $T_{\rm reh}$ can be separately constrained. 

As a final comment, the scalar power spectrum has been measured at CMB scales. Since we know that the GR signal should exist, a non-detection of the corresponding amplified spectrum would constrain the value of $\alpha$. 

\section{Conclusions and Outlook} \label{sec:4}

This work is a study of the production of second-order tensor modes sourced by first-order scalar perturbations in the context of Palatini $f(R)$ gravity. We have focused on the radiation dominated era, where the only contribution to the trace of the energy-momentum tensor is the subdominant pressureless dust component. This allowed us to use a perturbative approach to modified gravity and succeed in obtaining an analytical expression for the density spectrum of scalar-induced gravitational waves, which turns out to be a simple scale-invariant enhancement of the General Relativity (GR) signal.

After giving an overview of Palatini $f(R)$ gravity in the first section, we developed the cosmological perturbation theory up to, and including, second order. Analytical expressions for the Riemann and Ricci tensors were obtained, as well as for the curvature scalar, without fixing a particular gauge (see App.~\ref{ap:second-order-calculations}). Then, by imposing the Newtonian gauge, we derived all the components of the Einstein equations to first order in the scalar perturbations, in addition to the traceless-transverse second-order component, turning our attention to the scalar contribution to the source. 

In order to make progress, we fixed $f(\hat{R})=\hat{R}+\alpha \hat{R}^2/(2\m^2)$ and studied the evolution of the perturbations during radiation domination. The effect of the higher-order curvature term in the Palatini formalism is to provide additional contributions to the matter sources of the Einstein equations. These contributions rely on the trace of the background fluid, meaning that, during radiation domination, they depend on the subdominant pressureless dust component (the only non-zero contribution to the trace). Motivated by this fact, we treated $\alpha \hat{R}^2/(2\m^2)$ as a perturbative effect on GR and obtained that the modified gravity part of the scalar perturbation satisfies an ordinary differential equation equivalent to that of GR, but with the GR scalar perturbation and density contrast acting as sources. This equation is analytically solvable, giving, after a few well-motivated approximations, that the solution is a simple re-scaling of the GR part. We additionally verified this claim by numerically solving the corresponding equations. 

Finally, we derived the kernel and the density spectrum of scalar-induced gravitational waves, both analytically and numerically. The final result implies that the resulting spectrum is enhanced by a factor of $\Delta(\alpha,T_{\rm reh})=2.76\alpha(T_{\rm reh}/\text{GeV})^2\cross 10^{-92}$. In order to comply with the fact that the modified gravity contribution to the scalar perturbation should be smaller than the GR part, we need that $\Delta < 1$. However, an $\mathcal{O}(10^{-1})$ enhancement can be achieved for $\alpha = 10^{61}$ and $T_{\rm reh}=T_{\rm GUT}\sim 10^{15}\text{GeV}$. Importantly, such a large value of $\alpha$ would lead to an unobservable tensor-to-scalar ratio $r$. The enhancement of the secondary signal is scale-invariant, and therefore, if present, it should be detectable for any of the future gravitational wave experiments such as BBO \cite{Harry:2006fi}, DECIGO, ET, LISA, and SKA.  This opens the door to gravitational wave astronomy not only probing pre-recombination physics, but also the theory of gravity itself. Indeed, the signal we obtained is not only different from that of GR (for large enough $\alpha$ and $T_{\rm reh}$), but also from $f(\hat{R})=\hat{R}+\alpha \hat{R}^2/(2\m^2)$ gravity in the metric formalism. Probing the degrees of freedom of the theory of gravity appears to be plausible and an exciting prospect.

Regarding future work, it would be interesting to extend our model to the matter dominated era, and adopt a realistic reheating history. Also, we would perform a study of the observability of the theory, exploring its parameter space, as well as considering other $f(R)$ functions, such as the Hu-Sawicki model \cite{Hu:2007nk}. This model is expected to introduce modifications to GR that become more relevant at late times, unlike $\hat{R}+\alpha \hat{R}^2/(2\m^2)$. One might expect models of this kind to have a large impact on the spectrum of scalar-induced gravitational waves, although it is likely that one might lose capacity to solve the dynamical equations analytically. Additionally, the effect of backreaction from the modified gravity terms on the GR evolution when $\alpha$ is larger than the bound under consideration (see Eq.~\eqref{eq:bound_on_alpha_GUT}) should be evaluated. Likewise, the presence of anisotropic stress from matter in the second-order tensor modes equations could be addressed in detail, as well as the possible differences depending on the formalism of gravity under consideration, or the modified gravity theory at large. We are not aware of previous considerations being made in the Palatini formalism. 

\vspace{10mm}

\textbf{Acknowledgements}

\vspace{4mm}

The authors would like to thank Dhiraj Kumar Hazra, Mindaugas Karčiauskas, and Eemeli Tomberg for useful discussions, as well as the authors of Ref. \cite{Kugarajh:2025rbt} for providing us with necessary data to plot the spectrum in the metric formalism in Fig. \ref{fig:comparisonmetric}. We are grateful to the anonymous referee for pointing out several weaknesses in the last version of the manuscript, and for the various interesting comments. SSL would like to thank the Indo-French Centre for the Promotion of Advanced Research (IFCPAR/CEFIPRA) for support of the proposal 6704-4 titled `Testing flavors of the early universe beyond vanilla models with cosmological observations’ under the Collaborative Scientific Research Programme. JJTD acknowledges the financial support provided through national funds by FCT-Fundação para a Ciência e Tecnologia, I.P., with DOI identifiers 10.54499/2023.11681.PEX, 10.54499/UIDB/04564/2020, 10.54499/UIDP/04564/2020 and by the project 2024.00249.CERN funded by measure RE-C06-i06.m02–``Reinforcement of funding for International Partnerships in Science, Technology and Innovation'' of the Recovery and Resilience Plan-RRP, within the framework of the financing contract signed between the Recover Portugal Mission Structure (EMRP) and the Foundation for Science and Technology I.P. (FCT), as an intermediate beneficiary.

\appendix
\section{Second-Order Cosmological Perturbations}
\label{ap:second-order-calculations}
To account for the scalar-induced gravitational waves, we need to develop the second-order formalism. The general line element that includes metric perturbations about the spatially-flat Friedmann-Lema\^{i}tre-Robertson-Walker background metric reads \cite{Baumann:2007zm,Acquaviva:2002ud} 
\begin{equation}
    \label{eq:line-element-general}\textrm{d}s^2 = -a^2\left[\left(1+2A^{(1)} + A^{(2)} \right) \textrm{d}\eta^2+2\left(B_i^{(1)} + \frac{1}{2} B_i^{(2)}\right) \textrm{d}x^{i} \textrm{d}\eta -\left(\delta_{ij} + 2C_{ij}^{(1)} +C_{ij}^{(2)} \right)\textrm{d}x^{i} \textrm{d}x^{j}\right],
\end{equation} 
where $a=a(\eta)$ is the scale factor, and ($r=1,2$)
\begin{align}
    \label{eq:B_vector_pert_decomposition}&B_i^{(r)} \equiv \partial_i B^{(r)} + S_i^{(r)},\\
    \label{eq:C_tensor_decomposition}&C_{ij}^{(r)} \equiv -\psi^{(r)} \delta_{ij}+\partial_i \partial_j E^{(r)} +\partial_{\left(i\right.}F^{(r)}_{\left.j\right)}+\frac{1}{2} h_{ij}^{(r)}.
\end{align}
All the perturbations are spacetime-dependent. $\eta$ denotes the conformal time, such that $\textrm{d}t = a(\eta) \textrm{d}\eta$, where $t$ is the coordinate time. Latin indices are lowered and raised with the comoving spatial background metric, $\delta_{ij}$ and $\delta^{ij}$, respectively, given that the perturbations are assumed to live on the background spacetime.

The components of the perturbed metric are then 
\begin{align}
&g_{00} = -a^2\left(1+2A^{(1)} + A^{(2)} \right),\\
&g_{0i} = g_{i0} = -a^2\left(B_i^{(1)} + \frac{1}{2} B_i^{(2)} \right),\\
&g_{ij} = a^2\left(\delta_{ij} + 2C^{(1)}_{ij} + C_{ij}^{(2)} \right),
\end{align} 
as can be read off from the line element \eqref{eq:line-element-general}. The inverse metric components are determined using the relation $g_{\mu\sigma} g^{\sigma\nu} = \delta_{\mu}^{\nu}$ order by order
\begin{align}
\label{eq:comp-inverse-beginning}&g^{00} = -a^{-2} \left[1-2A^{(1)} -A^{(2)} +4(A^{(1)})^2 -B_i^{(1)} B^{(1)i}\right],\\
&g^{0i} = g^{i0} = -a^{-2} \left(B^{(1)i}+\frac{1}{2} B^{(2)i}-2A^{(1)} B^{(1)i} -2\tensor{C}{^{(1)i}_{j}} B^{(1)j}\right),\\
\label{eq:comp-inverse-end}&g^{ij} = a^{-2} \left(\delta^{ij} -2C^{(1)ij}-C^{(2)ij}-B^{(1)i}B^{(1)j}+4\tensor{C}{^{(1)i}_{k}}C^{(1)kj}\right).
\end{align}

Along with the metric tensor, the energy-momentum (or stress-energy) one is a fundamental quantity in the gravity theory. The general decomposition into scalar, vector and tensor quantities of the symmetric tensor is 
\begin{equation} 
\label{eq:em-tensor-decomposition}T_{\mu\nu} = \rho u_{\mu} u_{\nu} + P h_{\mu\nu} + u_{\mu} q_{\nu} + q_{\mu} u_{\nu} +\Pi_{\mu\nu}~,
\end{equation} 
$h_{\mu\nu} \equiv g_{\mu\nu} + u_{\mu} u_{\nu}$ being the covariant components of the projection tensor onto the space-like hyper-surfaces. The four-velocity vector $u_{\mu}$ satisfies the normalisation condition $u_{\mu} u^{\mu} = -1$, and $u^{\mu} q_{\mu} = 0$ and $u^{\mu} \Pi_{\mu\nu} = 0$. The components of this vector can be parameterised as follows 
\begin{equation} 
\label{eq:u-four-velocity-def}u^{\mu} = \frac{1}{a} \left(\delta^{\mu}_{0} + u^{(1)\mu} + \frac{1}{2} u^{(2)\mu} \right).
\end{equation} 
Using the normalisation condition and the perturbed components of the metric tensor, we have 
\begin{align}
&u^{0} = a^{-1} \left[1-A^{(1)} -\frac{1}{2} A^{(2)} +\frac{3}{2} (A^{(1)})^2 -B_i^{(1)} u^{(1)i} +\frac{1}{2} u_i^{(1)} u^{(1)i}\right],\\
\label{eq:comp-ui}&u^{i} = a^{-1} \left(u^{(1)i} +\frac{1}{2} u^{(2)i}\right).
\end{align} 
The covariant components are 
\begin{align}
&u_0 = -a\left[1+A^{(1)}+\frac{1}{2} A^{(2)} -\frac{1}{2} (A^{(1)})^2 +\frac{1}{2} u_i^{(1)} u^{(1)i}\right],\\
&u_i = -a\left[B_i^{(1)} -u_i^{(1)} +\frac{1}{2}\left(B_i^{(2)} -u_i^{(2)}\right) -B_i^{(1)} A^{(1)} -2C_{ij}^{(1)} u^{(1)j}\right].
\end{align}

Regarding the energy flux vector $q^{\mu}$, we parameterise it in the following way: 
\begin{equation} 
\label{eq:def-energy-flux-vector}q^{\mu} = \frac{1}{a} \left(q^{(1)\mu} + \frac{1}{2} q^{(2)\mu} \right).
\end{equation} 
The condition $u_{\mu} q^{\mu} = 0$ yields 
\begin{align}
&q^{0} = a^{-1}\left(u_i^{(1)} - B_i^{(1)} \right)q^{(1)i},\\
&q^{i} = a^{-1} \left(q^{(1)i} +\frac{1}{2} q^{(2)i} \right),
\end{align} 
while the covariant components read 
\begin{align}
&q_0 =-au_i^{(1)}q^{(1)i}, \\
&q_i = a\left(q^{(1)}_i +\frac{1}{2} q^{(2)}_i +2C^{(1)}_{ij} q^{(1)j}\right). 
\end{align} 

The anisotropic stress tensor $\Pi_{\mu\nu}$ has vanishing background components as well, and it may be written as 
\begin{equation} 
\label{eq:def-anisotropic-stress-tensor}\Pi_{\mu\nu} = a^2\left(\Pi_{\mu\nu}^{(1)}+\frac{1}{2} \Pi^{(2)}_{\mu\nu}\right).
\end{equation} 
This tensor is symmetric, and we can identify some components using the traceless property $g^{\mu\nu}\Pi_{\mu\nu} = \tensor{\Pi}{_{\mu}^{\mu}}=0$, and the relation $u^{\nu} \Pi_{\mu\nu} = 0$, such that
\begin{align}
&\Pi_{00} = 0~,\\
&\Pi_{0i} = \Pi_{i0} = -a^2\Pi^{(1)}_{ij}u^{(1)j}, \\
&\Pi_{ij} = a^2\left(\Pi^{(1)}_{ij} +\frac{1}{2} \Pi^{(2)}_{ij}\right),
\end{align} 
where $\tensor{\Pi}{^{(1)}_{i}^{i}} = 0$, while 
\begin{equation} 
\label{eq:trace-anisotropic-stress}\tensor{\Pi}{_i^{i}}=\frac{a^2}{2} \tensor{\Pi}{^{(2)}_{i}^{i}} = 2a^2C^{(1)}_{ij} \Pi^{(1)ij}.
\end{equation}  

The energy density and pressure are expanded as \begin{align}
&\rho(\eta,\mathbf{x}) = \rho^{(0)}(\eta)+\rho^{(1)}(\eta,\mathbf{x}) + \frac{1}{2} \rho^{(2)}(\eta,\mathbf{x})~,\\
&P(\eta,\mathbf{x}) = P^{(0)}(\eta)+P^{(1)}(\eta,\mathbf{x}) + \frac{1}{2} P^{(2)}(\eta,\mathbf{x})~,
\end{align} 
and the Equation-of-State parameter is defined as $w\equiv P^{(0)}/\rho^{(0)}$. Knowing all this, we can compute the mixed components of the energy-momentum tensor $\tensor{T}{_{\mu}^{\nu}}$, which read  
\begin{align}
\nonumber&\tensor{T}{_{0}^{0}} = -\left\{\rho^{(0)}+\rho^{(1)}+\frac{1}{2} \rho^{(2)}-B_i^{(1)} \left[q^{(1)i}+\left(\rho^{(0)}+P^{(0)}\right)u^{(1)i}\right]+2u_i^{(1)}\left[q^{(1)i}+\right.\right.\\
\label{eq:T-mixed-00}&\left.\left.+\frac{1}{2} \left(\rho^{(0)}+P^{(0)}\right)u^{(1)i}\right]\right\},\\
\nonumber&\tensor{T}{_{i}^{0}} = q^{(1)}_i + \left(\rho^{(0)}+P^{(0)}\right)\left(u_i^{(1)}-B_i^{(1)}\right)+\frac{1}{2} \left[q_i^{(2)}+\left(\rho^{(0)}+P^{(0)}\right)\left(u_i^{(2)}-B_i^{(2)}\right)\right]-\\
\nonumber&-A^{(1)} \left[q_i^{(1)}+\left(\rho^{(0)}+P^{(0)}\right)\left(u_i^{(1)}-2B_i^{(1)}\right)\right]+\left(\rho^{(1)}+P^{(1)}\right)\left(u_i^{(1)}-B_i^{(1)}\right)+\\
&+2C_{ij}^{(1)}\left[q^{(1)j}+\left(\rho^{(0)}+P^{(0)}\right)u^{(1)j}\right]+ \Pi^{(1)}_{ij} \left(u^{(1)j}-B^{(1)j}\right),\\
\nonumber&\tensor{T}{_{0}^{i}} = -\left\{q^{(1)i}+\left(\rho^{(0)}+P^{(0)}\right)u^{(1)i}+\frac{1}{2}\left[q^{(2)i}+\left(\rho^{(0)}+P^{(0)}\right)u^{(2)i}\right]+\right.\\
&\left.+A^{(1)}\left[q^{(1)i}+\left(\rho^{(0)}+P^{(0)}\right)u^{(1)i}\right]+\left(\rho^{(1)}+P^{(1)}\right)u^{(1)i}+u_j^{(1)} \Pi^{(1)ij}\right\},\\
\nonumber&\tensor{T}{_{i}^{j}} = P^{(0)} \delta_i^j+P^{(1)} \delta_i^j +\tensor{\Pi}{^{(1)}_i^j}+\frac{1}{2}P^{(2)} \delta_i^j +\frac{1}{2} \tensor{\Pi}{^{(2)}_i^j}+q_i^{(1)} u^{(1)j}+\left(u_i^{(1)}-B_i^{(1)} \right)\left[q^{(1)j}+\right.\\
\label{eq:T-mixed-ij}&\left.+\left(\rho^{(0)}+P^{(0)}\right)u^{(1)j}\right]-2\Pi^{(1)}_{ik} C^{(1)kj}.
\end{align} 
The trace is found to be 
\begin{equation}
\label{eq:trace-energy-momentum-tensor}T= -\left[\rho^{(0)}-3P^{(0)} +\rho^{(1)}-3P^{(1)}+\frac{1}{2}\left(\rho^{(2)}-3P^{(2)} \right) \right],
\end{equation} 
up to and including the second order. 

Using the metric components, we can compute the non-vanishing Christoffel symbols of the Levi-Civita connection
\begin{equation}
    \Gamma^{\alpha}_{\mu\nu}= \frac{1}{2}g^{\alpha\beta} \left(\partial_{\mu}g_{\beta\nu}+\partial_{\nu}g_{\beta\mu}-\partial_{\beta}g_{\mu\nu}\right),
\end{equation}
which read 
\begin{align}
\label{eq:Chris-beginning}&\Gamma^{0}_{00} = \mathcal{H}+A^{(1)'}+\frac{1}{2} A^{(2)'}-2A^{(1)} A^{(1)'}+\left(B_i^{(1)'}+\mathcal{H}B_i^{(1)}-\partial_i A^{(1)}\right)B^{(1)i},\\
\nonumber&\Gamma^{0}_{i0} = \Gamma^{0}_{0i} = \partial_i A^{(1)}-\mathcal{H} B_i^{(1)}+\frac{1}{2} \left(\partial_i A^{(2)}-\mathcal{H} B_i^{(2)}\right)-2A^{(1)}\left(\partial_i A^{(1)}-\mathcal{H} B_i^{(1)}\right)+\\
&+\left[\frac{1}{2} \left(\partial_i B_j^{(1)}-\partial_j B_i^{(1)}\right)-C_{ij}^{(1)'}\right]B^{(1)j},\\
\nonumber&\Gamma^{i}_{00} = -B^{(1)i'}-\mathcal{H} B^{(1)i}+\partial^{i} A^{(1)}-\frac{1}{2} \left(B^{(2)i'}+\mathcal{H} B^{(2)i}-\partial^{i} A^{(2)}\right)+A^{(1)'}B^{(1)i}+\\
&+2\left(B^{(1)'}_j +\mathcal{H} B_j^{(1)}-\partial_j A^{(1)}\right)C^{(1)ji},\\
\nonumber&\Gamma^{0}_{ij} = \mathcal{H} \delta_{ij} -2\mathcal{H} A^{(1)} \delta_{ij}+\frac{1}{2} \left(\partial_i B_j^{(1)}+\partial_j B_i^{(1)}\right)+C_{ij}^{(1)'}+2\mathcal{H}C_{ij}^{(1)}-\mathcal{H} A^{(2)} \delta_{ij}+\\
\nonumber&+\frac{1}{4} \left(\partial_i B_j^{(2)}+\partial_j B_i^{(2)}\right)+\frac{1}{2}C_{ij}^{(2)'}+\mathcal{H} C_{ij}^{(2)}+2A^{(1)} \left[2\mathcal{H} A^{(1)}\delta_{ij}-\frac{1}{2} \left(\partial_i B_j^{(1)}+\partial_j B_i^{(1)}\right)-\right.\\
&\left.-C_{ij}^{(1)'}-2\mathcal{H} C_{ij}^{(1)}\right]-\left(\mathcal{H} B_k^{(1)} \delta_{ij}+\partial_i C_{kj}^{(1)} + \partial_j C_{ki}^{(1)}-\partial_k C_{ij}^{(1)}\right)B^{(1)k},\\
\nonumber&\Gamma^{i}_{0j} = \Gamma^{i}_{j0} = \mathcal{H} \delta^i_{j} +\frac{1}{2} \left(\partial^{i} B_j^{(1)}-\partial_j B^{(1)i} \right)+\tensor{C}{^{(1)i}_j^{'}}+\frac{1}{4}\left(\partial^{i} B_j^{(2)}-\partial_j B^{(2)i} \right)+\frac{1}{2} \tensor{C}{^{(2)i}_{j}^{'}}-\\
&-B^{(1)i}\left(\mathcal{H} B_j^{(1)}-\partial_j A^{(1)} \right)-2C^{(1)ik} \left[\frac{1}{2} \left(\partial_k B_j^{(1)}-\partial_j B_k^{(1)} \right)+C_{kj}^{(1)'}\right],\\
\nonumber&\Gamma^{k}_{ij} = \mathcal{H} \delta_{ij} B^{(1)k} +\partial_i \tensor{C}{^{(1)k}_{j}}+\partial_j \tensor{C}{^{(1)k}_{i}}-\partial^{k} C_{ij}^{(1)}+\frac{1}{2} \mathcal{H} \delta_{ij} B^{(2)k}+\frac{1}{2} \left(\partial_i \tensor{C}{^{(2)k}_{j}}+\right.\\
\nonumber&\left.+\partial_j \tensor{C}{^{(2)k}_{i}}-\partial^{k} C_{ij}^{(2)}\right)-\left[2\mathcal{H} A^{(1)}\delta_{ij}-\frac{1}{2} \left(\partial_i B_j^{(1)}+\partial_j B_i^{(1)}\right)-C_{ij}^{(1)'}-2\mathcal{H} C_{ij}^{(1)}\right] B^{(1)k}-\\
\label{eq:Chris-end}&-2\left(\mathcal{H} \delta_{ij} B^{(1)}_s+\partial_i C_{js}^{(1)}+\partial_j C_{is}^{(1)}-\partial_s C_{ij}^{(1)}\right)C^{(1)sk}, 
\end{align}
$\mathcal{H} \equiv a^{'}/a$ being the comoving expansion rate (Hubble) parameter. Overprimes denote derivatives with respect to the conformal time. 

The Christoffel symbols can be employed to calculate the Riemann tensor components 
\begin{equation}
    \tensor{R}{^{\alpha}_{\beta\mu\nu}} \equiv \partial_{\mu} \Gamma^{\alpha}_{\nu\beta} - \partial_{\nu} \Gamma^{\alpha}_{\mu\beta} + \Gamma^{\alpha}_{\mu\lambda} \Gamma^{\lambda}_{\nu\beta} - \Gamma^{\alpha}_{\nu\lambda} \Gamma^{\lambda}_{\mu\beta}~.
\end{equation}
The non-zero ones read
\begin{align}
\nonumber&\tensor{R}{^{0}_{00i}} = -\tensor{R}{^{0}_{0i0}} = -\mathcal{H}^{'} B^{(1)}_i -\frac{1}{2} \mathcal{H}^{'} B_i^{(2)}+\left(\mathcal{H} A^{(1)'} +2\mathcal{H}^{'} A^{(1)} \right) B_i^{(1)} +B^{(1)j} \left[\partial_i \partial_j A^{(1)} -\right.\\
&\left.-\frac{1}{2}\left(\partial_i B_j^{(1)'} +\partial_j B_i^{(1)'}\right) -\frac{1}{2}\mathcal{H} \left(\partial_i B_j^{(1)}+\partial_j B_i^{(1)}\right)-C_{ij}^{(1)''}-\mathcal{H} C_{ij}^{(1)'}\right],\\
\nonumber&\tensor{R}{^{0}_{i0j}} = -\tensor{R}{^{0}_{ij0}} = \mathcal{H}^{'} \delta_{ij} -\left(\mathcal{H} A^{(1)'} +2\mathcal{H}^{'} A^{(1)} \right)\delta_{ij} -\partial_i \partial_j A^{(1)} +\frac{1}{2} \left[\partial_i B_j^{(1)'} + \partial_j B_i^{(1)'} +\right.\\
\nonumber&\left.+\mathcal{H} \left(\partial_i B_j^{(1)} + \partial_j B_i^{(1)} \right)\right]+C_{ij}^{(1)''} + 2\mathcal{H}^{'} C_{ij}^{(1)} + \mathcal{H} C_{ij}^{(1)'} -\frac{1}{2} \left(\mathcal{H} A^{(2)'} + 2\mathcal{H}^{'} A^{(2)}\right) \delta_{ij} -\\
\nonumber&-\frac{1}{2} \partial_i \partial_j A^{(2)}+\frac{1}{4} \left[\partial_i B_j^{(2)'} +\partial_j B_i^{(2)'} +\mathcal{H} \left(\partial_i B_j^{(2)}+\partial_j B_i^{(2)} \right)\right]+\frac{1}{2} C_{ij}^{(2)''} +\mathcal{H}^{'} C_{ij}^{(2)} +\\
\nonumber&+ \frac{1}{2} \mathcal{H} C_{ij}^{(2)'} +2A^{(1)}\left[2\left(\mathcal{H} A^{(1)'} + \mathcal{H}^{'} A^{(1)}\right) \delta_{ij} +\partial_i \partial_j A^{(1)} \right] +\left(\partial_i A^{(1)} - \mathcal{H} B_i^{(1)} \right) \partial_j A^{(1)}-\\
\nonumber&-A^{(1)} \left(\partial_i B_j^{(1)'} + \partial_j B_i^{(1)'} \right)-\frac{1}{2} \left(A^{(1)'} + 2\mathcal{H} A^{(1)}\right) \left(\partial_i B_j^{(1)} + \partial_j B_i^{(1)} \right)-\frac{1}{2} B^{(1)k} \left(\partial_i \partial_j B^{(1)}_k -\right.\\
\nonumber&\left.-\partial_k \partial_j B_i^{(1)} \right)-B^{(1)k} \left(\mathcal{H} B_k^{(1)'} + \mathcal{H}^{'} B_k^{(1)} \right) \delta_{ij}-\frac{1}{4} \left(\partial_i B_k^{(1)}-\partial_k B_i^{(1)} \right) \left(\partial_j B^{(1)k}-\partial^{k} B_j^{(1)} \right)-\\
\nonumber&-\left(B^{(1)k'} +\mathcal{H} B^{(1)k} -\partial^{k} A^{(1)} \right)\left(\partial_i C_{kj}^{(1)} + \partial_j C_{ki}^{(1)} - \partial_k C_{ij}^{(1)} \right)-A^{(1)'} \left(C_{ij}^{(1)'} +2\mathcal{H} C_{ij}^{(1)} \right)-\\
\nonumber&-2A^{(1)} \left(C_{ij}^{(1)''} +\mathcal{H} C_{ij}^{(1)'} + 2\mathcal{H}^{'} C_{ij}^{(1)} \right)+\frac{1}{2} \left[C_{ik}^{(1)'} \left(\partial_j B^{(1)k} - \partial^{k} B_j^{(1)} \right) + \left(\partial_i B^{(1)k}-\right.\right.\\
&\left.\left.- \partial^{k}B_i^{(1)}\right) C_{kj}^{(1)'} \right]-B^{(1)k}\left(\partial_i C_{kj}^{(1)'}-\partial_k C_{ij}^{(1)'}\right)-\tensor{C}{^{(1)}_i^{k'}}C^{(1)'}_{kj},\\
\nonumber&\tensor{R}{^{0}_{0ij}} = \tensor{R}{^{0}_{i0j}}-\tensor{R}{^{0}_{j0i}} = \mathcal{H} \left(\partial_i A^{(1)} B_j^{(1)} - \partial_j A^{(1)} B_i^{(1)} \right) -\frac{1}{2} B^{(1)k} \left(\partial_k \partial_i B_j^{(1)} -\partial_k \partial_j B_i^{(1)} \right)-\\
&-B^{(1)k} \left(\partial_i C_{kj}^{(1)'} -\partial_j C_{ki}^{(1)'} \right),\\
\nonumber&\tensor{R}{^{0}_{ijk}} =-\mathcal{H} \partial_j A^{(1)} \delta_{ik} +\frac{1}{2} \partial_i \partial_j B_k^{(1)} +\partial_j C_{ik}^{(1)'}-\frac{1}{2} \mathcal{H} \partial_j A^{(2)} \delta_{ik} +\frac{1}{4} \partial_i \partial_j B_k^{(2)} +\frac{1}{2} \partial_j C_{ik}^{(2)'} +\\
\nonumber&+A^{(1)} \left[2\left(2\mathcal{H} \partial_j A^{(1)} \delta_{ik} - \partial_j C_{ik}^{(1)'} \right) -\partial_i \partial_j B_k^{(1)} \right]-\partial_j A^{(1)} \left[\frac{1}{2} \left(\partial_i B_k^{(1)} +\partial_k B_i^{(1)} \right) + C_{ik}^{(1)'} +\right.\\
\nonumber&\left.+ 2\mathcal{H} C_{ik}^{(1)} \right]-B^{(1)s} \left(\mathcal{H} \partial_j B_s^{(1)} \delta_{ik}+\partial_i \partial_j C_{sk}^{(1)} - \partial_s \partial_j C_{ik}^{(1)} \right)+\frac{1}{2} \left(\partial_s B^{(1)}_j - \partial_j B_s^{(1)} \right)\left(\partial_i \tensor{C}{^{(1)s}_k}+\right.\\
&\left.+\partial_k \tensor{C}{^{(1)s}_i}-\partial^{s} C_{ik}^{(1)}\right)+\left(\partial_i \tensor{C}{^{(1)s}_{k}}+\partial_k \tensor{C}{^{(1)s}_i}-\partial^{s} C_{ik}^{(1)} \right)C_{sj}^{(1)'}-(j\leftrightarrow k)~,\\
\nonumber&\tensor{R}{^{i}_{00j}} = - \tensor{R}{^{i}_{0j0}} = \mathcal{H}^{'} \delta_j^{i} -\mathcal{H} A^{(1)'} \delta_j^{i}-\partial^{i} \partial_j A^{(1)}+\frac{1}{2} \left[\partial^{i} B^{(1)'}_j +\partial_j B^{(1)i'}+\mathcal{H} \left(\partial^{i} B^{(1)}_j +\right.\right.\\
\nonumber&\left.\left.+\partial_j B^{(1)i}\right)\right]+\tensor{C}{^{(1)i}_j^{''}}+\mathcal{H}\tensor{C}{^{(1)i}_{j}^{'}} -\frac{1}{2} \mathcal{H} A^{(2)'} \delta_j^{i}-\frac{1}{2} \partial^{i} \partial_j A^{(2)}+\frac{1}{4} \left[\partial^{i} B_j^{(2)'} + \partial_j B^{(2)i'} +\right.\\
\nonumber&\left.+\mathcal{H} \left(\partial^{i} B_j^{(2)} + \partial_j B^{(2)i} \right) \right]+\frac{1}{2} \tensor{C}{^{(2)i}_j^{''}}+\frac{1}{2} \mathcal{H} \tensor{C}{^{(2)i}_j^{'}}+A^{(1)'} \left[2\mathcal{H} A^{(1)} \delta_j^{i}-\frac{1}{2} \left(\partial^{i}B^{(1)}_j +\right.\right.\\
\nonumber&\left.\left.+\partial_j B^{(1)i} \right)-\tensor{C}{^{(1)i}_j^{'}}\right]+\left(B^{(1)k'} + \mathcal{H} B^{(1)k} - \partial^{k} A^{(1)} \right)\left(\partial_k \tensor{C}{^{(1)i}_j}-\partial^{i} C_{jk}^{(1)} -\partial_j \tensor{C}{^{(1)i}_k}-\right.\\
\nonumber&\left.-\mathcal{H} B_k^{(1)} \delta_j^{i} \right)+\left(\mathcal{H}^2-\mathcal{H}^{'} \right) B^{(1)i} B_j^{(1)}-\frac{1}{4} \left(\partial^{i} B_k^{(1)} - \partial_k B^{(1)i} \right) \left(\partial_j B^{(1)k}-\partial^{k} B_j^{(1)} \right)+\\
\nonumber&+\partial^{i} A^{(1)} \partial_j A^{(1)} -\mathcal{H} \left(\partial^{i} A^{(1)} B_j^{(1)} +B^{(1)i} \partial_j A^{(1)} \right)-C^{(1)ik'} C_{kj}^{(1)'} +2C^{(1)ik} \left\{\partial_k \partial_j A^{(1)}-\right.\\
\nonumber&\left.-\frac{1}{2} \left[\partial_k B_j^{(1)'} + \partial_j B_k^{(1)'} +\mathcal{H} \left(\partial_k B_j^{(1)} + \partial_j B_k^{(1)} \right)\right] -\mathcal{H} C_{kj}^{(1)'} -C_{kj}^{(1)''}\right\}+\frac{1}{2} \left[\tensor{C}{^{(1)i}_{k}^{'}}\left(\partial_j B^{(1)k}-\right.\right.\\
&\left.\left.-\partial^{k} B_j^{(1)}\right)+\tensor{C}{^{(1)k}_j^{'}}\left(\partial^{i} B_k^{(1)}-\partial_k B^{(1)i}\right)\right],\\
\nonumber&\tensor{R}{^{i}_{j0k}} = - \tensor{R}{^{i}_{jk0}}=\mathcal{H} \left[\left(\partial^{i} A^{(1)} -\mathcal{H} B^{(1)i}\right) \delta_{jk} - \left(\partial_j A^{(1)}-\mathcal{H} B_j^{(1)}\right)\delta_k^{i} \right]+\mathcal{H}^{'} \delta_{jk} B^{(1)i}-\\
\nonumber&-\frac{1}{2} \left(\partial_k \partial^{i} B_j^{(1)} - \partial_k \partial_j B^{(1)i} \right) +\partial_j \tensor{C}{^{(1)i}_{k}^{'}}-\partial^{i} C_{jk}^{(1)'}+\frac{1}{2} \mathcal{H} \left[\left(\partial^{i} A^{(2)} - \mathcal{H} B^{(2)i} \right) \delta_{jk}-\right.\\
\nonumber&\left.-\delta_k^{i}\left(\partial_j A^{(2)} - \mathcal{H} B_{j}^{(2)} \right)\right]+\frac{1}{2} \mathcal{H}^{'} \delta_{jk} B^{(2)i}-\frac{1}{4} \left(\partial_k \partial^{i} B_{j}^{(2)} - \partial_k \partial_j B^{(2)i} \right)+\frac{1}{2} \left(\partial_j \tensor{C}{^{(2)i}_{k}^{'}}-\right.\\
\nonumber&\left.-\partial^{i} C^{(2)'}_{jk}\right)-2\mathcal{H} A^{(1)}\left[\left(\partial^{i} A^{(1)} - \mathcal{H} B^{(1)i} \right) \delta_{jk} - \left(\partial_j A^{(1)}-\mathcal{H} B_j^{(1)}\right) \delta_k^{i}\right]-B^{(1)i} \left[\partial_j \partial_k A^{(1)} +\right.\\
\nonumber&\left.+\left(2\mathcal{H}^{'} A^{(1)} +\mathcal{H} A^{(1)'} \right)\delta_{jk}\right]+\frac{1}{2} \left[\partial^{i} A^{(1)} \left(\partial_j B_k^{(1)} + \partial_k B_j^{(1)} \right) -\partial_j A^{(1)} \left(\partial^{i} B_k^{(1)} + \partial_k B^{(1)i} \right) \right]+\\
\nonumber&+\frac{1}{2} \mathcal{H} B^{(1)s} \left[\delta_k^{i} \left(\partial_s B_j^{(1)}-\partial_j B_s^{(1)} \right) +\delta_{jk} \left(\partial^{i} B_s^{(1)} - \partial_s B^{(1)i}\right) \right]+\frac{1}{2} B^{(1)i} \left(\partial_j B_k^{(1)'} + \partial_k B_j^{(1)'} \right)+\\
\nonumber&+\frac{1}{2} \mathcal{H} B_j^{(1)} \left(\partial^{i} B_k^{(1)}+\partial_k B^{(1)i} \right)-\tensor{C}{^{(1)i}_{k}^{'}}\left(\partial_j A^{(1)} -\mathcal{H} B_j^{(1)} \right)+\partial^{i} A^{(1)} \left(C_{jk}^{(1)'}+2\mathcal{H} C_{jk}^{(1)} \right)-\\
\nonumber&-C^{(1)is} \left[2\mathcal{H} \delta_{jk} \left(\partial_s A^{(1)} -\mathcal{H} B_s^{(1)} \right)-\partial_k \partial_s B_j^{(1)} + \partial_j \partial_k B_s^{(1)} +2\mathcal{H}^{'} \delta_{jk} B_s^{(1)}+2\left(\partial_j C_{sk}^{(1)'}-\right.\right.\\
\nonumber&\left.\left.-\partial_s C_{jk}^{(1)'}\right)\right]-\mathcal{H}B^{(1)}_s\left(\delta_{jk} C^{(1)si'} -\delta_k^{i} \tensor{C}{^{(1)s}_{j}^{'}}\right)+B^{(1)i} \left[C_{jk}^{(1)''}+2\left(\mathcal{H}^{'} -\mathcal{H}^2 \right)C_{jk}^{(1)}\right]+\\
\nonumber&+\frac{1}{2} \left[\left(\partial^{i} B_s^{(1)}-\partial_s B^{(1)i} \right) \left(\partial_j \tensor{C}{^{(1)s}_{k}}+\partial_k \tensor{C}{^{(1)s}_{j}}-\partial^{s} C_{jk}^{(1)}\right)+\left(\partial_s B_j^{(1)}-\partial_j B_s^{(1)} \right)\left(\partial^i \tensor{C}{^{(1)s}_{k}}+\right.\right.\\
\nonumber&\left.\left.+\partial_k C^{(1)si}-\partial^{s} \tensor{C}{^{(1)i}_k}\right)\right]+\left(\partial^{i} \tensor{C}{^{(1)}_{k}^{s}}+\partial_k C^{(1)is}-\partial^s \tensor{C}{^{(1)i}_{k}}\right) C_{sj}^{(1)'}-C^{(1)is'} \left(\partial_j C_{sk}^{(1)} +\right.\\
&\left.+\partial_k C_{sj}^{(1)}-\partial_s C_{jk}^{(1)} \right),\\
\nonumber&\tensor{R}{^{i}_{0jk}} = \tensor{R}{^{i}_{j0k}}-\tensor{R}{^{i}_{k0j}} = -\mathcal{H} \left(\partial_j A^{(1)} - \mathcal{H} B_j^{(1)}\right)\delta_k^{i}-\frac{1}{2} \partial_k \partial^{i} B_j^{(1)} +\partial_j \tensor{C}{^{(1)i}_{k}^{'}}-\\
\nonumber&-\frac{1}{2} \mathcal{H} \delta_k^{i} \left(\partial_j A^{(2)} - \mathcal{H} B_j^{(2)}\right)-\frac{1}{4} \partial_k \partial^{i} B_j^{(2)}+\frac{1}{2} \partial_j \tensor{C}{^{(2)i}_{k}^{'}}+2\mathcal{H} A^{(1)}\delta_k^{i} \left(\partial_j A^{(1)} -\mathcal{H} B_j^{(1)} \right)-\\
\nonumber&-\frac{1}{2} \partial_j A^{(1)} \left(\partial^{i} B_k^{(1)}+\partial_k B^{(1)i} \right)+\frac{1}{2} \mathcal{H} B^{(1)s} \delta_k^{i} \left(\partial_s B_j^{(1)} - \partial_j B_s^{(1)} \right)+\frac{1}{2} \mathcal{H} B_j^{(1)} \left(\partial^{i} B_k^{(1)} +\right.\\
\nonumber&\left.+ \partial_k B^{(1)i} \right)-\tensor{C}{^{(1)i}_{k}^{'}}\left(\partial_j A^{(1)} -\mathcal{H} B_j^{(1)} \right)+C^{(1)is} \left(\partial_s \partial_k B_j^{(1)}-2\partial_j C_{sk}^{(1)'} \right)+\mathcal{H}B_s^{(1)} \delta_k^{i} \tensor{C}{^{(1)s}_{j}^{'}}+\\
\nonumber&+\frac{1}{2} \left(\partial_s B_j^{(1)}-\partial_j B_s^{(1)} \right) \left(\partial^{i} \tensor{C}{^{(1)s}_{k}} + \partial_k C^{(1)si} - \partial^{s} \tensor{C}{^{(1)i}_{k}}\right)+\left(\partial^{i} \tensor{C}{^{(1)}_k^{s}}+\partial_k C^{(1)is} -\right.\\
&\left.- \partial^{s} \tensor{C}{^{(1)i}_{k}}\right) C_{sj}^{(1)'} - (j\leftrightarrow k)~,\\
\nonumber&\tensor{R}{^{i}_{jkl}} = \mathcal{H}^2 \delta_k^{i} \delta_{jl} -2\mathcal{H}^2 A^{(1)} \delta_k^{i} \delta_{jl} +\frac{1}{2} \mathcal{H} \left[\delta_k^{i} \left(\partial_l B_j^{(1)} + \partial_j B_l^{(1)} \right) +\delta_{jl} \left(\partial^{i} B_k^{(1)} + \partial_k B^{(1)i} \right)\right]+\\
\nonumber&+2\mathcal{H}^2 \delta_k^{i} C_{jl}^{(1)} +\mathcal{H} \left(\tensor{C}{^{(1)i}_{k}^{'}}\delta_{jl} + \delta_k^{i} C_{jl}^{(1)'} \right)-\partial_k \partial^{i} C_{jl}^{(1)} +\partial_k \partial_j \tensor{C}{^{(1)i}_{l}}-\mathcal{H}^2 A^{(2)} \delta_k^{i} \delta_{jl}+\\
\nonumber&+\frac{1}{4} \mathcal{H} \left[\delta_k^{i} \left(\partial_l B_j^{(2)} + \partial_j B_l^{(2)} \right) + \delta_{jl} \left(\partial^{i} B_k^{(2)} + \partial_k B^{(2)i} \right)\right]+\mathcal{H}^2 \delta_{k}^{i} C_{jl}^{(2)}+\frac{1}{2} \mathcal{H} \left(\tensor{C}{^{(2)i}_{k}^{'}}\delta_{jl} +\right.\\
\nonumber&\left.+ \delta_{k}^{i} C_{jl}^{(2)'} \right)-\frac{1}{2} \left(\partial_k \partial^{i} C_{jl}^{(2)} -\partial_k \partial_j \tensor{C}{^{(2)i}_{l}}\right)+\mathcal{H} A^{(1)} \left[\left(4\mathcal{H} A^{(1)} \delta_{k}^{i} -\partial^{i} B_k^{(1)} - \partial_k B^{(1)i} \right) \delta_{jl}-\right.\\
\nonumber&\left.-\left(\partial_l B_j^{(1)} + \partial_j B_l^{(1)} \right) \delta_k^{i} \right]-B^{(1)i} \left(\mathcal{H} \partial_k A^{(1)} \delta_{jl} -\frac{1}{2} \partial_k \partial_j B_l^{(1)} \right)-\mathcal{H}^2 B_s^{(1)} B^{(1)s} \delta_k^{i} \delta_{jl}+\\
\nonumber&+\frac{1}{4} \left(\partial^{i} B_k^{(1)} + \partial_k B^{(1)i} \right)\left(\partial_l B_j^{(1)} + \partial_j B_l^{(1)} \right)-2\mathcal{H} A^{(1)} \left[\delta_k^{i}\left(C_{jl}^{(1)'}+2\mathcal{H} C_{jl}^{(1)}\right)+\delta_{jl} \tensor{C}{^{(1)i}_{k}^{'}}\right]+\\
\nonumber&+B^{(1)i} \partial_k C_{jl}^{(1)'} +\frac{1}{2} \left[\tensor{C}{^{(1)i}_{k}^{'}}\left(\partial_l B_j^{(1)} + \partial_j B_l^{(1)} \right)+C_{jl}^{(1)'} \left(\partial^{i} B_k^{(1)} + \partial_k B^{(1)i} \right) \right]-\\
\nonumber&-\mathcal{H}\left[C^{(1)is} \left(\partial_s B_k^{(1)} + \partial_k B_s^{(1)} \right) \delta_{jl}-C_{jl}^{(1)}\left(\partial^{i} B_k^{(1)} + \partial_k B^{(1)i} \right)\right]-\mathcal{H} B^{(1)s} \left[\delta_k^{i} \left(\partial_l C_{js}^{(1)} +\right.\right.\\
\nonumber&\left.\left.+ \partial_j C_{ls}^{(1)} -\partial_s C_{jl}^{(1)} \right)+\delta_{jl} \left(\partial^{i} C_{ks}^{(1)} + \partial_k \tensor{C}{^{(1)i}_s}-\partial_s \tensor{C}{^{(1)i}_{k}}\right) \right]+\tensor{C}{^{(1)i}_{k}^{'}}\left( C_{jl}^{(1)'}+2\mathcal{H} C_{jl}^{(1)} \right)-\\
\nonumber&-2\mathcal{H} C^{(1)is} C_{sk}^{(1)'} \delta_{jl}-\left(\partial^{i} C_{ks}^{(1)} + \partial_k \tensor{C}{^{(1)i}_{s}}-\partial_s \tensor{C}{^{(1)i}_{k}}\right) \left(\partial_l \tensor{C}{^{(1)s}_{j}}+\partial_j \tensor{C}{^{(1)s}_{l}}-\partial^{s} C_{jl}^{(1)} \right)-\\
&-2\left(\partial_k \partial_j C_{ls}^{(1)}-\partial_k \partial_s C_{jl}^{(1)} \right) C^{(1)si}-(k\leftrightarrow l)~.
\end{align}

The Ricci tensor can be obtained from the Riemann one by taking the trace $\tensor{R}{^{\sigma}_{\mu\sigma\nu}}$. The mixed components read
\begin{align}
    \nonumber&\tensor{R}{_{0}^{0}}=a^{-2} \left[3\mathcal{H}^{'} -3\mathcal{H} A^{(1)'} -6\mathcal{H}^{'} A^{(1)} -\partial_i \partial^{i} A^{(1)} +\partial_i B^{(1)i'}+\mathcal{H} \partial_i B^{(1)i} +\tensor{C}{^{(1)i}_{i}^{''}}+\mathcal{H} \tensor{C}{^{(1)i}_{i}^{'}}-\right.\\
    \nonumber&\left.-\frac{3}{2} \mathcal{H} A^{(2)'}-3\mathcal{H}^{'} A^{(2)}-\frac{1}{2} \partial_i \partial^{i} A^{(2)}+\frac{1}{2} \partial_i B^{(2)i'} +\frac{1}{2} \mathcal{H} \partial_i B^{(2)i}+\frac{1}{2} \tensor{C}{^{(2)i}_{i}^{''}}+\frac{1}{2} \mathcal{H} \tensor{C}{^{(2)i}_{i}^{'}}+\right.\\
    \nonumber&\left.+2A^{(1)} \partial_i \partial^{i} A^{(1)}+12A^{(1)} \left(\mathcal{H} A^{(1)'}+\mathcal{H}^{'} A^{(1)}\right)+\partial_i A^{(1)}\left(\partial^{i} A^{(1)}-\mathcal{H} B^{(1)i} \right)-2A^{(1)} \partial_i B^{(1)i'}-\right.\\
    \nonumber&\left.-\left(A^{(1)'}+2\mathcal{H} A^{(1)}\right) \partial_i B^{(1)i}-3B_i^{(1)} \left(\mathcal{H} B^{(1)i'}+\mathcal{H}^{'} B^{(1)i}\right)-\frac{1}{2} B^{(1)i} \left(\partial_j \partial^{j} B_i^{(1)}-\partial_i \partial_j B^{(1)j} \right)-\right.\\
    \nonumber&\left.-\frac{1}{2} \partial^{i} B^{(1)j}\left(\partial_i B_j^{(1)}-\partial_j B^{(1)}_i\right)-\left(A^{(1)'} +2\mathcal{H} A^{(1)} \right) \tensor{C}{^{(1)i}_{i}^{'}}-2A^{(1)} \tensor{C}{^{(1)i}_{i}^{''}}-C^{(1)ij'}C_{ij}^{(1)'}-\right.\\
    \nonumber&\left.-\left(\partial^{i} A^{(1)}-\mathcal{H} B^{(1)i} -B^{(1)i'} \right)\left(\partial_i \tensor{C}{^{(1)j}_{j}}-2\partial_j \tensor{C}{^{(1)}_i^{j}}\right)+B^{(1)i} \left(\partial_i \tensor{C}{^{(1)j}_{j}^{'}}-\partial_j \tensor{C}{^{(1)}_i^{j'}}\right)+\right.\\
    \label{eq:R00-upt-second-order-full}&\left.+2C^{(1)ij} \left(\partial_i \partial_j A^{(1)}-\partial_i B^{(1)'}_j -\mathcal{H} \partial_i B^{(1)}_j-\mathcal{H} C_{ij}^{(1)'}-C_{ij}^{(1)''}\right)\right],\\
    \nonumber&\tensor{R}{_i^{0}}=a^{-2} \left[-2\mathcal{H} \partial_i A^{(1)}+\frac{1}{2} \left(\partial_i \partial_j B^{(1)j}-\partial_j \partial^{j} B_i^{(1)} \right)+\partial_i \tensor{C}{^{(1)j}_{j}^{'}}-\partial_j \tensor{C}{^{(1)}_{i}^{j'}}-\mathcal{H} \partial_i A^{(2)}+\right.\\
    \nonumber&\left.+\frac{1}{4} \left(\partial_i \partial_j B^{(2)j}-\partial_j \partial^{j} B_i^{(2)}\right)+\frac{1}{2} \left(\partial_i \tensor{C}{^{(2)j}_{j}^{'}}-\partial_j \tensor{C}{^{(2)}_i^{j'}}\right)+8\mathcal{H} A^{(1)} \partial_i A^{(1)}-\partial_i A^{(1)} \partial_j B^{(1)j}+\right.\\
    \nonumber&\left.+\frac{1}{2} \partial^{j} A^{(1)} \left(\partial_i B_j^{(1)}+\partial_j B^{(1)}_i\right)-2\mathcal{H} B^{(1)j} \partial_i B_j^{(1)}-A^{(1)} \left(\partial_i \partial_j B^{(1)j} - \partial_j \partial^{j} B_i^{(1)} \right)+C_{ij}^{(1)'} \partial^{j} A^{(1)}-\right.\\
    \nonumber&\left.-\tensor{C}{^{(1)j}_{j}^{'}}\partial_i A^{(1)}+2A^{(1)} \left(\partial_j \tensor{C}{^{(1)}_{i}^{j'}}-\partial_i \tensor{C}{^{(1)j}_{j}^{'}}\right)-C^{(1)jk} \left(\partial_i \partial_j B_k^{(1)}-\partial_j \partial_k B_i^{(1)} \right)+\right.\\
    \nonumber&\left.+B^{(1)j} \left(\partial_k \partial^{k} C_{ij}^{(1)}+\partial_i \partial_j \tensor{C}{^{(1)k}_k}-\partial_i \partial_k \tensor{C}{^{(1)k}_{j}}-\partial_j \partial_k \tensor{C}{^{(1)k}_{i}}\right)-C^{(1)jk'} \partial_i C_{jk}^{(1)}-\right.\\
    \nonumber&\left.-\frac{1}{2} \left(\partial_i B_j^{(1)} -\partial_j B_i^{(1)}\right) \left(2\partial^{k} \tensor{C}{^{(1)j}_{k}}-\partial^{j} \tensor{C}{^{(1)k}_{k}}\right)-\partial_k \tensor{C}{^{(1)}_i^{j}}\left(\partial_j B^{(1)k} - \partial^{k} B_j^{(1)} \right)+\right.\\
    &\left.+C_{ij}^{(1)'} \left(2\partial^{k} \tensor{C}{^{(1)j}_{k}}-\partial^{j} \tensor{C}{^{(1)k}_{k}}\right)-2C^{(1)jk} \left(\partial_i C_{jk}^{(1)'} - \partial_k C_{ij}^{(1)'} \right)\right],\\
    \nonumber&\tensor{R}{_{0}^{i}}=a^{-2} \left[2\mathcal{H} \partial^{i} A^{(1)}+2\left(\mathcal{H}^{'}-\mathcal{H}^2\right) B^{(1)i}-\frac{1}{2}\left(\partial^{i} \partial_j B^{(1)j}-\partial_j \partial^{j} B^{(1)i}\right)-\partial^{i} \tensor{C}{^{(1)j}_j^{'}}+\right.\\
    \nonumber&\left.+\partial_j C^{(1)ij'}+\mathcal{H} \partial^{i} A^{(2)}+\left(\mathcal{H}^{'}-\mathcal{H}^2\right)B^{(2)i}-\frac{1}{4}\left(\partial^{i} \partial_j B^{(2)j}-\partial_j \partial^{j} B^{(2)i}\right)-\frac{1}{2}\left(\partial^{i} \tensor{C}{^{(2)j}_{j}^{'}}-\right.\right.\\
    \nonumber&\left.\left.-\partial_j C^{(2)ij'}\right)-2\mathcal{H}A^{(1)'} B^{(1)i}-4\mathcal{H} A^{(1)} \partial^{i} A^{(1)}-4\left(\mathcal{H}^{'}-\mathcal{H}^2\right)A^{(1)} B^{(1)i}+\partial^{i} A^{(1)} \partial_j B^{(1)j}-\right.\\
    \nonumber&\left.-B^{(1)i}\partial_j \partial^{j} A^{(1)}+B^{(1)}_j \partial^{i} \partial^{j} A^{(1)}-\frac{1}{2} \partial^{j} A^{(1)} \left(\partial^{i} B_j^{(1)}+\partial_j B^{(1)i} \right)+B^{(1)i} \partial_j B^{(1)j'}-\right.\\
    \nonumber&\left.-\frac{1}{2} B^{(1)j}\left(\partial^{i} B_j^{(1)'}+\partial_j B^{(1)i'}\right)+\mathcal{H}B^{(1)j}\left(\partial^{i} B_j^{(1)}-\partial_j B^{(1)i}\right)+\tensor{C}{^{(1)j}_j^{'}}\partial^{i} A^{(1)}-\right.\\
    \nonumber&\left.-\left(C^{(1)ij'}+4\mathcal{H} C^{(1)ij}\right) \partial_j A^{(1)}+B^{(1)i} \tensor{C}{^{(1)j}_{j}^{''}}-B_j^{(1)} C^{(1)ij''}-2\mathcal{H}B_j^{(1)} C^{(1)ij'}-\right.\\
    \nonumber&\left.-4\left(\mathcal{H}^{'}-\mathcal{H}^2\right)C^{(1)ij}B_j^{(1)}+C^{(1)ij} \left(\partial_j \partial_k B^{(1)k}-\partial_k \partial^{k} B_j^{(1)}\right)+C^{(1)jk} \left(\partial^{i} \partial_j B_k^{(1)}-\right.\right.\\
    \nonumber&\left.\left.-\partial_j \partial_k B^{(1)i}\right)+\frac{1}{2} \left(\partial^{i} B^{(1)j} - \partial^{j} B^{(1)i} \right)\left(2\partial_k \tensor{C}{^{(1)}_j^{k}}-\partial_j \tensor{C}{^{(1)k}_{k}}\right)+\partial_k C^{(1)ij}\left(\partial_j B^{(1)k}-\right.\right.\\
    \nonumber&\left.\left.-\partial^{k} B^{(1)}_j\right)+C_{jk}^{(1)'} \partial^{i} C^{(1)jk}+2C^{(1)ij}\left(\partial_j \tensor{C}{^{(1)k}_k^{'}}-\partial_k \tensor{C}{^{(1)}_j^{k'}}\right)+2C^{(1)jk}\left(\partial^{i} C_{jk}^{(1)'}-\right.\right.\\
    &\left.\left.-\partial_k \tensor{C}{^{(1)i}_{j}^{'}}\right)-C^{(1)ij'}\left(2\partial_k \tensor{C}{^{(1)k}_{j}}-\partial_j \tensor{C}{^{(1)k}_{k}}\right)\right],\\
    \nonumber&\tensor{R}{_i^{j}}=a^{-2} \left[2\mathcal{H}^2 \delta_i^{j} + \mathcal{H}^{'} \delta_{i}^{j}-4\mathcal{H}^2 A^{(1)} \delta_i^{j}-2\mathcal{H}^{'} A^{(1)} \delta_i^{j}-\mathcal{H} A^{(1)'} \delta_i^{j}-\partial_i \partial^{j} A^{(1)}+\mathcal{H} \partial_k B^{(1)k} \delta_i^{j}+\right.\\
    \nonumber&\left.+\frac{1}{2} \left(\partial_i B^{(1)j'}+\partial^{j} B_i^{(1)'} \right)+\mathcal{H} \left(\partial_i B^{(1)j} + \partial^{j} B_i^{(1)} \right)+\tensor{C}{^{(1)}_i^{j''}}+2\mathcal{H} \tensor{C}{^{(1)}_i^{j'}}+\mathcal{H} \tensor{C}{^{(1)k}_k^{'}}\delta_i^{j}-\right.\\
    \nonumber&\left.-\partial_k \partial^{k} \tensor{C}{^{(1)}_i^{j}}-\partial_i \partial^{j} \tensor{C}{^{(1)k}_{k}}+\partial_i \partial_k C^{(1)kj}+\partial^{j} \partial_k \tensor{C}{^{(1)k}_{i}}-2\mathcal{H}^2 A^{(2)} \delta_i^{j}-\mathcal{H}^{'} A^{(2)} \delta_i^{j}-\right.\\
    \nonumber&\left.-\frac{1}{2} \mathcal{H} A^{(2)'} \delta_i^{j}-\frac{1}{2} \partial_i \partial^{j} A^{(2)}+\frac{1}{2} \mathcal{H} \partial_k B^{(2)k} \delta_i^{j}+\frac{1}{4} \left(\partial_i B^{(2)j'} + \partial^{j} B^{(2)'}_i\right)+\frac{1}{2} \mathcal{H} \left(\partial_i B^{(2)j} + \right.\right.\\
    \nonumber&\left.\left.+\partial^{j} B_i^{(2)}\right)+\frac{1}{2} \tensor{C}{^{(2)}_i^{j''}}+\mathcal{H} \tensor{C}{^{(2)}_i^{j'}}+\frac{1}{2} \mathcal{H} \tensor{C}{^{(2)k}_{k}^{'}}\delta_i^{j}-\frac{1}{2} \partial_k \partial^{k} \tensor{C}{^{(2)}_i^{j}}-\frac{1}{2} \partial_i \partial^{j} \tensor{C}{^{(2)k}_k}+\right.\\
    \nonumber&\left.+\frac{1}{2} \partial_i \partial^{k} \tensor{C}{^{(2)}_k^{j}}+\frac{1}{2} \partial^{j} \partial_k \tensor{C}{^{(2)k}_i}+8\mathcal{H}^2 \left(A^{(1)}\right)^2 \delta_i^{j}+4\mathcal{H}^{'}\left(A^{(1)}\right)^2 \delta_i^{j}+\partial_i A^{(1)} \partial^{j} A^{(1)}+\right.\\
    \nonumber&\left.+2A^{(1)} \partial_i \partial^{j} A^{(1)}+4\mathcal{H} A^{(1)} A^{(1)'} \delta_i^{j}-A^{(1)} \left(\partial_i B^{(1)j'} + \partial^{j} B^{(1)'}_i\right)-2\mathcal{H} \partial_i A^{(1)} B^{(1)j}-\right.\\
    \nonumber&\left.-\frac{1}{2} \left(A^{(1)'} + 4\mathcal{H} A^{(1)} \right)\left(\partial_i B^{(1)j} + \partial^{j} B_i^{(1)} \right)-\mathcal{H} B^{(1)k} \partial_k A^{(1)} \delta_i^{j}-2\mathcal{H}A^{(1)} \partial_k B^{(1)k} \delta_i^{j}-\right.\\
    \nonumber&\left.-B^{(1)k} \left(\mathcal{H} B_k^{(1)'} + \mathcal{H}^{'} B^{(1)}_k +2\mathcal{H}^2 B_k^{(1)} \right) \delta_i^{j}-B^{(1)k}\partial_i \partial^{j} B_k^{(1)}+\frac{1}{2} B^{(1)j} \partial_i \partial_k B^{(1)k}-\right.\\
    \nonumber&\left.-\frac{1}{2} B^{(1)j} \partial_k \partial^{k} B_i^{(1)}+\frac{1}{2} \partial_k B^{(1)k} \left(\partial_i B^{(1)j} + \partial^{j} B^{(1)}_i \right)+\frac{1}{2} B^{(1)k}\left(\partial_k \partial_i B^{(1)j} + \partial_k \partial^{j} B_i^{(1)} \right)-\right.\\
    \nonumber&\left.-\frac{1}{2} \left(\partial_i B_k^{(1)} \partial^{j} B^{(1)k} + \partial_k B_i^{(1)} \partial^{k} B^{(1)j} \right)-A^{(1)'} \tensor{C}{^{(1)}_i^{j'}}+2C^{(1)jk}\partial_i \partial_k A^{(1)}-\left(B^{(1)k'}+\right.\right.\\
    \nonumber&\left.\left.+2\mathcal{H} B^{(1)k}-\partial^{k} A^{(1)}\right)\left(\partial_i \tensor{C}{^{(1)}_k^{j}}+\partial^{j} C^{(1)}_{ki} -\partial_k \tensor{C}{^{(1)}_i^{j}}\right)-2A^{(1)}\left(\tensor{C}{^{(1)}_i^{j''}}+2\mathcal{H} \tensor{C}{^{(1)}_i^{j'}}+\right.\right.\\
    \nonumber&\left.\left.+\mathcal{H} \tensor{C}{^{(1)k}_{k}^{'}}\delta_i^{j}\right)-B^{(1)j} \partial^{k} C_{ik}^{(1)'}+B^{(1)j} \partial_i \tensor{C}{^{(1)k}_{k}^{'}}+\partial_k B^{(1)k} \tensor{C}{^{(1)}_i^{j'}}-2\mathcal{H} C^{(1)ks} \partial_k B_s^{(1)} \delta_i^{j}-\right.\\
    \nonumber&\left.-\tensor{C}{^{(1)k}_i^{'}}\partial_k B^{(1)j}-C^{(1)jk'} \partial_k B_i^{(1)}-C^{(1)jk}\left(\partial_i B_k^{(1)'} + \partial_k B_i^{(1)'} \right)+B^{(1)k} \left(2\partial_k \tensor{C}{^{(1)}_i^{j'}}-\right.\right.\\
    \nonumber&\left.\left.-\partial_i \tensor{C}{^{(1)}_k^{j'}}-\partial^{j} C_{ki}^{(1)'} \right)-2\mathcal{H} C^{(1)jk} \left(\partial_i B_k^{(1)}+\partial_k B_i^{(1)} \right)-\mathcal{H} B^{(1)k} \left(2\partial^{s} C_{sk}^{(1)} - \partial_k \tensor{C}{^{(1)s}_{s}}\right) \delta_i^{j}+\right.\\
    \nonumber&\left.+\frac{1}{2} \tensor{C}{^{(1)k}_{k}^{'}}\left(\partial_i B^{(1)j} + \partial^{j} B_i^{(1)} \right)+\partial_i \tensor{C}{^{(1)s}_{k}}\partial^{j} \tensor{C}{^{(1)k}_{s}}+\tensor{C}{^{(1)k}_{k}^{'}}\tensor{C}{^{(1)}_i^{j'}}-2C^{(1)jk} C_{ki}^{(1)''}-\right.\\
    \nonumber&\left.-4\mathcal{H} C^{(1)jk} C_{ki}^{(1)'}-2\tensor{C}{^{(1)}_i^{k'}}\tensor{C}{^{(1)}_k^{j'}}+2\partial_k \tensor{C}{^{(1)s}_i}\left(\partial^{k} \tensor{C}{^{(1)}_s^{j}}-\partial_s C^{(1)kj}\right)-2\mathcal{H} C^{(1)ks} C_{ks}^{(1)'} \delta_i^{j}+\right.\\
    \nonumber&\left.+2C^{(1)jk} \left(\partial_s \partial^{s} C_{ik}^{(1)}+\partial_i \partial_k \tensor{C}{^{(1)s}_{s}}-\partial_i \partial_s \tensor{C}{^{(1)s}_{k}}-\partial_k \partial_s \tensor{C}{^{(1)s}_{i}}\right)-2C^{(1)ks} \left(\partial_k \partial_i \tensor{C}{^{(1)}_s^{j}}+\right.\right.\\
    \nonumber&\left.\left.+\partial_k \partial^{j} C_{is}^{(1)}-\partial_i \partial^{j} C_{ks}^{(1)} -\partial_k \partial_s \tensor{C}{^{(1)}_i^{j}}\right)-\left(2\partial^{k} C_{ks}^{(1)}-\partial_s \tensor{C}{^{(1)k}_{k}}\right)\left(\partial_i C^{(1)sj} + \partial^{j} \tensor{C}{^{(1)s}_{i}}-\right.\right.\\
    \label{eq:Rij-upt-second-order-full}&\left.\left.-\partial^{s} \tensor{C}{^{(1)}_i^{j}}\right)\right].
\end{align}

Finally, the Ricci scalar, $R = \tensor{R}{_0^{0}} + \tensor{R}{_i^{i}}$, up to second order, is
\begin{align}
    \nonumber&R=a^{-2} \left[6\mathcal{H}^2+6\mathcal{H}^{'} -6\mathcal{H} A^{(1)'}-12\mathcal{H}^2 A^{(1)}-12\mathcal{H}^{'} A^{(1)}-2\partial_i \partial^{i} A^{(1)}+2\partial_i B^{(1)i'}+\right.\\
    \nonumber&\left.+6\mathcal{H} \partial_i B^{(1)i}+2\tensor{C}{^{(1)i}_{i}^{''}}+6\mathcal{H} \tensor{C}{^{(1)i}_{i}^{'}}+2\left(\partial_i \partial_j C^{(1)ij} - \partial_j \partial^{j} \tensor{C}{^{(1)i}_{i}}\right)-3\mathcal{H} A^{(2)'}-\right.\\
    \nonumber&\left.-6\mathcal{H}^2 A^{(2)} -6\mathcal{H}^{'} A^{(2)}-\partial_i \partial^{i} A^{(2)}+\partial_i B^{(2)i'}+3\mathcal{H} \partial_i B^{(2)i}+\tensor{C}{^{(2)i}_{i}^{''}}+3\mathcal{H} \tensor{C}{^{(2)i}_{i}^{'}}+\right.\\
    \nonumber&\left.+\partial_i \partial_j C^{(2)ij} - \partial_j \partial^{j} \tensor{C}{^{(2)i}_{i}}+24\mathcal{H} A^{(1)} A^{(1)'}+24\mathcal{H}^2 (A^{(1)})^2+24\mathcal{H}^{'} (A^{(1)})^2+\right.\\
    \nonumber&\left.+4A^{(1)} \partial_i \partial^{i} A^{(1)}+2\partial^i A^{(1)}\left(\partial_{i} A^{(1)}-3\mathcal{H} B^{(1)}_i \right)-2A^{(1)'} \partial_i B^{(1)i} -4A^{(1)} \left(\partial_i B^{(1)i'} +\right.\right.\\
    \nonumber&\left.\left.+3\mathcal{H} \partial_i B^{(1)i} \right)-6B^{(1)i} \left(\mathcal{H} B_i^{(1)'} + \mathcal{H}^2 B^{(1)}_i + \mathcal{H}^{'} B_i^{(1)} \right)+2B^{(1)i} \left(\partial_i \partial_j B^{(1)j}-\right.\right.\\
    \nonumber&\left.\left.-\partial_j \partial^{j} B_i^{(1)}\right)-\frac{1}{2} \partial^{i} B^{(1)j} \left(3\partial_i B_j^{(1)}-\partial_j B_i^{(1)} \right)+\left(\partial_i B^{(1)i} \right)^2-2A^{(1)'}\tensor{C}{^{(1)i}_{i}^{'}}-\right.\\
    \nonumber&\left.-4A^{(1)}\left(\tensor{C}{^{(1)i}_{i}^{''}}+3\mathcal{H} \tensor{C}{^{(1)i}_{i}^{'}}\right)-2\left(\partial^{i} A^{(1)} -B^{(1)i'}\right)\left(\partial_i \tensor{C}{^{(1)j}_{j}}-2\partial_j \tensor{C}{^{(1)}_i^{j}}\right)+\right.\\
    \nonumber&\left.+4B^{(1)i} \left(\partial_i \tensor{C}{^{(1)j}_{j}^{'}}-\partial_j \tensor{C}{^{(1)}_i^{j'}}\right)-2C^{(1)ij'} \partial_i B_j^{(1)}+2\tensor{C}{^{(1)i}_{i}^{'}}\partial_j B^{(1)j}+\right.\\
    \nonumber&\left.+6\mathcal{H} B^{(1)i} \left(\partial_i \tensor{C}{^{(1)j}_{j}}-2\partial_j \tensor{C}{^{(1)}_i^{j}}\right)+4C^{(1)ij} \left(\partial_i \partial_j A^{(1)}-\partial_i B^{(1)'}_j -3\mathcal{H} \partial_i B_j^{(1)} -\right.\right.\\
    \nonumber&\left.\left.-C_{ij}^{(1)''}-3\mathcal{H} C_{ij}^{(1)'}\right)+\left(\tensor{C}{^{(1)i}_{i}^{'}}\right)^2-3C^{(1)ij'} C_{ij}^{(1)'}+\partial_i \tensor{C}{^{(1)j}_{k}}\left(3\partial^{i} \tensor{C}{^{(1)}_j^{k}}-\right.\right.\\
    \nonumber&\left.\left.-2\partial_j C^{(1)ik}\right)-\left(\partial_i \tensor{C}{^{(1)j}_{j}}-2\partial^{j} C_{ij}^{(1)} \right) \left(\partial^{i} \tensor{C}{^{(1)k}_{k}}-2\partial^{k} \tensor{C}{^{(1)i}_{k}}\right)+\right.\\
    \label{eq:Ricci-scalar-upt-second-order-full}&\left.+4C^{(1)ij} \left(\partial_k \partial^{k} C^{(1)}_{ij}+\partial_i \partial_j \tensor{C}{^{(1)k}_{k}}-2\partial_i \partial_k \tensor{C}{^{(1)k}_{j}}\right)\right].
\end{align}

\section{Palatini $f(R)$ Gravity in the Einstein Frame}
\label{ap:Einstein-frame-discussion}
In this section, we bring the action in Eq.~\eqref{eq:mainaction} to the Einstein frame, where the gravitational sector takes on the form of the Einstein-Hilbert (EH) term of General Relativity (GR). The interest in the Einstein frame lies in the fact that the tensor perturbations of the metric are conformally invariant at all orders in perturbation theory \cite{Gong:2011qe} (see Refs.~\cite{White:2012ya,Kubota:2011re,Diaz:2023tma} too), meaning that one will have the same evolution of the scalar-induced ones in both the Jordan and Einstein frames. In order to recast the action as the Einstein frame one, we use the Legendre transform first, such that
\begin{equation}
    S= \int \textrm{d}^4x \sqrt{-g} \left\{\frac{m^2_{\textrm{P}}}{2}\left[f(\chi) +f_{,\chi}(\hat{R}-\chi)\right]+\mathcal{L}_{\textrm{M}}(g_{\mu\nu},\Psi)\right\}.
\end{equation}
A variation with respect to $\chi$ yields $\chi = \hat{R}$ unless $f_{,\chi\chi}=0$, and hence one recovers Eq.~\eqref{eq:mainaction}. We now perform a conformal rescaling of the metric \cite{Dabrowski:2008kx} (which does not alter the spacetime coordinates)
\begin{equation}
    \label{eq:conf-trans-fchi}\tilde{g}_{\mu\nu}(x) = f_{,\chi}(\chi(x)) g_{\mu\nu}(x)~,
\end{equation}
bearing in mind that $\tilde{\hat{R}}=\tilde{g}^{\mu\nu} \hat{R}_{\mu\nu} = f_{,\chi}^{-1} \hat{R}$. The action above takes on the following form:
\begin{equation}
    \label{eq:rescaled-action-Einstein}S = \int \textrm{d}^4x \sqrt{-\tilde{g}}\left\{\frac{m^2_{\textrm{P}}}{2}\tilde{\hat{R}}-U(\chi)+\tilde{\mathcal{L}}_{\textrm{M}}(f^{-1}_{,\chi}\tilde{g}_{\mu\nu},\Psi)\right\},
\end{equation}
where $U(\chi)$ is
\begin{equation}
   \label{eq:definition-U-chi} U(\chi) \equiv \frac{m^2_{\textrm{P}}(f_{,\chi}\chi-f)}{2f_{,\chi}^2}~,
\end{equation}
and $\tilde{\mathcal{L}}_{\textrm{M}} = f^{-2}_{,\chi} \mathcal{L}_{\textrm{M}}$. We now vary the action in the new frame with respect to $\chi$ (and set the variation to zero), and hence
\begin{equation}
    \label{eq:trace-equation-Einstein}m^2_{\textrm{P}}\chi^3 \left(\frac{f}{\chi^2}\right)_{,\chi} = T~,
\end{equation}
where
\begin{equation}
    \frac{\delta(\sqrt{-\tilde{g}}\tilde{\mathcal{L}}_{\textrm{M}})}{\delta \chi} = -\frac{\sqrt{-\tilde{g}}f_{,\chi\chi}T}{2f^3_{,\chi}}~.
\end{equation}
Eq.~\eqref{eq:trace-equation-Einstein} is nothing but the trace equation in Eq.~\eqref{trace-eq-fR-Palatini} given that $\chi = \hat{R}$. Notice that, for conformal matter fields ($T=0$), one has $f(\chi) \propto \chi^2$, which implies $f(\hat{R}) \propto \hat{R}^2$ in the Jordan frame \cite{Sotiriou:2008rp}. This is reasonable because, in this case, the gravitational sector is conformally invariant too.

In the case of $\hat{R}+\frac{\alpha}{2\m^2}\hat{R}^2$ Gravity, where
\begin{equation}
    f(\chi) = \chi + \frac{\alpha}{2m_{\textrm{P}}^2}\chi^2~,
\end{equation}
such that $f_{,\chi}\chi = 2f-\chi$, one might solve the equation \eqref{eq:trace-equation-Einstein} for $\chi$ readily, obtaining 
\begin{equation}
    \chi = -\frac{T}{m^2_{\textrm{P}}}~,
\end{equation}
meaning that one has the GR action, a function of the trace $U(T)$
\begin{equation}
    U(T) = \frac{\alpha}{4m^4_{\textrm{P}}}\frac{T^2}{\left(1-\frac{\alpha}{m^4_{\textrm{P}}}T\right)^2}~,
\end{equation}
and non-minimally coupled matter fields.  

We derive the metric field equations from action \eqref{eq:rescaled-action-Einstein}. The connection field equations imply that the connection, in this frame, is the Levi-Civita one (with respect to the rescaled metric $\tilde{g}_{\mu\nu}$) plus a contribution from an arbitrary vector field which is removed by a projective transformation \cite{Bernal:2016lhq} and, in fact, does not affect the Ricci scalar even if present. So we have
\begin{equation}
    \tilde{\hat{R}} = \tilde{R} = f_{,\chi}^{-1} \hat{R}~,
\end{equation}    
where $\tilde{R}$ is built upon $\tilde{g}_{\mu\nu}$ and its derivatives solely. The metric field equations (obtained by varying with respect to the inverse metric $\tilde{g}^{\mu\nu}$) are
\begin{equation}
    \tilde{R}_{\mu\nu}-\frac{1}{2} \tilde{g}_{\mu\nu}\left(\tilde{R}-\frac{2U}{m^2_{\textrm{P}}}\right) = \frac{1}{m^2_{\textrm{P}}}\tilde{T}_{\mu\nu}~.
\end{equation}
Noticing that $\tilde{T}_{\mu\nu} = f^{-1}_{,\chi} T_{\mu\nu}$ and $ \tilde{R}_{\mu\nu}=\tilde{\hat{R}}_{(\mu\nu)} =\hat{R}_{(\mu\nu)}$ (because the Ricci tensor is not affected by the conformal transformation in the Palatini formalism), and since $\tilde{R} = f_{,\chi}^{-1}\hat{R} = f_{,\chi}^{-1}\chi$, using Eqs.~\eqref{eq:definition-U-chi} and \eqref{eq:conf-trans-fchi}, we arrive at
\begin{equation}
    f_{,\chi}\hat{R}_{(\mu\nu)}-\frac{1}{2}g_{\mu\nu}f = \frac{1}{m^2_{\textrm{P}}}T_{\mu\nu}~, 
\end{equation}
which is Eq.~\eqref{eq.metric.fR.Palatini} in the Jordan frame (where $f_{,\chi} = f_{,\hat{R}} = F$), as expected.
\newpage

\bibliography{bibliography}

@article{Baumann:2007zm,
    author = "Baumann, Daniel and Steinhardt, Paul J. and Takahashi, Keitaro and Ichiki, Kiyotomo",
    title = "{Gravitational Wave Spectrum Induced by Primordial Scalar Perturbations}",
    eprint = "hep-th/0703290",
    archivePrefix = "arXiv",
    doi = "10.1103/PhysRevD.76.084019",
    journal = "Phys. Rev. D",
    volume = "76",
    pages = "084019",
    year = "2007"
}

@article{Malik:2008im,
    author = "Malik, Karim A. and Wands, David",
    title = "{Cosmological perturbations}",
    eprint = "0809.4944",
    archivePrefix = "arXiv",
    primaryClass = "astro-ph",
    doi = "10.1016/j.physrep.2009.03.001",
    journal = "Phys. Rept.",
    volume = "475",
    pages = "1--51",
    year = "2009"
}

@article{Acquaviva:2002ud,
    author = "Acquaviva, Viviana and Bartolo, Nicola and Matarrese, Sabino and Riotto, Antonio",
    title = "{Second order cosmological perturbations from inflation}",
    eprint = "astro-ph/0209156",
    archivePrefix = "arXiv",
    reportNumber = "DFPD-A-02-21",
    doi = "10.1016/S0550-3213(03)00550-9",
    journal = "Nucl. Phys. B",
    volume = "667",
    pages = "119--148",
    year = "2003"
}

@article{Zhou:2024doz,
    author = {Zhou, Jing-Zhi and Kuang, Yu-Ting and Wu, Di and Chen, Fei-Yu and L\"u, H. and Chang, Zhe},
    title = "{Scalar induced gravitational waves in f(R) gravity}",
    eprint = "2409.07702",
    archivePrefix = "arXiv",
    primaryClass = "gr-qc",
    doi = "10.1088/1475-7516/2024/12/021",
    journal = "JCAP",
    volume = "12",
    pages = "021",
    year = "2024"
}

@article{Kohri:2018awv,
    author = "Kohri, Kazunori and Terada, Takahiro",
    title = "{Semianalytic calculation of gravitational wave spectrum nonlinearly induced from primordial curvature perturbations}",
    eprint = "1804.08577",
    archivePrefix = "arXiv",
    primaryClass = "gr-qc",
    reportNumber = "KEK-TH-2046, KEK-COSMO-223",
    doi = "10.1103/PhysRevD.97.123532",
    journal = "Phys. Rev. D",
    volume = "97",
    number = "12",
    pages = "123532",
    year = "2018"
}

@article{Saito:2008jc,
    author = "Saito, Ryo and Yokoyama, Jun'ichi",
    title = "{Gravitational wave background as a probe of the primordial black hole abundance}",
    eprint = "0812.4339",
    archivePrefix = "arXiv",
    primaryClass = "astro-ph",
    reportNumber = "RESCEU-63-08",
    doi = "10.1103/PhysRevLett.102.161101",
    journal = "Phys. Rev. Lett.",
    volume = "102",
    pages = "161101",
    year = "2009",
    note = "[Erratum: Phys.Rev.Lett. 107, 069901 (2011)]"
}

@article{Matarrese:1997ay,
    author = "Matarrese, Sabino and Mollerach, Silvia and Bruni, Marco",
    title = "{Second order perturbations of the Einstein-de Sitter universe}",
    eprint = "astro-ph/9707278",
    archivePrefix = "arXiv",
    reportNumber = "SISSA-83-97-A",
    doi = "10.1103/PhysRevD.58.043504",
    journal = "Phys. Rev. D",
    volume = "58",
    pages = "043504",
    year = "1998"
}

@article{Matarrese:1993zf,
    author = "Matarrese, Sabino and Pantano, Ornella and Saez, Diego",
    title = "{General relativistic dynamics of irrotational dust: Cosmological implications}",
    eprint = "astro-ph/9310036",
    archivePrefix = "arXiv",
    reportNumber = "DFPD-93-A-67",
    doi = "10.1103/PhysRevLett.72.320",
    journal = "Phys. Rev. Lett.",
    volume = "72",
    pages = "320--323",
    year = "1994"
}

@article{Inomata:2018epa,
    author = "Inomata, Keisuke and Nakama, Tomohiro",
    title = "{Gravitational waves induced by scalar perturbations as probes of the small-scale primordial spectrum}",
    eprint = "1812.00674",
    archivePrefix = "arXiv",
    primaryClass = "astro-ph.CO",
    reportNumber = "IPMU 18-0200",
    doi = "10.1103/PhysRevD.99.043511",
    journal = "Phys. Rev. D",
    volume = "99",
    number = "4",
    pages = "043511",
    year = "2019"
}

@article{Domenech:2021ztg,
    author = "Dom\`enech, Guillem",
    title = "{Scalar Induced Gravitational Waves Review}",
    eprint = "2109.01398",
    archivePrefix = "arXiv",
    primaryClass = "gr-qc",
    doi = "10.3390/universe7110398",
    journal = "Universe",
    volume = "7",
    number = "11",
    pages = "398",
    year = "2021"
}

@article{Dimopoulos:2020pas,
    author = "Dimopoulos, Konstantinos and S\'anchez L\'opez, Samuel",
    title = "{Quintessential inflation in Palatini $f(R)$ gravity}",
    eprint = "2012.06831",
    archivePrefix = "arXiv",
    primaryClass = "gr-qc",
    doi = "10.1103/PhysRevD.103.043533",
    journal = "Phys. Rev. D",
    volume = "103",
    number = "4",
    pages = "043533",
    year = "2021"
}

@article{DeFelice:2010aj,
    author = "De Felice, Antonio and Tsujikawa, Shinji",
    title = "{f(R) theories}",
    eprint = "1002.4928",
    archivePrefix = "arXiv",
    primaryClass = "gr-qc",
    doi = "10.12942/lrr-2010-3",
    journal = "Living Rev. Rel.",
    volume = "13",
    pages = "3",
    year = "2010"
}

@book{Baumann:2022mni,
    author = "Baumann, Daniel",
    title = "{Cosmology}",
    doi = "10.1017/9781108937092",
    isbn = "978-1-108-93709-2, 978-1-108-83807-8",
    publisher = "Cambridge University Press",
    month = "7",
    year = "2022"
}

@article{Koivisto:2005yc,
    author = "Koivisto, Tomi and Kurki-Suonio, Hannu",
    title = "{Cosmological perturbations in the palatini formulation of modified gravity}",
    eprint = "astro-ph/0509422",
    archivePrefix = "arXiv",
    reportNumber = "HIP-2005-38-TH",
    doi = "10.1088/0264-9381/23/7/009",
    journal = "Class. Quant. Grav.",
    volume = "23",
    pages = "2355--2369",
    year = "2006"
}

@article{NANOGrav:2023gor,
    author = "Agazie, Gabriella and others",
    collaboration = "NANOGrav",
    title = "{The NANOGrav 15 yr Data Set: Evidence for a Gravitational-wave Background}",
    eprint = "2306.16213",
    archivePrefix = "arXiv",
    primaryClass = "astro-ph.HE",
    doi = "10.3847/2041-8213/acdac6",
    journal = "Astrophys. J. Lett.",
    volume = "951",
    number = "1",
    pages = "L8",
    year = "2023"
}

@article{NANOGrav:2023hde,
    author = "Agazie, Gabriella and others",
    collaboration = "NANOGrav",
    title = "{The NANOGrav 15 yr Data Set: Observations and Timing of 68 Millisecond Pulsars}",
    eprint = "2306.16217",
    archivePrefix = "arXiv",
    primaryClass = "astro-ph.HE",
    doi = "10.3847/2041-8213/acda9a",
    journal = "Astrophys. J. Lett.",
    volume = "951",
    number = "1",
    pages = "L9",
    year = "2023"
}

@article{NANOGrav:2023hvm,
    author = "Afzal, Adeela and others",
    collaboration = "NANOGrav",
    title = "{The NANOGrav 15 yr Data Set: Search for Signals from New Physics}",
    eprint = "2306.16219",
    archivePrefix = "arXiv",
    primaryClass = "astro-ph.HE",
    reportNumber = "FERMILAB-PUB-23-589-T",
    doi = "10.3847/2041-8213/acdc91",
    journal = "Astrophys. J. Lett.",
    volume = "951",
    number = "1",
    pages = "L11",
    year = "2023",
    note = "[Erratum: Astrophys.J.Lett. 971, L27 (2024), Erratum: Astrophys.J. 971, L27 (2024)]"
}

@article{EPTA:2023fyk,
    author = "Antoniadis, J. and others",
    collaboration = "EPTA, InPTA:",
    title = "{The second data release from the European Pulsar Timing Array - III. Search for gravitational wave signals}",
    eprint = "2306.16214",
    archivePrefix = "arXiv",
    primaryClass = "astro-ph.HE",
    doi = "10.1051/0004-6361/202346844",
    journal = "Astron. Astrophys.",
    volume = "678",
    pages = "A50",
    year = "2023"
}

@article{EPTA:2023sfo,
    author = "Antoniadis, J. and others",
    collaboration = "EPTA",
    title = "{The second data release from the European Pulsar Timing Array - I. The dataset and timing analysis}",
    eprint = "2306.16224",
    archivePrefix = "arXiv",
    primaryClass = "astro-ph.HE",
    doi = "10.1051/0004-6361/202346841",
    journal = "Astron. Astrophys.",
    volume = "678",
    pages = "A48",
    year = "2023"
}

@article{EPTA:2023xxk,
    author = "Antoniadis, J. and others",
    collaboration = "EPTA, InPTA",
    title = "{The second data release from the European Pulsar Timing Array - IV. Implications for massive black holes, dark matter, and the early Universe}",
    eprint = "2306.16227",
    archivePrefix = "arXiv",
    primaryClass = "astro-ph.CO",
    doi = "10.1051/0004-6361/202347433",
    journal = "Astron. Astrophys.",
    volume = "685",
    pages = "A94",
    year = "2024"
}

@article{Zic:2023gta,
    author = "Zic, Andrew and others",
    title = "{The Parkes Pulsar Timing Array third data release}",
    eprint = "2306.16230",
    archivePrefix = "arXiv",
    primaryClass = "astro-ph.HE",
    doi = "10.1017/pasa.2023.36",
    journal = "Publ. Astron. Soc. Austral.",
    volume = "40",
    pages = "e049",
    year = "2023"
}

@article{Reardon:2023zen,
    author = "Reardon, Daniel J. and others",
    title = "{The Gravitational-wave Background Null Hypothesis: Characterizing Noise in Millisecond Pulsar Arrival Times with the Parkes Pulsar Timing Array}",
    eprint = "2306.16229",
    archivePrefix = "arXiv",
    primaryClass = "astro-ph.HE",
    doi = "10.3847/2041-8213/acdd03",
    journal = "Astrophys. J. Lett.",
    volume = "951",
    number = "1",
    pages = "L7",
    year = "2023"
}

@article{Xu:2023wog,
    author = "Xu, Heng and others",
    title = "{Searching for the Nano-Hertz Stochastic Gravitational Wave Background with the Chinese Pulsar Timing Array Data Release I}",
    eprint = "2306.16216",
    archivePrefix = "arXiv",
    primaryClass = "astro-ph.HE",
    doi = "10.1088/1674-4527/acdfa5",
    journal = "Res. Astron. Astrophys.",
    volume = "23",
    number = "7",
    pages = "075024",
    year = "2023"
}

@article{Loc:2024qbz,
    author = "Loc, Ngo Phuc Duc",
    title = "{Gravitational waves from burdened primordial black holes dark matter}",
    eprint = "2410.17544",
    archivePrefix = "arXiv",
    primaryClass = "gr-qc",
    doi = "10.1103/PhysRevD.111.023509",
    journal = "Phys. Rev. D",
    volume = "111",
    number = "2",
    pages = "023509",
    year = "2025"
}

@article{Picard:2024ekd,
    author = {Picard, Rapha{\"e}l and Davies, Matthew W.},
    title = "{Effects of scalar non-Gaussianity on induced scalar-tensor gravitational waves}",
    eprint = "2410.17819",
    archivePrefix = "arXiv",
    primaryClass = "astro-ph.CO",
    doi = "10.1088/1475-7516/2025/02/037",
    journal = "JCAP",
    volume = "02",
    pages = "037",
    year = "2025"
}

@article{TerenteDiaz:2023kgc,
    author = "Terente D\'\i{}az, Jos\'e Jaime and Dimopoulos, Konstantinos and Kar\v{c}iauskas, Mindaugas and Racioppi, Antonio",
    title = "{Quintessence in the Weyl-Gauss-Bonnet model}",
    eprint = "2310.08128",
    archivePrefix = "arXiv",
    primaryClass = "gr-qc",
    doi = "10.1088/1475-7516/2024/02/040",
    journal = "JCAP",
    volume = "02",
    pages = "040",
    year = "2024"
}

@article{Kubota:2020ehu,
    author = "Kubota, Mio and Oda, Kin-Ya and Shimada, Keigo and Yamaguchi, Masahide",
    title = "{Cosmological Perturbations in Palatini Formalism}",
    eprint = "2010.07867",
    archivePrefix = "arXiv",
    primaryClass = "hep-th",
    doi = "10.1088/1475-7516/2021/03/006",
    journal = "JCAP",
    volume = "03",
    pages = "006",
    year = "2021"
}

@article{Dimopoulos:2022tvn,
    author = "Dimopoulos, Konstantinos and Karam, Alexandros and S\'anchez L\'opez, Samuel and Tomberg, Eemeli",
    title = "{Modelling Quintessential Inflation in Palatini-Modified Gravity}",
    eprint = "2203.05424",
    archivePrefix = "arXiv",
    primaryClass = "gr-qc",
    doi = "10.3390/galaxies10020057",
    journal = "Galaxies",
    volume = "10",
    number = "2",
    pages = "57",
    year = "2022"
}

@article{Dimopoulos:2022rdp,
    author = "Dimopoulos, Konstantinos and Karam, Alexandros and S\'anchez L\'opez, Samuel and Tomberg, Eemeli",
    title = "{Palatini R $^{2}$ quintessential inflation}",
    eprint = "2206.14117",
    archivePrefix = "arXiv",
    primaryClass = "gr-qc",
    doi = "10.1088/1475-7516/2022/10/076",
    journal = "JCAP",
    volume = "10",
    pages = "076",
    year = "2022"
}

@article{SanchezLopez:2023ixx,
    author = "S\'anchez L\'opez, Samuel and Dimopoulos, Konstantinos and Karam, Alexandros and Tomberg, Eemeli",
    title = "{Observable gravitational waves from hyperkination in Palatini gravity and beyond}",
    eprint = "2305.01399",
    archivePrefix = "arXiv",
    primaryClass = "gr-qc",
    doi = "10.1140/epjc/s10052-023-12332-x",
    journal = "Eur. Phys. J. C",
    volume = "83",
    number = "12",
    pages = "1152",
    year = "2023"
}

@article{He:2024luf,
    author = "He, Xin-Chen and Cai, Yi-Fu and Ma, Xiao-Han and Papanikolaou, Theodoros and Saridakis, Emmanuel N. and Sasaki, Misao",
    title = "{Gravitational waves from primordial black hole isocurvature: the effect of non-Gaussianities}",
    eprint = "2409.11333",
    archivePrefix = "arXiv",
    primaryClass = "astro-ph.CO",
    reportNumber = "YITP-24-93",
    doi = "10.1088/1475-7516/2024/12/039",
    journal = "JCAP",
    volume = "12",
    pages = "039",
    year = "2024"
}

@article{Picard:2023sbz,
    author = "Picard, Raphael and Malik, Karim A.",
    title = "{Induced gravitational waves: the effect of first order tensor perturbations}",
    eprint = "2311.14513",
    archivePrefix = "arXiv",
    primaryClass = "astro-ph.CO",
    doi = "10.1088/1475-7516/2024/10/010",
    journal = "JCAP",
    volume = "10",
    pages = "010",
    year = "2024"
}

@article{Domenech:2024rks,
    author = "Dom\`enech, Guillem and Pi, Shi and Wang, Ao and Wang, Jianing",
    title = "{Induced gravitational wave interpretation of PTA data: a complete study for general equation of state}",
    eprint = "2402.18965",
    archivePrefix = "arXiv",
    primaryClass = "astro-ph.CO",
    doi = "10.1088/1475-7516/2024/08/054",
    journal = "JCAP",
    volume = "08",
    pages = "054",
    year = "2024"
}

@article{Iovino:2024sgs,
    author = "Iovino, A. J. and Matarrese, S. and Perna, G. and Ricciardone, A. and Riotto, A.",
    title = "{How Well Do We Know the Scalar-Induced Gravitational Waves?}",
    eprint = "2412.06764",
    archivePrefix = "arXiv",
    primaryClass = "astro-ph.CO",
    month = "12",
    year = "2024"
}

@article{Kugarajh:2025pjl,
    author = "Kugarajh, Anjali Abirami",
    title = "{Gauge-dependence of Scalar Induced Gravitational Waves}",
    eprint = "2503.00083",
    archivePrefix = "arXiv",
    primaryClass = "gr-qc",
    doi = "10.1088/1361-6382/ade2b3",
    journal = "Class. Quant. Grav.",
    volume = "42",
    number = "12",
    pages = "127001",
    year = "2025"
}

@article{Domenech:2024drm,
    author = "Dom\`enech, Guillem and Ganz, Alexander",
    title = "{Enhanced induced gravitational waves in Horndeski gravity}",
    eprint = "2406.19950",
    archivePrefix = "arXiv",
    primaryClass = "gr-qc",
    doi = "10.1088/1475-7516/2025/01/020",
    journal = "JCAP",
    volume = "01",
    pages = "020",
    year = "2025"
}

@article{Clifton:2011jh,
    author = "Clifton, Timothy and Ferreira, Pedro G. and Padilla, Antonio and Skordis, Constantinos",
    title = "{Modified Gravity and Cosmology}",
    eprint = "1106.2476",
    archivePrefix = "arXiv",
    primaryClass = "astro-ph.CO",
    doi = "10.1016/j.physrep.2012.01.001",
    journal = "Phys. Rept.",
    volume = "513",
    pages = "1--189",
    year = "2012"
}

@article{Lima:2016npg,
    author = "Lima, Nelson A. and Smer-Barreto, Vanessa and Lombriser, Lucas",
    title = "{Constraints on decaying early modified gravity from cosmological observations}",
    eprint = "1603.05239",
    archivePrefix = "arXiv",
    primaryClass = "astro-ph.CO",
    doi = "10.1103/PhysRevD.94.083507",
    journal = "Phys. Rev. D",
    volume = "94",
    number = "8",
    pages = "083507",
    year = "2016"
}

@article{Kugarajh:2025rbt,
    author = "Kugarajh, Anjali Abirami and Traforetti, Marisol and Maselli, Andrea and Matarrese, Sabino and Ricciardone, Angelo",
    title = "{Scalar-Induced Gravitational Waves in Modified Gravity}",
    eprint = "2502.20137",
    archivePrefix = "arXiv",
    primaryClass = "gr-qc",
    doi = "10.1088/1475-7516/2025/07/022",
    month = "2",
    year = "2025"
}

@article{Bertacca:2011wu,
    author = "Bertacca, Daniele and Bartolo, Nicola and Matarrese, Sabino",
    title = "{A new approach to cosmological perturbations in f(R) models}",
    eprint = "1109.2082",
    archivePrefix = "arXiv",
    primaryClass = "astro-ph.CO",
    doi = "10.1088/1475-7516/2012/08/021",
    journal = "JCAP",
    volume = "08",
    pages = "021",
    year = "2012"
}

@article{Maggiore:1999vm,
    author = "Maggiore, Michele",
    title = "{Gravitational wave experiments and early universe cosmology}",
    eprint = "gr-qc/9909001",
    archivePrefix = "arXiv",
    reportNumber = "IFUP-TH-20-99",
    doi = "10.1016/S0370-1573(99)00102-7",
    journal = "Phys. Rept.",
    volume = "331",
    pages = "283--367",
    year = "2000"
}

@article{Adshead:2021hnm,
    author = "Adshead, Peter and Lozanov, Kaloian D. and Weiner, Zachary J.",
    title = "{Non-Gaussianity and the induced gravitational wave background}",
    eprint = "2105.01659",
    archivePrefix = "arXiv",
    primaryClass = "astro-ph.CO",
    doi = "10.1088/1475-7516/2021/10/080",
    journal = "JCAP",
    volume = "10",
    pages = "080",
    year = "2021"
}

@article{Planck:2018vyg,
    author = "Aghanim, N. and others",
    collaboration = "Planck",
    title = "{Planck 2018 results. VI. Cosmological parameters}",
    eprint = "1807.06209",
    archivePrefix = "arXiv",
    primaryClass = "astro-ph.CO",
    doi = "10.1051/0004-6361/201833910",
    journal = "Astron. Astrophys.",
    volume = "641",
    pages = "A6",
    year = "2020",
    note = "[Erratum: Astron.Astrophys. 652, C4 (2021)]"
}

@article{Bhaumik:2025kuj,
    author = "Bhaumik, Arko and Papanikolaou, Theodoros and Ghoshal, Anish",
    title = "{Vector induced Gravitational Waves sourced by Primordial Magnetic Fields}",
    eprint = "2504.10477",
    archivePrefix = "arXiv",
    primaryClass = "astro-ph.CO",
    month = "4",
    year = "2025"
}

@article{Pi:2020otn,
    author = "Pi, Shi and Sasaki, Misao",
    title = "{Gravitational Waves Induced by Scalar Perturbations with a Lognormal Peak}",
    eprint = "2005.12306",
    archivePrefix = "arXiv",
    primaryClass = "gr-qc",
    reportNumber = "YITP-20-75, YITP-75, IPMU20-0054",
    doi = "10.1088/1475-7516/2020/09/037",
    journal = "JCAP",
    volume = "09",
    pages = "037",
    year = "2020"
}

@article{Bernal:2016lhq,
    author = "Bernal, Antonio N. and Janssen, Bert and Jimenez-Cano, Alejandro and Orejuela, Jose Alberto and Sanchez, Miguel and Sanchez-Moreno, Pablo",
    title = "{On the (non-)uniqueness of the Levi-Civita solution in the Einstein\textendash{}Hilbert\textendash{}Palatini formalism}",
    eprint = "1606.08756",
    archivePrefix = "arXiv",
    primaryClass = "gr-qc",
    doi = "10.1016/j.physletb.2017.03.001",
    journal = "Phys. Lett. B",
    volume = "768",
    pages = "280--287",
    year = "2017"
}

@article{Bejarano:2019zco,
    author = "Bejarano, Cecilia and Delhom, Adria and Jim\'enez-Cano, Alejandro and Olmo, Gonzalo J. and Rubiera-Garcia, Diego",
    title = "{Geometric inequivalence of metric and Palatini formulations of General Relativity}",
    eprint = "1907.04137",
    archivePrefix = "arXiv",
    primaryClass = "gr-qc",
    doi = "10.1016/j.physletb.2020.135275",
    journal = "Phys. Lett. B",
    volume = "802",
    pages = "135275",
    year = "2020"
}

@article{Dadhich:2012htv,
    author = "Dadhich, Naresh and Pons, Josep M.",
    title = "{On the equivalence of the Einstein-Hilbert and the Einstein-Palatini formulations of general relativity for an arbitrary connection}",
    eprint = "1010.0869",
    archivePrefix = "arXiv",
    primaryClass = "gr-qc",
    doi = "10.1007/s10714-012-1393-9",
    journal = "Gen. Rel. Grav.",
    volume = "44",
    pages = "2337--2352",
    year = "2012"
}

@article{Comeau:2023mxi,
    author = "Comeau, Vincent",
    title = "{Gauge-Invariant Scalar-Induced Gravitational Waves from Physical Observables}",
    eprint = "2309.14624",
    archivePrefix = "arXiv",
    primaryClass = "gr-qc",
    month = "9",
    year = "2023"
}

@article{Ali:2023moi,
    author = "Ali, Arshad and Hu, Ya-Peng and Sabir, Mudassar and Sui, Taotao",
    title = "{On the gauge dependence of scalar induced secondary gravitational waves during radiation and matter domination eras}",
    eprint = "2308.04713",
    archivePrefix = "arXiv",
    primaryClass = "gr-qc",
    doi = "10.1007/s11433-022-2118-5",
    journal = "Sci. China Phys. Mech. Astron.",
    volume = "66",
    number = "9",
    pages = "290411",
    year = "2023"
}

@article{Chiba:2008ia,
    author = "Chiba, Takeshi and Yamaguchi, Masahide",
    title = "{Extended Slow-Roll Conditions and Rapid-Roll Conditions}",
    eprint = "0807.4965",
    archivePrefix = "arXiv",
    primaryClass = "astro-ph",
    doi = "10.1088/1475-7516/2008/10/021",
    journal = "JCAP",
    volume = "10",
    pages = "021",
    year = "2008"
}

@article{Kuusk:2016rso,
    author = {Kuusk, Piret and R\"unkla, Mihkel and Saal, Margus and Vilson, Ott},
    title = "{Invariant slow-roll parameters in scalar\textendash{}tensor theories}",
    eprint = "1605.07033",
    archivePrefix = "arXiv",
    primaryClass = "gr-qc",
    doi = "10.1088/0264-9381/33/19/195008",
    journal = "Class. Quant. Grav.",
    volume = "33",
    number = "19",
    pages = "195008",
    year = "2016"
}

@article{Karciauskas:2022jzd,
    author = "Kar\v{c}iauskas, Mindaugas and D\'\i{}az, Jos\'e Jaime Terente",
    title = "{Slow-roll inflation in the Jordan frame}",
    eprint = "2206.08677",
    archivePrefix = "arXiv",
    primaryClass = "gr-qc",
    doi = "10.1103/PhysRevD.106.083526",
    journal = "Phys. Rev. D",
    volume = "106",
    number = "8",
    pages = "083526",
    year = "2022"
}

@article{Gong:2011qe,
    author = "Gong, Jinn-Ouk and Hwang, Jai-chan and Park, Wan-Il and Sasaki, Misao and Song, Yong-Seon",
    title = "{Conformal invariance of curvature perturbation}",
    eprint = "1107.1840",
    archivePrefix = "arXiv",
    primaryClass = "gr-qc",
    reportNumber = "CERN-PH-TH-2011-123, YITP-11-57",
    doi = "10.1088/1475-7516/2011/09/023",
    journal = "JCAP",
    volume = "09",
    pages = "023",
    year = "2011"
}

@article{White:2012ya,
    author = "White, Jonathan and Minamitsuji, Masato and Sasaki, Misao",
    title = "{Curvature perturbation in multi-field inflation with non-minimal coupling}",
    eprint = "1205.0656",
    archivePrefix = "arXiv",
    primaryClass = "astro-ph.CO",
    reportNumber = "YITP-12-37",
    doi = "10.1088/1475-7516/2012/07/039",
    journal = "JCAP",
    volume = "07",
    pages = "039",
    year = "2012"
}

@article{Kubota:2011re,
    author = "Kubota, Takahiro and Misumi, Nobuhiko and Naylor, Wade and Okuda, Naoya",
    title = "{The Conformal Transformation in General Single Field Inflation with Non-Minimal Coupling}",
    eprint = "1112.5233",
    archivePrefix = "arXiv",
    primaryClass = "gr-qc",
    reportNumber = "OU-HET-737",
    doi = "10.1088/1475-7516/2012/02/034",
    journal = "JCAP",
    volume = "02",
    pages = "034",
    year = "2012"
}

@article{Diaz:2023tma,
    author = "D\'\i{}az, Jos\'e Jaime Terente and Kar\v{c}iauskas, Mindaugas",
    title = "{Comoving curvature perturbation in Jordan and Einstein frames}",
    eprint = "2305.15326",
    archivePrefix = "arXiv",
    primaryClass = "gr-qc",
    doi = "10.1103/PhysRevD.108.083535",
    journal = "Phys. Rev. D",
    volume = "108",
    number = "8",
    pages = "083535",
    year = "2023"
}

@article{Dabrowski:2008kx,
    author = "Dabrowski, Mariusz P. and Garecki, Janusz and Blaschke, David B.",
    title = "{Conformal transformations and conformal invariance in gravitation}",
    eprint = "0806.2683",
    archivePrefix = "arXiv",
    primaryClass = "gr-qc",
    doi = "10.1002/andp.200810331",
    journal = "Annalen Phys.",
    volume = "18",
    pages = "13--32",
    year = "2009"
}

@article{Fakir:1992cg,
    author = "Fakir, Redouane and Habib, Salman and Unruh, William",
    title = "{Cosmological density perturbations with modified gravity}",
    reportNumber = "LA-UR-91-4047",
    doi = "10.1086/171591",
    journal = "Astrophys. J.",
    volume = "394",
    pages = "396",
    year = "1992"
}

@article{Hwang:1990re,
    author = "Hwang, J. C.",
    title = "{Cosmological perturbations in generalized gravity theories: Formulation}",
    doi = "10.1088/0264-9381/7/9/013",
    journal = "Class. Quant. Grav.",
    volume = "7",
    pages = "1613--1631",
    year = "1990"
}

@article{Hwang:1990jh,
    author = "Hwang, J. C.",
    title = "{Cosmological perturbations in generalized gravity theories: Solutions}",
    doi = "10.1103/PhysRevD.42.2601",
    journal = "Phys. Rev. D",
    volume = "42",
    pages = "2601--2606",
    year = "1990"
}

@article{Hwang:1996xh,
    author = "Hwang, Jai-chan and Noh, Hyerim",
    title = "{Cosmological perturbations in generalized gravity theories}",
    reportNumber = "PRINT-96-116 (KYUNGPOOK)",
    doi = "10.1103/PhysRevD.54.1460",
    journal = "Phys. Rev. D",
    volume = "54",
    pages = "1460--1473",
    year = "1996"
}

@article{Hwang:1996np,
    author = "Hwang, Jai-chan",
    title = "{Cosmological perturbations in generalized gravity theories: Conformal transformation}",
    eprint = "gr-qc/9605024",
    archivePrefix = "arXiv",
    doi = "10.1088/0264-9381/14/7/029",
    journal = "Class. Quant. Grav.",
    volume = "14",
    pages = "1981--1991",
    year = "1997"
}

@article{Gordon:2000hv,
    author = "Gordon, Christopher and Wands, David and Bassett, Bruce A. and Maartens, Roy",
    title = "{Adiabatic and entropy perturbations from inflation}",
    eprint = "astro-ph/0009131",
    archivePrefix = "arXiv",
    doi = "10.1103/PhysRevD.63.023506",
    journal = "Phys. Rev. D",
    volume = "63",
    pages = "023506",
    year = "2000"
}

@article{Weinberg:2003sw,
    author = "Weinberg, Steven",
    title = "{Adiabatic modes in cosmology}",
    eprint = "astro-ph/0302326",
    archivePrefix = "arXiv",
    reportNumber = "UTTG-12-02",
    doi = "10.1103/PhysRevD.67.123504",
    journal = "Phys. Rev. D",
    volume = "67",
    pages = "123504",
    year = "2003"
}

@article{Lyth:1984gv,
    author = "Lyth, D. H.",
    title = "{Large Scale Energy Density Perturbations and Inflation}",
    reportNumber = "Print-84-0373 (LANCASTER)",
    doi = "10.1103/PhysRevD.31.1792",
    journal = "Phys. Rev. D",
    volume = "31",
    pages = "1792--1798",
    year = "1985"
}

@article{ACT:2025fju, author = "Louis, Thibaut and others", collaboration = "ACT", title = "{The Atacama Cosmology Telescope: DR6 Power Spectra, Likelihoods and CDM Parameters}", eprint = "2503.14452", archivePrefix = "arXiv", primaryClass = "astro-ph.CO", reportNumber = "FERMILAB-PUB-25-0071-PPD", month = "3", year = "2025" }

@article{Bartolo:2016ami,
    author = "Bartolo, Nicola and others",
    title = "{Science with the space-based interferometer LISA. IV: Probing inflation with gravitational waves}",
    eprint = "1610.06481",
    archivePrefix = "arXiv",
    primaryClass = "astro-ph.CO",
    reportNumber = "ACFI-T16-19, UMN-TH-3608-16, CERN-TH-2016-222, KCL-PH-TH-2016-58, IFT-UAM-CSIC-16-104",
    doi = "10.1088/1475-7516/2016/12/026",
    journal = "JCAP",
    volume = "12",
    pages = "026",
    year = "2016"
}

@article{Caprini:2019pxz,
    author = "Caprini, Chiara and Figueroa, Daniel G. and Flauger, Raphael and Nardini, Germano and Peloso, Marco and Pieroni, Mauro and Ricciardone, Angelo and Tasinato, Gianmassimo",
    title = "{Reconstructing the spectral shape of a stochastic gravitational wave background with LISA}",
    eprint = "1906.09244",
    archivePrefix = "arXiv",
    primaryClass = "astro-ph.CO",
    reportNumber = "LISA-CosWG-19-02",
    doi = "10.1088/1475-7516/2019/11/017",
    journal = "JCAP",
    volume = "11",
    pages = "017",
    year = "2019"
}

@article{LISACosmologyWorkingGroup:2022jok,
    author = "Auclair, Pierre and others",
    collaboration = "LISA Cosmology Working Group",
    title = "{Cosmology with the Laser Interferometer Space Antenna}",
    eprint = "2204.05434",
    archivePrefix = "arXiv",
    primaryClass = "astro-ph.CO",
    reportNumber = "LISA CosWG-22-03, FERMILAB-PUB-22-349-SCD",
    doi = "10.1007/s41114-023-00045-2",
    journal = "Living Rev. Rel.",
    volume = "26",
    number = "1",
    pages = "5",
    year = "2023"
}

@article{Janssen:2014dka,
    author = "Janssen, Gemma and others",
    editor = "Bourke, Tyler L. and others",
    title = "{Gravitational wave astronomy with the SKA}",
    eprint = "1501.00127",
    archivePrefix = "arXiv",
    primaryClass = "astro-ph.IM",
    doi = "10.22323/1.215.0037",
    journal = "PoS",
    volume = "AASKA14",
    pages = "037",
    year = "2015"
}

@article{Punturo:2010zz,
    author = "Punturo, M. and others",
    editor = "Ricci, Fulvio",
    title = "{The Einstein Telescope: A third-generation gravitational wave observatory}",
    doi = "10.1088/0264-9381/27/19/194002",
    journal = "Class. Quant. Grav.",
    volume = "27",
    pages = "194002",
    year = "2010"
}

@article{Hild:2010id,
    author = "Hild, S. and others",
    title = "{Sensitivity Studies for Third-Generation Gravitational Wave Observatories}",
    eprint = "1012.0908",
    archivePrefix = "arXiv",
    primaryClass = "gr-qc",
    doi = "10.1088/0264-9381/28/9/094013",
    journal = "Class. Quant. Grav.",
    volume = "28",
    pages = "094013",
    year = "2011"
}

@article{Harry:2010zz,
    author = "Harry, Gregory M.",
    editor = "Marka, Zsuzsa and Marka, Szabolcs",
    collaboration = "LIGO Scientific",
    title = "{Advanced LIGO: The next generation of gravitational wave detectors}",
    doi = "10.1088/0264-9381/27/8/084006",
    journal = "Class. Quant. Grav.",
    volume = "27",
    pages = "084006",
    year = "2010"
}

@article{VIRGO:2014yos,
    author = "Acernese, F. and others",
    collaboration = "VIRGO",
    title = "{Advanced Virgo: a second-generation interferometric gravitational wave detector}",
    eprint = "1408.3978",
    archivePrefix = "arXiv",
    primaryClass = "gr-qc",
    doi = "10.1088/0264-9381/32/2/024001",
    journal = "Class. Quant. Grav.",
    volume = "32",
    number = "2",
    pages = "024001",
    year = "2015"
}

@article{LIGOScientific:2014pky,
    author = "Aasi, J. and others",
    collaboration = "LIGO Scientific",
    title = "{Advanced LIGO}",
    eprint = "1411.4547",
    archivePrefix = "arXiv",
    primaryClass = "gr-qc",
    doi = "10.1088/0264-9381/32/7/074001",
    journal = "Class. Quant. Grav.",
    volume = "32",
    pages = "074001",
    year = "2015"
}

@article{LIGOScientific:2019lzm,
    author = "Abbott, Rich and others",
    collaboration = "LIGO Scientific, Virgo",
    title = "{Open data from the first and second observing runs of Advanced LIGO and Advanced Virgo}",
    eprint = "1912.11716",
    archivePrefix = "arXiv",
    primaryClass = "gr-qc",
    reportNumber = "LIGO-P1900206",
    doi = "10.1016/j.softx.2021.100658",
    journal = "SoftwareX",
    volume = "13",
    pages = "100658",
    year = "2021"
}

@article{KAGRA:2020tym,
    author = "Akutsu, T. and others",
    collaboration = "KAGRA",
    title = "{Overview of KAGRA: Detector design and construction history}",
    eprint = "2005.05574",
    archivePrefix = "arXiv",
    primaryClass = "physics.ins-det",
    doi = "10.1093/ptep/ptaa125",
    journal = "PTEP",
    volume = "2021",
    number = "5",
    pages = "05A101",
    year = "2021"
}

@article{Kawamura:2006up,
    author = "Kawamura, S. and others",
    editor = "Mio, N.",
    title = "{The Japanese space gravitational wave antenna DECIGO}",
    doi = "10.1088/0264-9381/23/8/S17",
    journal = "Class. Quant. Grav.",
    volume = "23",
    pages = "S125--S132",
    year = "2006"
}

@article{Kawamura:2011zz,
    author = "Kawamura, Seiji and others",
    editor = "Buchman, Sasha and Sun, Ke-Xun",
    title = "{The Japanese space gravitational wave antenna: DECIGO}",
    doi = "10.1088/0264-9381/28/9/094011",
    journal = "Class. Quant. Grav.",
    volume = "28",
    pages = "094011",
    year = "2011"
}

@article{Kawamura:2020pcg,
    author = "Kawamura, Seiji and others",
    title = "{Current status of space gravitational wave antenna DECIGO and B-DECIGO}",
    eprint = "2006.13545",
    archivePrefix = "arXiv",
    primaryClass = "gr-qc",
    doi = "10.1093/ptep/ptab019",
    journal = "PTEP",
    volume = "2021",
    number = "5",
    pages = "05A105",
    year = "2021"
}

@article{Harry:2006fi,
    author = "Harry, G. M. and Fritschel, P. and Shaddock, D. A. and Folkner, W. and Phinney, E. S.",
    title = "{Laser interferometry for the big bang observer}",
    doi = "10.1088/0264-9381/23/15/008",
    journal = "Class. Quant. Grav.",
    volume = "23",
    pages = "4887--4894",
    year = "2006",
    note = "[Erratum: Class.Quant.Grav. 23, 7361 (2006)]"
}

@article{Enckell:2018hmo,
    author = "Enckell, Vera-Maria and Enqvist, Kari and Rasanen, Syksy and Wahlman, Lumi-Pyry",
    title = "{Inflation with $R^2$ term in the Palatini formalism}",
    eprint = "1810.05536",
    archivePrefix = "arXiv",
    primaryClass = "gr-qc",
    reportNumber = "HIP-2018-19/TH",
    doi = "10.1088/1475-7516/2019/02/022",
    journal = "JCAP",
    volume = "02",
    pages = "022",
    year = "2019"
}

@article{Antoniadis:2018ywb,
    author = "Antoniadis, I. and Karam, A. and Lykkas, A. and Tamvakis, K.",
    title = "{Palatini inflation in models with an $R^2$ term}",
    eprint = "1810.10418",
    archivePrefix = "arXiv",
    primaryClass = "gr-qc",
    doi = "10.1088/1475-7516/2018/11/028",
    journal = "JCAP",
    volume = "11",
    pages = "028",
    year = "2018"
}

@article{Antoniadis:2018yfq,
    author = "Antoniadis, I. and Karam, A. and Lykkas, A. and Pappas, T. and Tamvakis, K.",
    title = "{Rescuing Quartic and Natural Inflation in the Palatini Formalism}",
    eprint = "1812.00847",
    archivePrefix = "arXiv",
    primaryClass = "gr-qc",
    doi = "10.1088/1475-7516/2019/03/005",
    journal = "JCAP",
    volume = "03",
    pages = "005",
    year = "2019"
}

@article{Hu:2007nk,
    author = "Hu, Wayne and Sawicki, Ignacy",
    title = "{Models of f(R) Cosmic Acceleration that Evade Solar-System Tests}",
    eprint = "0705.1158",
    archivePrefix = "arXiv",
    primaryClass = "astro-ph",
    doi = "10.1103/PhysRevD.76.064004",
    journal = "Phys. Rev. D",
    volume = "76",
    pages = "064004",
    year = "2007"
}

@article{LIGOScientific:2016aoc,
    author = "Abbott, B. P. and others",
    collaboration = "LIGO Scientific, Virgo",
    title = "{Observation of Gravitational Waves from a Binary Black Hole Merger}",
    eprint = "1602.03837",
    archivePrefix = "arXiv",
    primaryClass = "gr-qc",
    reportNumber = "LIGO-P150914",
    doi = "10.1103/PhysRevLett.116.061102",
    journal = "Phys. Rev. Lett.",
    volume = "116",
    number = "6",
    pages = "061102",
    year = "2016"
}

@article{LIGOScientific:2016sjg,
    author = "Abbott, B. P. and others",
    collaboration = "LIGO Scientific, Virgo",
    title = "{GW151226: Observation of Gravitational Waves from a 22-Solar-Mass Binary Black Hole Coalescence}",
    eprint = "1606.04855",
    archivePrefix = "arXiv",
    primaryClass = "gr-qc",
    reportNumber = "LIGO-P151226",
    doi = "10.1103/PhysRevLett.116.241103",
    journal = "Phys. Rev. Lett.",
    volume = "116",
    number = "24",
    pages = "241103",
    year = "2016"
}

@article{LIGOScientific:2017vwq,
    author = "Abbott, B. P. and others",
    collaboration = "LIGO Scientific, Virgo",
    title = "{GW170817: Observation of Gravitational Waves from a Binary Neutron Star Inspiral}",
    eprint = "1710.05832",
    archivePrefix = "arXiv",
    primaryClass = "gr-qc",
    reportNumber = "LIGO-P170817",
    doi = "10.1103/PhysRevLett.119.161101",
    journal = "Phys. Rev. Lett.",
    volume = "119",
    number = "16",
    pages = "161101",
    year = "2017"
}

@article{LIGOScientific:2017ync,
    author = "Abbott, B. P. and others",
    collaboration = "LIGO Scientific, Virgo, Fermi GBM, INTEGRAL, IceCube, AstroSat Cadmium Zinc Telluride Imager Team, IPN, Insight-Hxmt, ANTARES, Swift, AGILE Team, 1M2H Team, Dark Energy Camera GW-EM, DES, DLT40, GRAWITA, Fermi-LAT, ATCA, ASKAP, Las Cumbres Observatory Group, OzGrav, DWF (Deeper Wider Faster Program), AST3, CAASTRO, VINROUGE, MASTER, J-GEM, GROWTH, JAGWAR, CaltechNRAO, TTU-NRAO, NuSTAR, Pan-STARRS, MAXI Team, TZAC Consortium, KU, Nordic Optical Telescope, ePESSTO, GROND, Texas Tech University, SALT Group, TOROS, BOOTES, MWA, CALET, IKI-GW Follow-up, H.E.S.S., LOFAR, LWA, HAWC, Pierre Auger, ALMA, Euro VLBI Team, Pi of Sky, Chandra Team at McGill University, DFN, ATLAS Telescopes, High Time Resolution Universe Survey, RIMAS, RATIR, SKA South Africa/MeerKAT",
    title = "{Multi-messenger Observations of a Binary Neutron Star Merger}",
    eprint = "1710.05833",
    archivePrefix = "arXiv",
    primaryClass = "astro-ph.HE",
    reportNumber = "LIGO-P1700294, VIR-0802A-17, FERMILAB-PUB-17-478-A-AE-CD",
    doi = "10.3847/2041-8213/aa91c9",
    journal = "Astrophys. J. Lett.",
    volume = "848",
    number = "2",
    pages = "L12",
    year = "2017"
}

@article{LIGOScientific:2017zic,
    author = "Abbott, B. P. and others",
    collaboration = "LIGO Scientific, Virgo, Fermi-GBM, INTEGRAL",
    title = "{Gravitational Waves and Gamma-rays from a Binary Neutron Star Merger: GW170817 and GRB 170817A}",
    eprint = "1710.05834",
    archivePrefix = "arXiv",
    primaryClass = "astro-ph.HE",
    reportNumber = "LIGO-P1700308",
    doi = "10.3847/2041-8213/aa920c",
    journal = "Astrophys. J. Lett.",
    volume = "848",
    number = "2",
    pages = "L13",
    year = "2017"
}

@article{Tomita:1967wkp,
    author = "Tomita, Kenji",
    title = "{Non-Linear Theory of Gravitational Instability in the Expanding Universe}",
    doi = "10.1143/PTP.37.831",
    journal = "Prog. Theor. Phys.",
    volume = "37",
    number = "5",
    pages = "831--846",
    year = "1967"
}

@article{Starobinsky:1979ty,
    author = "Starobinsky, Alexei A.",
    editor = "Khalatnikov, I. M. and Mineev, V. P.",
    title = "{Spectrum of relict gravitational radiation and the early state of the universe}",
    journal = "JETP Lett.",
    volume = "30",
    pages = "682--685",
    year = "1979"
}

@article{Reardon:2023gzh,
    author = "Reardon, Daniel J. and others",
    title = "{Search for an Isotropic Gravitational-wave Background with the Parkes Pulsar Timing Array}",
    eprint = "2306.16215",
    archivePrefix = "arXiv",
    primaryClass = "astro-ph.HE",
    doi = "10.3847/2041-8213/acdd02",
    journal = "Astrophys. J. Lett.",
    volume = "951",
    number = "1",
    pages = "L6",
    year = "2023"
}

@article{Guzzetti:2016mkm,
    author = "Guzzetti, M. C. and Bartolo, N. and Liguori, M. and Matarrese, S.",
    title = "{Gravitational waves from inflation}",
    eprint = "1605.01615",
    archivePrefix = "arXiv",
    primaryClass = "astro-ph.CO",
    doi = "10.1393/ncr/i2016-10127-1",
    journal = "Riv. Nuovo Cim.",
    volume = "39",
    number = "9",
    pages = "399--495",
    year = "2016"
}

@article{Caprini:2018mtu,
    author = "Caprini, Chiara and Figueroa, Daniel G.",
    title = "{Cosmological Backgrounds of Gravitational Waves}",
    eprint = "1801.04268",
    archivePrefix = "arXiv",
    primaryClass = "astro-ph.CO",
    doi = "10.1088/1361-6382/aac608",
    journal = "Class. Quant. Grav.",
    volume = "35",
    number = "16",
    pages = "163001",
    year = "2018"
}

@article{CosmoVerse:2025txj,
    author = "Di Valentino, Eleonora and others",
    collaboration = "CosmoVerse",
    title = "{The CosmoVerse White Paper: Addressing observational tensions in cosmology with systematics and fundamental physics}",
    eprint = "2504.01669",
    archivePrefix = "arXiv",
    primaryClass = "astro-ph.CO",
    doi = "10.1016/j.dark.2025.101965",
    month = "4",
    year = "2025"
}

@article{Huang:2023chx,
    author = "Huang, Hai-Long and Cai, Yong and Jiang, Jun-Qian and Zhang, Jun and Piao, Yun-Song",
    title = "{Supermassive Primordial Black Holes for Nano-Hertz Gravitational Waves and High-redshift JWST Galaxies}",
    eprint = "2306.17577",
    archivePrefix = "arXiv",
    primaryClass = "gr-qc",
    doi = "10.1088/1674-4527/ad683d",
    journal = "Res. Astron. Astrophys.",
    volume = "24",
    number = "9",
    pages = "091001",
    year = "2024"
}

@article{Depta:2023uhy,
    author = "Depta, Paul Frederik and Schmidt-Hoberg, Kai and Schwaller, Pedro and Tasillo, Carlo",
    title = "{Signals of merging supermassive black holes in pulsar timing arrays}",
    eprint = "2306.17836",
    archivePrefix = "arXiv",
    primaryClass = "astro-ph.CO",
    reportNumber = "DESY-23-093, MITP-23-036",
    doi = "10.1103/PhysRevResearch.7.013196",
    journal = "Phys. Rev. Res.",
    volume = "7",
    number = "1",
    pages = "013196",
    year = "2025"
}

@article{Gouttenoire:2023nzr,
    author = "Gouttenoire, Yann and Trifinopoulos, Sokratis and Valogiannis, Georgios and Vanvlasselaer, Miguel",
    title = "{Scrutinizing the primordial black hole interpretation of PTA gravitational waves and JWST early galaxies}",
    eprint = "2307.01457",
    archivePrefix = "arXiv",
    primaryClass = "astro-ph.CO",
    doi = "10.1103/PhysRevD.109.123002",
    journal = "Phys. Rev. D",
    volume = "109",
    number = "12",
    pages = "123002",
    year = "2024"
}

@article{Matarrese:1992rp,
    author = "Matarrese, Sabino and Pantano, Ornella and Saez, Diego",
    title = "{A General relativistic approach to the nonlinear evolution of collisionless matter}",
    reportNumber = "DFPD-92-A-39",
    doi = "10.1103/PhysRevD.47.1311",
    journal = "Phys. Rev. D",
    volume = "47",
    pages = "1311--1323",
    year = "1993"
}

@article{Antoniadis:2022cqh,
    author = "Antoniadis, Ignatios and Guillen, Anthony and Tamvakis, Kyriakos",
    title = "{Late time acceleration in Palatini gravity}",
    eprint = "2207.13732",
    archivePrefix = "arXiv",
    primaryClass = "gr-qc",
    doi = "10.1007/JHEP11(2022)144",
    journal = "JHEP",
    volume = "11",
    pages = "144",
    year = "2022"
}

@article{Achucarro:2022qrl,
    author = "Ach\'ucarro, Ana and others",
    title = "{Inflation: Theory and Observations}",
    eprint = "2203.08128",
    archivePrefix = "arXiv",
    primaryClass = "astro-ph.CO",
    month = "3",
    year = "2022"
}

@article{Planck:2018jri,
    author = "Akrami, Y. and others",
    collaboration = "Planck",
    title = "{Planck 2018 results. X. Constraints on inflation}",
    eprint = "1807.06211",
    archivePrefix = "arXiv",
    primaryClass = "astro-ph.CO",
    doi = "10.1051/0004-6361/201833887",
    journal = "Astron. Astrophys.",
    volume = "641",
    pages = "A10",
    year = "2020"
}

@article{Baumann:2018muz,
    author = "Baumann, Daniel",
    title = "{Primordial Cosmology}",
    eprint = "1807.03098",
    archivePrefix = "arXiv",
    primaryClass = "hep-th",
    doi = "10.22323/1.305.0009",
    journal = "PoS",
    volume = "TASI2017",
    pages = "009",
    year = "2018"
}

@article{PhysRevD.83.083521,
  title = {Constraints on the induced gravitational wave background from primordial black holes},
  author = {Bugaev, Edgar and Klimai, Peter},
  journal = {Phys. Rev. D},
  volume = {83},
  issue = {8},
  pages = {083521},
  numpages = {10},
  year = {2011},
  month = {Apr},
  publisher = {American Physical Society},
  doi = {10.1103/PhysRevD.83.083521},
  url = {https://link.aps.org/doi/10.1103/PhysRevD.83.083521}
}

@article{Callan:1970ze,
    author = "Callan, Jr., Curtis G. and Coleman, Sidney R. and Jackiw, Roman",
    title = "{A New improved energy - momentum tensor}",
    doi = "10.1016/0003-4916(70)90394-5",
    journal = "Annals Phys.",
    volume = "59",
    pages = "42--73",
    year = "1970"
}

@article{Allen:1983dg,
    author = "Allen, Bruce",
    title = "{Phase Transitions in de Sitter Space}",
    reportNumber = "Print-83-0020 (CAMBRIDGE)",
    doi = "10.1016/0550-3213(83)90470-4",
    journal = "Nucl. Phys. B",
    volume = "226",
    pages = "228--252",
    year = "1983"
}

@article{Hrycyna:2015vvs,
    author = "Hrycyna, Orest",
    title = "{What $\xi$? Cosmological constraints on the non-minimal coupling constant}",
    eprint = "1511.08736",
    archivePrefix = "arXiv",
    primaryClass = "astro-ph.CO",
    doi = "10.1016/j.physletb.2017.02.062",
    journal = "Phys. Lett. B",
    volume = "768",
    pages = "218--227",
    year = "2017"
}

@article{Joyce:2014kja,
    author = "Joyce, Austin and Jain, Bhuvnesh and Khoury, Justin and Trodden, Mark",
    title = "{Beyond the Cosmological Standard Model}",
    eprint = "1407.0059",
    archivePrefix = "arXiv",
    primaryClass = "astro-ph.CO",
    doi = "10.1016/j.physrep.2014.12.002",
    journal = "Phys. Rept.",
    volume = "568",
    pages = "1--98",
    year = "2015"
}

@article{delaCruz-Dombriz:2008ium,
    author = "de la Cruz-Dombriz, A. and Dobado, A. and Maroto, Antonio Lopez",
    title = "{On the evolution of density perturbations in f(R) theories of gravity}",
    eprint = "0802.2999",
    archivePrefix = "arXiv",
    primaryClass = "astro-ph",
    doi = "10.1103/PhysRevD.77.123515",
    journal = "Phys. Rev. D",
    volume = "77",
    pages = "123515",
    year = "2008"
}

@article{Esposito-Farese:2000pbo,
    author = "Esposito-Farese, Gilles and Polarski, D.",
    title = "{Scalar tensor gravity in an accelerating universe}",
    eprint = "gr-qc/0009034",
    archivePrefix = "arXiv",
    reportNumber = "CPT-00-PE-4053",
    doi = "10.1103/PhysRevD.63.063504",
    journal = "Phys. Rev. D",
    volume = "63",
    pages = "063504",
    year = "2001"
}

@article{Boisseau:2000pr,
    author = "Boisseau, B. and Esposito-Farese, Gilles and Polarski, D. and Starobinsky, Alexei A.",
    title = "{Reconstruction of a scalar tensor theory of gravity in an accelerating universe}",
    eprint = "gr-qc/0001066",
    archivePrefix = "arXiv",
    reportNumber = "CPT-99-P-3917",
    doi = "10.1103/PhysRevLett.85.2236",
    journal = "Phys. Rev. Lett.",
    volume = "85",
    pages = "2236",
    year = "2000"
}

@article{Palatini:1919ffw,
    author = "Palatini, Attilio",
    title = "{Deduzione invariantiva delle equazioni gravitazionali dal principio di Hamilton}",
    doi = "10.1007/BF03014670",
    journal = "Rend. Circ. Mat. Palermo",
    volume = "43",
    number = "1",
    pages = "203--212",
    year = "1919"
}

@article{Ferraris:1982wci,
    author = "Ferraris, M. and Francaviglia, M. and Reina, C.",
    title = "{Variational formulation of general relativity from 1915 to 1925 \textquotedblleft{}Palatini's method\textquotedblright{} discovered by Einstein in 1925}",
    doi = "10.1007/BF00756060",
    journal = "Gen. Rel. Grav.",
    volume = "14",
    number = "3",
    pages = "243--254",
    year = "1982"
}

@article{Bauer:2008zj,
      author         = "Bauer, Florian and Demir, Durmus A.",
      title          = "{Inflation with Non-Minimal Coupling: Metric versus
                        Palatini Formulations}",
      journal        = "Phys. Lett.",
      volume         = "B665",
      year           = "2008",
      pages          = "222-226",
      doi            = "10.1016/j.physletb.2008.06.014",
      eprint         = "0803.2664",
      archivePrefix  = "arXiv",
      primaryClass   = "hep-ph",
      reportNumber   = "DESY-08-033, IZTECH-P-08-02",
      SLACcitation   = "%%CITATION = ARXIV:0803.2664;%%"
}

@article{Bauer:2010bu,
    author = "Bauer, Florian",
    title = "{Filtering out the cosmological constant in the Palatini formalism of modified gravity}",
    eprint = "1007.2546",
    archivePrefix = "arXiv",
    primaryClass = "gr-qc",
    doi = "10.1007/s10714-011-1153-2",
    journal = "Gen. Rel. Grav.",
    volume = "43",
    pages = "1733--1757",
    year = "2011"
}

@article{Olmo:2011uz,
    author = "Olmo, Gonzalo J.",
    title = "{Palatini Approach to Modified Gravity: f(R) Theories and Beyond}",
    eprint = "1101.3864",
    archivePrefix = "arXiv",
    primaryClass = "gr-qc",
    doi = "10.1142/S0218271811018925",
    journal = "Int. J. Mod. Phys. D",
    volume = "20",
    pages = "413--462",
    year = "2011"
}

@article{Rasanen:2017ivk,
    author = "Rasanen, Syksy and Wahlman, Pyry",
    title = "{Higgs inflation with loop corrections in the Palatini formulation}",
    eprint = "1709.07853",
    archivePrefix = "arXiv",
    primaryClass = "astro-ph.CO",
    reportNumber = "HIP-2017-23-TH, KOBE-COSMO-17-12",
    doi = "10.1088/1475-7516/2017/11/047",
    journal = "JCAP",
    volume = "11",
    pages = "047",
    year = "2017"
}

@article{Tenkanen:2017jih,
    author = "Tenkanen, Tommi",
    title = "{Resurrecting Quadratic Inflation with a non-minimal coupling to gravity}",
    eprint = "1710.02758",
    archivePrefix = "arXiv",
    primaryClass = "astro-ph.CO",
    doi = "10.1088/1475-7516/2017/12/001",
    journal = "JCAP",
    volume = "12",
    pages = "001",
    year = "2017"
}

@article{Racioppi:2021jai,
    author = "Racioppi, Antonio and Vasar, Martin",
    title = "{On the number of e-folds in the Jordan and Einstein frames}",
    eprint = "2111.09677",
    archivePrefix = "arXiv",
    primaryClass = "gr-qc",
    doi = "10.1140/epjp/s13360-022-02853-x",
    journal = "Eur. Phys. J. Plus",
    volume = "137",
    number = "5",
    pages = "637",
    year = "2022"
}

@article{Sotiriou:2008rp,
    author = "Sotiriou, Thomas P. and Faraoni, Valerio",
    title = "{f(R) Theories Of Gravity}",
    eprint = "0805.1726",
    archivePrefix = "arXiv",
    primaryClass = "gr-qc",
    doi = "10.1103/RevModPhys.82.451",
    journal = "Rev. Mod. Phys.",
    volume = "82",
    pages = "451--497",
    year = "2010"
}

@article{Koivisto:2005yk,
    author = "Koivisto, Tomi",
    title = "{Covariant conservation of energy momentum in modified gravities}",
    eprint = "gr-qc/0505128",
    archivePrefix = "arXiv",
    doi = "10.1088/0264-9381/23/12/N01",
    journal = "Class. Quant. Grav.",
    volume = "23",
    pages = "4289--4296",
    year = "2006"
}

@article{Starobinsky:1980te,
    author = "Starobinsky, Alexei A.",
    editor = "Khalatnikov, I. M. and Mineev, V. P.",
    title = "{A New Type of Isotropic Cosmological Models Without Singularity}",
    doi = "10.1016/0370-2693(80)90670-X",
    journal = "Phys. Lett. B",
    volume = "91",
    pages = "99--102",
    year = "1980"
}

@article{Guth:1980zm,
    author = "Guth, Alan H.",
    editor = "Fang, Li-Zhi and Ruffini, R.",
    title = "{The Inflationary Universe: A Possible Solution to the Horizon and Flatness Problems}",
    reportNumber = "SLAC-PUB-2576",
    doi = "10.1103/PhysRevD.23.347",
    journal = "Phys. Rev. D",
    volume = "23",
    pages = "347--356",
    year = "1981"
}

@article{Linde:1981mu,
    author = "Linde, Andrei D.",
    editor = "Fang, Li-Zhi and Ruffini, R.",
    title = "{A New Inflationary Universe Scenario: A Possible Solution of the Horizon, Flatness, Homogeneity, Isotropy and Primordial Monopole Problems}",
    reportNumber = "LEBEDEV-81-229",
    doi = "10.1016/0370-2693(82)91219-9",
    journal = "Phys. Lett. B",
    volume = "108",
    pages = "389--393",
    year = "1982"
}

@article{Linde:1983gd,
    author = "Linde, Andrei D.",
    title = "{Chaotic Inflation}",
    doi = "10.1016/0370-2693(83)90837-7",
    journal = "Phys. Lett. B",
    volume = "129",
    pages = "177--181",
    year = "1983"
}

@article{Poisson:2023tja,
    author = "Poisson, Arthur and Timiryasov, Inar and Zell, Sebastian",
    title = "{Critical points in Palatini Higgs inflation with small non-minimal coupling}",
    eprint = "2306.03893",
    archivePrefix = "arXiv",
    primaryClass = "hep-ph",
    doi = "10.1007/JHEP03(2024)130",
    journal = "JHEP",
    volume = "03",
    pages = "130",
    year = "2024"
}

@article{Giovannini:2019mgk,
    author = "Giovannini, Massimo",
    title = "{Post-inflationary phases stiffer than radiation and Palatini formulation}",
    eprint = "1905.06182",
    archivePrefix = "arXiv",
    primaryClass = "gr-qc",
    reportNumber = "CERN-TH-2019-108",
    doi = "10.1088/1361-6382/ab52a8",
    journal = "Class. Quant. Grav.",
    volume = "36",
    number = "23",
    pages = "235017",
    year = "2019"
}

@article{Rubio:2019ypq,
      author         = "Rubio, Javier and Tomberg, Eemeli S.",
      title          = "{Preheating in Palatini Higgs inflation}",
      journal        = "JCAP",
      volume         = "1904",
      year           = "2019",
      number         = "04",
      pages          = "021",
      doi            = "10.1088/1475-7516/2019/04/021",
      eprint         = "1902.10148",
      archivePrefix  = "arXiv",
      primaryClass   = "hep-ph",
      SLACcitation   = "%%CITATION = ARXIV:1902.10148;%%"
}

@article{Gialamas:2020vto,
    author = "Gialamas, Ioannis D. and Karam, Alexandros and Lykkas, Angelos and Pappas, Thomas D.",
    title = "{Palatini-Higgs inflation with nonminimal derivative coupling}",
    eprint = "2008.06371",
    archivePrefix = "arXiv",
    primaryClass = "gr-qc",
    doi = "10.1103/PhysRevD.102.063522",
    journal = "Phys. Rev. D",
    volume = "102",
    number = "6",
    pages = "063522",
    year = "2020"
}

@article{Mikura:2020qhc,
    author = "Mikura, Yusuke and Tada, Yuichiro and Yokoyama, Shuichiro",
    title = "{Conformal inflation in the metric-affine geometry}",
    eprint = "2008.00628",
    archivePrefix = "arXiv",
    primaryClass = "hep-th",
    doi = "10.1209/0295-5075/132/39001",
    journal = "EPL",
    volume = "132",
    number = "3",
    pages = "39001",
    year = "2020"
}

@article{Verner:2020gfa,
    author = "Verner, Sarunas",
    title = "{Quintessential Inflation in Palatini Gravity}",
    eprint = "2010.11201",
    archivePrefix = "arXiv",
    primaryClass = "gr-qc",
    doi = "10.1088/1475-7516/2021/04/001",
    journal = "JCAP",
    volume = "04",
    year = "2021"
}

@article{Enckell:2020lvn,
    author = {Enckell, Vera-Maria and Nurmi, Sami and R{\"a}s{\"a}nen, Syksy and Tomberg, Eemeli},
    title = "{Critical point Higgs inflation in the Palatini formulation}",
    eprint = "2012.03660",
    archivePrefix = "arXiv",
    primaryClass = "astro-ph.CO",
    reportNumber = "HIP-2020-33/TH",
    doi = "10.1007/JHEP04(2021)059",
    journal = "JHEP",
    volume = "04",
    pages = "059",
    year = "2021"
}

@article{Reyimuaji:2020goi,
    author = "Reyimuaji, Yakefu and Zhang, Xinyi",
    title = "{Natural inflation with a nonminimal coupling to gravity}",
    eprint = "2012.14248",
    archivePrefix = "arXiv",
    primaryClass = "astro-ph.CO",
    doi = "10.1088/1475-7516/2021/03/059",
    journal = "JCAP",
    volume = "03",
    pages = "059",
    year = "2021"
}

@article{Karam:2021wzz,
    author = "Karam, Alexandros and Karamitsos, Sotirios and Saal, Margus",
    title = "{\ensuremath{\beta}-function reconstruction of Palatini inflationary attractors}",
    eprint = "2103.01182",
    archivePrefix = "arXiv",
    primaryClass = "gr-qc",
    doi = "10.1088/1475-7516/2021/10/068",
    journal = "JCAP",
    volume = "10",
    pages = "068",
    year = "2021"
}

@article{Mikura:2021ldx,
    author = "Mikura, Yusuke and Tada, Yuichiro and Yokoyama, Shuichiro",
    title = "{Minimal $k$-inflation in light of the conformal metric-affine geometry}",
    eprint = "2103.13045",
    archivePrefix = "arXiv",
    primaryClass = "hep-th",
    doi = "10.1103/PhysRevD.103.L101303",
    journal = "Phys. Rev. D",
    volume = "103",
    number = "10",
    pages = "L101303",
    year = "2021"
}

@article{Gomez:2021roj,
    author = "G{\'o}mez, Diego S{\'a}ez-Chill{\'o}n",
    title = "{3+1 decomposition in modified gravities within the Palatini formalism and some applications}",
    eprint = "2103.16319",
    archivePrefix = "arXiv",
    primaryClass = "gr-qc",
    doi = "10.1103/PhysRevD.104.024029",
    journal = "Phys. Rev. D",
    volume = "104",
    number = "2",
    pages = "024029",
    year = "2021"
}

@article{Bekov:2020dww,
    author = "Bekov, Sabit and Myrzakulov, Kairat and Myrzakulov, Ratbay and G\'omez, Diego S\'aez-Chill\'on",
    title = "{General slow-roll inflation in $f(R)$ gravity under the Palatini approach}",
    eprint = "2010.12360",
    archivePrefix = "arXiv",
    primaryClass = "gr-qc",
    doi = "10.3390/sym12121958",
    journal = "Symmetry",
    volume = "12",
    number = "12",
    pages = "1958",
    year = "2020"
}

@article{Gomez:2020rnq,
    author = "G\'omez, Diego S\'aez-Chill\'on",
    title = "{Variational principle and boundary terms in gravity \`{a} la Palatini}",
    eprint = "2011.11568",
    archivePrefix = "arXiv",
    primaryClass = "gr-qc",
    doi = "10.1016/j.physletb.2021.136103",
    journal = "Phys. Lett. B",
    volume = "814",
    pages = "136103",
    year = "2021"
}

@article{Karam:2021sno,
    author = {Karam, Alexandros and Tomberg, Eemeli and Veerm\"ae, Hardi},
    title = "{Tachyonic preheating in Palatini R 2 inflation}",
    eprint = "2102.02712",
    archivePrefix = "arXiv",
    primaryClass = "astro-ph.CO",
    doi = "10.1088/1475-7516/2021/06/023",
    journal = "JCAP",
    volume = "06",
    pages = "023",
    year = "2021"
}

@article{Annala:2021zdt,
    author = "Annala, Jaakko and Rasanen, Syksy",
    title = "{Inflation with R (\ensuremath{\alpha}\ensuremath{\beta}) terms in the Palatini formulation}",
    eprint = "2106.12422",
    archivePrefix = "arXiv",
    primaryClass = "astro-ph.CO",
    reportNumber = "HIP-2021-22/TH",
    doi = "10.1088/1475-7516/2021/09/032",
    journal = "JCAP",
    volume = "09",
    pages = "032",
    year = "2021"
}

@article{Lykkas:2021vax,
    author = "Lykkas, Angelos and Tamvakis, Kyriakos",
    title = "{Extended interactions in the Palatini-$R^2$ inflation}",
    eprint = "2103.10136",
    archivePrefix = "arXiv",
    primaryClass = "gr-qc",
    doi = "10.1088/1475-7516/2021/08/043",
    journal = "JCAP",
    volume = "08",
    number = "043",
    pages = "",
    year = "2021"
}

@article{Gialamas:2021enw,
    author = "Gialamas, Ioannis D. and Karam, Alexandros and Pappas, Thomas D. and Spanos, Vassilis C.",
    title = "{Scale-invariant quadratic gravity and inflation in the Palatini formalism}",
    eprint = "2104.04550",
    archivePrefix = "arXiv",
    primaryClass = "astro-ph.CO",
    doi = "10.1103/PhysRevD.104.023521",
    journal = "Phys. Rev. D",
    volume = "104",
    number = "2",
    pages = "023521",
    year = "2021"
}

@article{Mikura:2021clt,
    author = "Mikura, Yusuke and Tada, Yuichiro",
    title = "{On UV-completion of Palatini-Higgs inflation}",
    eprint = "2110.03925",
    archivePrefix = "arXiv",
    primaryClass = "hep-ph",
    doi = "10.1088/1475-7516/2022/05/035",
    journal = "JCAP",
    volume = "05",
    number = "05",
    pages = "035",
    year = "2022"
}

@article{AlHallak:2021hwb,
    author = "AlHallak, Mahmoud and AlRakik, Amer and Chamoun, Nidal and El-Daher, Moustafa Sayem",
    title = "{Palatini f(R) Gravity and Variants of k-/Constant Roll/Warm Inflation within Variation of Strong Coupling Scenario}",
    eprint = "2111.05075",
    archivePrefix = "arXiv",
    primaryClass = "astro-ph.CO",
    doi = "10.3390/universe8020126",
    journal = "Universe",
    volume = "8",
    number = "2",
    pages = "126",
    year = "2022"
}

@article{Dioguardi:2021fmr,
    author = "Dioguardi, Christian and Racioppi, Antonio and Tomberg, Eemeli",
    title = "{Slow-roll inflation in Palatini F(R) gravity}",
    eprint = "2112.12149",
    archivePrefix = "arXiv",
    primaryClass = "gr-qc",
    doi = "10.1007/JHEP06(2022)106",
    journal = "JHEP",
    volume = "06",
    pages = "106",
    year = "2022"
}

@article{Tamanini:2010uq,
    author = "Tamanini, Nicola and Contaldi, Carlo R.",
    title = "{Inflationary Perturbations in Palatini Generalised Gravity}",
    eprint = "1010.0689",
    archivePrefix = "arXiv",
    primaryClass = "gr-qc",
    doi = "10.1103/PhysRevD.83.044018",
    journal = "Phys. Rev. D",
    volume = "83",
    pages = "044018",
    year = "2011"
}

@article{Bauer:2010jg,
    author = "Bauer, Florian and Demir, Durmus A.",
    title = "{Higgs-Palatini Inflation and Unitarity}",
    eprint = "1012.2900",
    archivePrefix = "arXiv",
    primaryClass = "hep-ph",
    reportNumber = "IZTECH-P-10-06, ICCUB-10-200",
    doi = "10.1016/j.physletb.2011.03.042",
    journal = "Phys. Lett. B",
    volume = "698",
    pages = "425--429",
    year = "2011"
}

@Article{Green2016,
  author        = {Green, Anne M.},
  title         = {{Microlensing and dynamical constraints on primordial black hole dark matter with an extended mass function}},
  journal       = {Phys. Rev.},
  year          = {2016},
  volume        = {D94},
  number        = {6},
  pages         = {063530},
  archiveprefix = {arXiv},
  doi           = {10.1103/PhysRevD.94.063530},
  eprint        = {1609.01143},
  primaryclass  = {astro-ph.CO},
  slaccitation  = {%%CITATION = ARXIV:1609.01143;%%},
}

@article{Kozak:2018vlp,
    author = "Kozak, Aleksander and Borowiec, Andrzej",
    title = "{Palatini frames in scalar\textendash{}tensor theories of gravity}",
    eprint = "1808.05598",
    archivePrefix = "arXiv",
    primaryClass = "hep-th",
    doi = "10.1140/epjc/s10052-019-6836-y",
    journal = "Eur. Phys. J. C",
    volume = "79",
    number = "4",
    pages = "335",
    year = "2019"
}

@article{Borowiec:2020lfx,
    author = "Borowiec, Andrzej and Kozak, Aleksander",
    title = "{New class of hybrid metric-Palatini scalar-tensor theories of gravity}",
    eprint = "2003.02741",
    archivePrefix = "arXiv",
    primaryClass = "gr-qc",
    doi = "10.1088/1475-7516/2020/07/003",
    journal = "JCAP",
    volume = "07",
    pages = "003",
    year = "2020"
}

@article{Racioppi:2017spw,
    author = "Racioppi, Antonio",
    title = "{Coleman-Weinberg linear inflation: metric vs. Palatini formulation}",
    eprint = "1710.04853",
    archivePrefix = "arXiv",
    primaryClass = "astro-ph.CO",
    doi = "10.1088/1475-7516/2017/12/041",
    journal = "JCAP",
    volume = "12",
    pages = "041",
    year = "2017"
}

@article{Markkanen:2017tun,
    author = {Markkanen, Tommi and Tenkanen, Tommi and Vaskonen, Ville and Veerm\"ae, Hardi},
    title = "{Quantum corrections to quartic inflation with a non-minimal coupling: metric vs. Palatini}",
    eprint = "1712.04874",
    archivePrefix = "arXiv",
    primaryClass = "gr-qc",
    reportNumber = "CERN-TH-2017-267, IMPERIAL-TP-2017-TM-03",
    doi = "10.1088/1475-7516/2018/03/029",
    journal = "JCAP",
    volume = "03",
    pages = "029",
    year = "2018"
}

@article{Jarv:2017azx,
    author = {J\"arv, Laur and Racioppi, Antonio and Tenkanen, Tommi},
    title = "{Palatini side of inflationary attractors}",
    eprint = "1712.08471",
    archivePrefix = "arXiv",
    primaryClass = "gr-qc",
    doi = "10.1103/PhysRevD.97.083513",
    journal = "Phys. Rev. D",
    volume = "97",
    number = "8",
    pages = "083513",
    year = "2018"
}

@article{Fu:2017iqg,
    author = "Fu, Chengjie and Wu, Puxun and Yu, Hongwei",
    title = "{Inflationary dynamics and preheating of the nonminimally coupled inflaton field in the metric and Palatini formalisms}",
    eprint = "1801.04089",
    archivePrefix = "arXiv",
    primaryClass = "gr-qc",
    doi = "10.1103/PhysRevD.96.103542",
    journal = "Phys. Rev. D",
    volume = "96",
    number = "10",
    pages = "103542",
    year = "2017"
}

@article{Bombacigno:2018tyw,
    author = "Bombacigno, Flavio and Montani, Giovanni",
    title = "{Big bounce cosmology for Palatini $R^2$ gravity with a Nieh\textendash{}Yan term}",
    eprint = "1809.07563",
    archivePrefix = "arXiv",
    primaryClass = "gr-qc",
    doi = "10.1140/epjc/s10052-019-6918-x",
    journal = "Eur. Phys. J. C",
    volume = "79",
    number = "5",
    pages = "405",
    year = "2019"
}

@article{Rasanen:2018fom,
    author = "Rasanen, Syksy and Tomberg, Eemeli",
    title = "{Planck scale black hole dark matter from Higgs inflation}",
    eprint = "1810.12608",
    archivePrefix = "arXiv",
    primaryClass = "astro-ph.CO",
    reportNumber = "HIP-2018-23/TH",
    doi = "10.1088/1475-7516/2019/01/038",
    journal = "JCAP",
    volume = "01",
    pages = "038",
    year = "2019"
}

@article{Rasanen:2018ihz,
    author = "Rasanen, Syksy",
    title = "{Higgs inflation in the Palatini formulation with kinetic terms for the metric}",
    eprint = "1811.09514",
    archivePrefix = "arXiv",
    primaryClass = "gr-qc",
    reportNumber = "HIP-2018-27/TH",
    doi = "10.21105/astro.1811.09514",
    journal = "Open J. Astrophys.",
    volume = "2",
    number = "1",
    pages = "1",
    year = "2019"
}

@article{Almeida:2018oid,
    author = "Almeida, Juan P. Beltr\'an and Bernal, Nicol\'as and Rubio, Javier and Tenkanen, Tommi",
    title = "{Hidden Inflaton Dark Matter}",
    eprint = "1811.09640",
    archivePrefix = "arXiv",
    primaryClass = "hep-ph",
    reportNumber = "PI/UAN-2018-641FT, HIP-2018-29/TH",
    doi = "10.1088/1475-7516/2019/03/012",
    journal = "JCAP",
    volume = "03",
    pages = "012",
    year = "2019"
}

@article{Shimada:2018lnm,
    author = "Shimada, Keigo and Aoki, Katsuki and Maeda, Kei-ichi",
    title = "{Metric-affine Gravity and Inflation}",
    eprint = "1812.03420",
    archivePrefix = "arXiv",
    primaryClass = "gr-qc",
    reportNumber = "WU-AP/1808/18",
    doi = "10.1103/PhysRevD.99.104020",
    journal = "Phys. Rev. D",
    volume = "99",
    number = "10",
    pages = "104020",
    year = "2019"
}

@article{Takahashi:2018brt,
    author = "Takahashi, Tomo and Tenkanen, Tommi",
    title = "{Towards distinguishing variants of non-minimal inflation}",
    eprint = "1812.08492",
    archivePrefix = "arXiv",
    primaryClass = "astro-ph.CO",
    doi = "10.1088/1475-7516/2019/04/035",
    journal = "JCAP",
    volume = "04",
    pages = "035",
    year = "2019"
}

@article{Jinno:2018jei,
    author = "Jinno, Ryusuke and Kaneta, Kunio and Oda, Kin-ya and Park, Seong Chan",
    title = "{Hillclimbing inflation in metric and Palatini formulations}",
    eprint = "1812.11077",
    archivePrefix = "arXiv",
    primaryClass = "gr-qc",
    reportNumber = "CTPU-18-44, LDU2018-06, OU-HET-1002, UMN--TH--3913/19,
  FTPI--MINN--19/04",
    doi = "10.1016/j.physletb.2019.03.012",
    journal = "Phys. Lett. B",
    volume = "791",
    pages = "396--402",
    year = "2019"
}

@article{Tenkanen:2019jiq,
    author = "Tenkanen, Tommi",
    title = "{Minimal Higgs inflation with an $R^2$ term in Palatini gravity}",
    eprint = "1901.01794",
    archivePrefix = "arXiv",
    primaryClass = "astro-ph.CO",
    doi = "10.1103/PhysRevD.99.063528",
    journal = "Phys. Rev. D",
    volume = "99",
    number = "6",
    pages = "063528",
    year = "2019"
}

@article{Edery:2019txq,
    author = "Edery, Ariel and Nakayama, Yu",
    title = "{Palatini formulation of pure $R^2$ gravity yields Einstein gravity with no massless scalar}",
    eprint = "1902.07876",
    archivePrefix = "arXiv",
    primaryClass = "hep-th",
    reportNumber = "RUP-19-6",
    doi = "10.1103/PhysRevD.99.124018",
    journal = "Phys. Rev. D",
    volume = "99",
    number = "12",
    pages = "124018",
    year = "2019"
}

@article{Tenkanen:2019xzn,
    author = "Tenkanen, Tommi and Visinelli, Luca",
    title = "{Axion dark matter from Higgs inflation with an intermediate $H_*$}",
    eprint = "1906.11837",
    archivePrefix = "arXiv",
    primaryClass = "astro-ph.CO",
    doi = "10.1088/1475-7516/2019/08/033",
    journal = "JCAP",
    volume = "08",
    pages = "033",
    year = "2019"
}

@article{Tenkanen:2019wsd,
    author = "Tenkanen, Tommi",
    title = "{Trans-Planckian censorship, inflation, and dark matter}",
    eprint = "1910.00521",
    archivePrefix = "arXiv",
    primaryClass = "astro-ph.CO",
    doi = "10.1103/PhysRevD.101.063517",
    journal = "Phys. Rev. D",
    volume = "101",
    number = "6",
    pages = "063517",
    year = "2020"
}

@article{Gialamas:2019nly,
    author = "Gialamas, Ioannis D. and Lahanas, A.B.",
    title = "{Reheating in $R^2$ Palatini inflationary models}",
    eprint = "1911.11513",
    archivePrefix = "arXiv",
    primaryClass = "gr-qc",
    doi = "10.1103/PhysRevD.101.084007",
    journal = "Phys. Rev. D",
    volume = "101",
    number = "8",
    pages = "084007",
    year = "2020"
}

@article{Racioppi:2019jsp,
    author = "Racioppi, Antonio",
    title = "{Non-Minimal (Self-)Running Inflation: Metric vs. Palatini Formulation}",
    eprint = "1912.10038",
    archivePrefix = "arXiv",
    primaryClass = "hep-ph",
    doi = "10.1007/JHEP01(2021)011",
    journal = "JHEP",
    volume = "21",
    pages = "011",
    year = "2020"
}

@article{Antoniadis:2019jnz,
    author = "Antoniadis, Ignatios and Karam, Alexandros and Lykkas, Angelos and Pappas, Thomas and Tamvakis, Kyriakos",
    title = "{Single-field inflation in models with an $R^2$ term}",
    eprint = "1912.12757",
    archivePrefix = "arXiv",
    primaryClass = "gr-qc",
    doi = "10.22323/1.376.0073",
    journal = "PoS",
    volume = "CORFU2019",
    pages = "073",
    year = "2020"
}

@article{Tenkanen:2020dge,
    author = "Tenkanen, Tommi",
    title = "{Tracing the high energy theory of gravity: an introduction to Palatini inflation}",
    eprint = "2001.10135",
    archivePrefix = "arXiv",
    primaryClass = "astro-ph.CO",
    doi = "10.1007/s10714-020-02682-2",
    journal = "Gen. Rel. Grav.",
    volume = "52",
    number = "4",
    pages = "33",
    year = "2020"
}

@article{Tenkanen:2020cvw,
    author = "Tenkanen, Tommi and Tomberg, Eemeli",
    title = "{Initial conditions for plateau inflation: a case study}",
    eprint = "2002.02420",
    archivePrefix = "arXiv",
    primaryClass = "astro-ph.CO",
    reportNumber = "HIP-2020-2/TH",
    doi = "10.1088/1475-7516/2020/04/050",
    journal = "JCAP",
    volume = "04",
    pages = "050",
    year = "2020"
}

@article{Shaposhnikov:2020fdv,
    author = "Shaposhnikov, Mikhail and Shkerin, Andrey and Zell, Sebastian",
    title = "{Quantum Effects in Palatini Higgs Inflation}",
    eprint = "2002.07105",
    archivePrefix = "arXiv",
    primaryClass = "hep-ph",
    doi = "10.1088/1475-7516/2020/07/064",
    journal = "JCAP",
    volume = "07",
    pages = "064",
    year = "2020"
}

@article{Lloyd-Stubbs:2020pvx,
    author = "Lloyd-Stubbs, Amy and McDonald, John",
    title = "{Sub-Planckian $\phi^2$ inflation in the Palatini formulation of gravity with an $R^2$ term}",
    eprint = "2002.08324",
    archivePrefix = "arXiv",
    primaryClass = "hep-ph",
    doi = "10.1103/PhysRevD.101.123515",
    journal = "Phys. Rev. D",
    volume = "101",
    number = "12",
    pages = "123515",
    year = "2020"
}

@article{Antoniadis:2020dfq,
    author = "Antoniadis, Ignatios and Lykkas, Angelos and Tamvakis, Kyriakos",
    title = "{Constant-roll in the Palatini-$R^2$ models}",
    eprint = "2002.12681",
    archivePrefix = "arXiv",
    primaryClass = "gr-qc",
    doi = "10.1088/1475-7516/2020/04/033",
    journal = "JCAP",
    volume = "04",
    number = "04",
    pages = "033",
    year = "2020"
}

@article{Ghilencea:2020piz,
    author = "Ghilencea, D. M.",
    title = "{Palatini quadratic gravity: spontaneous breaking of gauged scale symmetry and inflation}",
    eprint = "2003.08516",
    archivePrefix = "arXiv",
    primaryClass = "hep-th",
    doi = "10.1140/epjc/s10052-020-08722-0",
    journal = "Eur. Phys. J. C",
    volume = "80",
    number = "12",
    pages = "1147",
    month = "4",
    year = "2020"
}

@article{Das:2020kff,
    author = "Das, Nayan and Panda, Sukanta",
    title = "{Inflation and Reheating in f(R,h) theory formulated in the Palatini formalism}",
    eprint = "2005.14054",
    archivePrefix = "arXiv",
    primaryClass = "gr-qc",
    doi = "10.1088/1475-7516/2021/05/019",
    journal = "JCAP",
    volume = "05",
    pages = "019",
    year = "2021"
}

@article{Jarv:2020qqm,
    author = {J\"arv, Laur and Karam, Alexandros and Kozak, Aleksander and Lykkas, Angelos and Racioppi, Antonio and Saal, Margus},
    title = "{Equivalence of inflationary models between the metric and Palatini formulation of scalar-tensor theories}",
    eprint = "2005.14571",
    archivePrefix = "arXiv",
    primaryClass = "gr-qc",
    doi = "10.1103/PhysRevD.102.044029",
    journal = "Phys. Rev. D",
    volume = "102",
    number = "4",
    pages = "044029",
    year = "2020"
}

@article{Gialamas:2020snr,
    author = "Gialamas, Ioannis D. and Karam, Alexandros and Racioppi, Antonio",
    title = "{Dynamically induced Planck scale and inflation in the Palatini formulation}",
    eprint = "2006.09124",
    archivePrefix = "arXiv",
    primaryClass = "gr-qc",
    doi = "10.1088/1475-7516/2020/11/014",
    journal = "JCAP",
    volume = "11",
    pages = "014",
    year = "2020"
}

@article{Karam:2020rpa,
    author = "Karam, Alexandros and Raidal, Martti and Tomberg, Eemeli",
    title = "{Gravitational dark matter production in Palatini preheating}",
    eprint = "2007.03484",
    archivePrefix = "arXiv",
    primaryClass = "astro-ph.CO",
    doi = "10.1088/1475-7516/2021/03/064",
    journal = "JCAP",
    volume = "03",
    pages = "064",
    year = "2021"
}

@article{McDonald:2020lpz,
    author = "McDonald, J.",
    title = "{Does Palatini Higgs Inflation Conserve Unitarity?}",
    eprint = "2007.04111",
    archivePrefix = "arXiv",
    primaryClass = "hep-ph",
    doi = "10.1088/1475-7516/2021/04/069",
    journal = "JCAP",
    volume = "04",
    pages = "069",
    year = "2021"
}

@article{Langvik:2020nrs,
    author = {L{\r{a}}ngvik, Miklos and Ojanper{\"a}, Juha-Matti and Raatikainen, Sami and Rasanen, Syksy},
    title = "{Higgs inflation with the Holst and the Nieh{\textendash}Yan term}",
    eprint = "2007.12595",
    archivePrefix = "arXiv",
    primaryClass = "astro-ph.CO",
    reportNumber = "HIP-2020-22/TH",
    doi = "10.1103/PhysRevD.103.083514",
    journal = "Phys. Rev. D",
    volume = "103",
    number = "8",
    pages = "083514",
    year = "2021"
}

@article{Ghilencea:2020rxc,
    author = "Ghilencea, D. M.",
    title = "{Gauging scale symmetry and inflation: Weyl versus Palatini gravity}",
    eprint = "2007.14733",
    archivePrefix = "arXiv",
    primaryClass = "hep-th",
    doi = "10.1140/epjc/s10052-021-09226-1",
    journal = "Eur. Phys. J. C",
    volume = "81",
    number = "6",
    pages = "510",
    year = "2021"
}

@article{Shaposhnikov:2020gts,
    author = "Shaposhnikov, Mikhail and Shkerin, Andrey and Timiryasov, Inar and Zell, Sebastian",
    title = "{Higgs inflation in Einstein-Cartan gravity}",
    eprint = "2007.14978",
    archivePrefix = "arXiv",
    primaryClass = "hep-ph",
    doi = "10.1088/1475-7516/2021/10/E01",
    journal = "JCAP",
    volume = "02",
    pages = "008",
    year = "2021",
    note = "[Erratum: JCAP 10, E01 (2021)]"
}

@article{Shaposhnikov:2020frq,
    author = "Shaposhnikov, Mikhail and Shkerin, Andrey and Timiryasov, Inar and Zell, Sebastian",
    title = "{Einstein-Cartan gravity, matter, and scale-invariant generalization}",
    eprint = "2007.16158",
    archivePrefix = "arXiv",
    primaryClass = "hep-th",
    doi = "10.1007/JHEP10(2020)177",
    journal = "JHEP",
    volume = "10",
    pages = "177",
    year = "2020"
}

@article{Bostan:2019wsd,
    author = "Bostan, Nilay",
    title = "{Quadratic, Higgs and hilltop potentials in the Palatini gravity}",
    eprint = "1908.09674",
    archivePrefix = "arXiv",
    primaryClass = "astro-ph.CO",
    doi = "10.1088/1572-9494/ab7ecb",
    journal = "Commun. Theor. Phys.",
    volume = "72",
    pages = "085401",
    year = "2020"
}

@article{Bostan:2019uvv,
    author = "Bostan, Nilay",
    title = "{Non-minimally coupled quartic inflation with Coleman-Weinberg one-loop corrections in the Palatini formulation}",
    eprint = "1907.13235",
    archivePrefix = "arXiv",
    primaryClass = "gr-qc",
    doi = "10.1016/j.physletb.2020.135954",
    journal = "Phys. Lett. B",
    volume = "811",
    pages = "135954",
    year = "2020"
}

@article{Racioppi:2018zoy,
    author = "Racioppi, Antonio",
    title = "{New universal attractor in nonminimally coupled gravity: Linear inflation}",
    eprint = "1801.08810",
    archivePrefix = "arXiv",
    primaryClass = "astro-ph.CO",
    doi = "10.1103/PhysRevD.97.123514",
    journal = "Phys. Rev. D",
    volume = "97",
    number = "12",
    pages = "123514",
    year = "2018"
}

@article{Carrilho:2018ffi,
    author = "Carrilho, Pedro and Mulryne, David and Ronayne, John and Tenkanen, Tommi",
    title = "{Attractor Behaviour in Multifield Inflation}",
    eprint = "1804.10489",
    archivePrefix = "arXiv",
    primaryClass = "astro-ph.CO",
    doi = "10.1088/1475-7516/2018/06/032",
    journal = "JCAP",
    volume = "06",
    pages = "032",
    year = "2018"
}

@article{Copeland:1997et,
    author = "Copeland, Edmund J. and Liddle, Andrew R and Wands, David",
    title = "{Exponential potentials and cosmological scaling solutions}",
    eprint = "gr-qc/9711068",
    archivePrefix = "arXiv",
    reportNumber = "SUSX-TH-97-022, SUSSEX-AST-97-11-1, PU-RCG-97-20",
    doi = "10.1103/PhysRevD.57.4686",
    journal = "Phys. Rev. D",
    volume = "57",
    pages = "4686--4690",
    year = "1998"
}

@article{Bari:2023rcw,
    author = "Bari, Pritha and Bartolo, Nicola and Dom\`enech, Guillem and Matarrese, Sabino",
    title = "{Gravitational waves induced by scalar-tensor mixing}",
    eprint = "2307.05404",
    archivePrefix = "arXiv",
    primaryClass = "astro-ph.CO",
    doi = "10.1103/PhysRevD.109.023509",
    journal = "Phys. Rev. D",
    volume = "109",
    number = "2",
    pages = "023509",
    year = "2024"
}

@article{Wu:2024qdb,
    author = "Wu, Di and Zhou, Jing-Zhi and Kuang, Yu-Ting and Li, Zhi-Chao and Chang, Zhe and Huang, Qing-Guo",
    title = "{Can tensor-scalar induced GWs dominate PTA observations?}",
    eprint = "2501.00228",
    archivePrefix = "arXiv",
    primaryClass = "astro-ph.CO",
    doi = "10.1088/1475-7516/2025/03/045",
    journal = "JCAP",
    volume = "03",
    pages = "045",
    year = "2025"
}

@article{Perna:2024ehx,
    author = "Perna, Gabriele and Testini, Chiara and Ricciardone, Angelo and Matarrese, Sabino",
    title = "{Fully non-Gaussian Scalar-Induced Gravitational Waves}",
    eprint = "2403.06962",
    archivePrefix = "arXiv",
    primaryClass = "astro-ph.CO",
    doi = "10.1088/1475-7516/2024/05/086",
    journal = "JCAP",
    volume = "05",
    pages = "086",
    year = "2024"
}

@article{LISACosmologyWorkingGroup:2025vdz,
    author = "Gammal, Jonas El and others",
    collaboration = "LISA Cosmology Working Group",
    title = "{Reconstructing primordial curvature perturbations via scalar-induced gravitational waves with LISA}",
    eprint = "2501.11320",
    archivePrefix = "arXiv",
    primaryClass = "astro-ph.CO",
    reportNumber = "CERN-TH-2024-217",
    doi = "10.1088/1475-7516/2025/05/062",
    journal = "JCAP",
    volume = "05",
    pages = "062",
    year = "2025"
}

@article{Papanikolaou:2021uhe,
    author = "Papanikolaou, Theodoros and Tzerefos, Charalampos and Basilakos, Spyros and Saridakis, Emmanuel N.",
    title = "{Scalar induced gravitational waves from primordial black hole Poisson fluctuations in f(R) gravity}",
    eprint = "2112.15059",
    archivePrefix = "arXiv",
    primaryClass = "astro-ph.CO",
    doi = "10.1088/1475-7516/2022/10/013",
    journal = "JCAP",
    volume = "10",
    pages = "013",
    year = "2022"
}

@article{Cai:2018dig,
    author = "Cai, Rong-gen and Pi, Shi and Sasaki, Misao",
    title = "{Gravitational Waves Induced by non-Gaussian Scalar Perturbations}",
    eprint = "1810.11000",
    archivePrefix = "arXiv",
    primaryClass = "astro-ph.CO",
    reportNumber = "IPMU18-0172, YITP-18-114",
    doi = "10.1103/PhysRevLett.122.201101",
    journal = "Phys. Rev. Lett.",
    volume = "122",
    number = "20",
    pages = "201101",
    year = "2019"
}

@article{Kuralkar:2025hoz,
    author = "Kuralkar, Hardik Jitendra and Panda, Sukanta and Vidyarthi, Archit",
    title = "{Observable primordial gravitational waves from non-minimally coupled R $^{2}$ Palatini modified gravity}",
    eprint = "2502.03573",
    archivePrefix = "arXiv",
    primaryClass = "astro-ph.CO",
    doi = "10.1088/1475-7516/2025/05/073",
    journal = "JCAP",
    volume = "05",
    pages = "073",
    year = "2025"
}

@article{Bruton:2025dyr,
    author = "Bruton, Jenna and Dunsby, Peter and de la Cruz-Dombriz, Alvaro",
    title = "{Close Hyperbolic Encounters In f(R) Gravity}",
    eprint = "2506.20787",
    archivePrefix = "arXiv",
    primaryClass = "gr-qc",
    month = "6",
    year = "2025"
}

@article{Weinberg:2003ur,
    author = "Weinberg, Steven",
    title = "{Damping of tensor modes in cosmology}",
    eprint = "astro-ph/0306304",
    archivePrefix = "arXiv",
    reportNumber = "UTTG-02-03",
    doi = "10.1103/PhysRevD.69.023503",
    journal = "Phys. Rev. D",
    volume = "69",
    pages = "023503",
    year = "2004"
}

@article{Bond:1983hb,
    author = "Bond, J. R. and Szalay, A. S.",
    title = "{The Collisionless Damping of Density Fluctuations in an Expanding Universe}",
    doi = "10.1086/161460",
    journal = "Astrophys. J.",
    volume = "274",
    pages = "443--468",
    year = "1983"
}

@article{Mangilli:2008bw,
    author = "Mangilli, Anna and Bartolo, Nicola and Matarrese, Sabino and Riotto, Antonio",
    title = "{The impact of cosmic neutrinos on the gravitational-wave background}",
    eprint = "0805.3234",
    archivePrefix = "arXiv",
    primaryClass = "astro-ph",
    doi = "10.1103/PhysRevD.78.083517",
    journal = "Phys. Rev. D",
    volume = "78",
    pages = "083517",
    year = "2008"
}

@article{Saga:2014jca,
    author = "Saga, Shohei and Ichiki, Kiyotomo and Sugiyama, Naoshi",
    title = "{Impact of anisotropic stress of free-streaming particles on gravitational waves induced by cosmological density perturbations}",
    eprint = "1412.1081",
    archivePrefix = "arXiv",
    primaryClass = "astro-ph.CO",
    doi = "10.1103/PhysRevD.91.024030",
    journal = "Phys. Rev. D",
    volume = "91",
    number = "2",
    pages = "024030",
    year = "2015"
}

@article{Zhang:2022dgx,
    author = "Zhang, Xukun and Zhou, Jing-Zhi and Chang, Zhe",
    title = "{Impact of the free-streaming neutrinos to the second order induced gravitational waves}",
    eprint = "2208.12948",
    archivePrefix = "arXiv",
    primaryClass = "astro-ph.CO",
    doi = "10.1140/epjc/s10052-022-10742-x",
    journal = "Eur. Phys. J. C",
    volume = "82",
    number = "9",
    pages = "781",
    year = "2022"
}

@article{Bartolo:2010qu,
    author = "Bartolo, N. and Matarrese, S. and Riotto, A.",
    title = "{Non-Gaussianity and the Cosmic Microwave Background Anisotropies}",
    eprint = "1001.3957",
    archivePrefix = "arXiv",
    primaryClass = "astro-ph.CO",
    reportNumber = "CERN-PH-TH-2010-006",
    doi = "10.1155/2010/157079",
    journal = "Adv. Astron.",
    volume = "2010",
    pages = "157079",
    year = "2010"
}

@article{Domenech:2025bvr,
    author = "Dom{\`e}nech, Guillem and Chluba, Jens",
    title = "{Regularizing the induced GW spectrum with dissipative effects}",
    eprint = "2503.13670",
    archivePrefix = "arXiv",
    primaryClass = "gr-qc",
    doi = "10.1088/1475-7516/2025/07/034",
    journal = "JCAP",
    volume = "07",
    pages = "034",
    year = "2025"
}

@article{Bardeen:1980kt,
    author = "Bardeen, James M.",
    title = "{Gauge Invariant Cosmological Perturbations}",
    doi = "10.1103/PhysRevD.22.1882",
    journal = "Phys. Rev. D",
    volume = "22",
    pages = "1882--1905",
    year = "1980"
}

@inproceedings{Baumann:2009ds,
    author = "Baumann, Daniel",
    title = "{Inflation}",
    booktitle = "{Theoretical Advanced Study Institute in Elementary Particle Physics}: {Physics of the Large and the Small}",
    eprint = "0907.5424",
    archivePrefix = "arXiv",
    primaryClass = "hep-th",
    reportNumber = "TASI-2009",
    doi = "10.1142/9789814327183_0010",
    pages = "523--686",
    year = "2011"
}

@article{Sato:1980yn,
    author = "Sato, K.",
    title = "{First Order Phase Transition of a Vacuum and Expansion of the Universe}",
    reportNumber = "NORDITA-80-29",
    journal = "Mon. Not. Roy. Astron. Soc.",
    volume = "195",
    pages = "467--479",
    year = "1981"
}

@article{Sasaki:2018dmp,
    author = "Sasaki, Misao and Suyama, Teruaki and Tanaka, Takahiro and Yokoyama, Shuichiro",
    title = "{Primordial black holes{\textemdash}perspectives in gravitational wave astronomy}",
    eprint = "1801.05235",
    archivePrefix = "arXiv",
    primaryClass = "astro-ph.CO",
    doi = "10.1088/1361-6382/aaa7b4",
    journal = "Class. Quant. Grav.",
    volume = "35",
    number = "6",
    pages = "063001",
    year = "2018"
}

@article{Carr:2020gox,
    author = "Carr, Bernard and Kohri, Kazunori and Sendouda, Yuuiti and Yokoyama, Jun'ichi",
    title = "{Constraints on primordial black holes}",
    eprint = "2002.12778",
    archivePrefix = "arXiv",
    primaryClass = "astro-ph.CO",
    reportNumber = "RESCEU-03/20; KEK-Cosmo-249; KEK-TH-2199; IPMU20-0024",
    doi = "10.1088/1361-6633/ac1e31",
    journal = "Rept. Prog. Phys.",
    volume = "84",
    number = "11",
    pages = "116902",
    year = "2021"
}

@article{Yuan:2021qgz,
    author = "Yuan, Chen and Huang, Qing-Guo",
    title = "{A topic review on probing primordial black hole dark matter with scalar induced gravitational waves}",
    eprint = "2103.04739",
    archivePrefix = "arXiv",
    primaryClass = "astro-ph.GA",
    doi = "10.1016/j.isci.2021.102860",
    journal = "iScience",
    volume = "24",
    pages = "102860",
    year = "2021"
}

@article{Byrnes:2018txb,
    author = "Byrnes, Christian T. and Cole, Philippa S. and Patil, Subodh P.",
    title = "{Steepest growth of the power spectrum and primordial black holes}",
    eprint = "1811.11158",
    archivePrefix = "arXiv",
    primaryClass = "astro-ph.CO",
    doi = "10.1088/1475-7516/2019/06/028",
    journal = "JCAP",
    volume = "06",
    pages = "028",
    year = "2019"
}

@article{Carr:2025kdk,
    author = "Carr, Bernard and Kuhnel, Florian",
    title = "{Primordial Black Holes}",
    eprint = "2502.15279",
    archivePrefix = "arXiv",
    primaryClass = "astro-ph.CO",
    month = "2",
    year = "2025"
}

@article{Ozsoy:2019lyy,
    author = {{\"O}zsoy, Ogan and Tasinato, Gianmassimo},
    title = "{On the slope of the curvature power spectrum in non-attractor inflation}",
    eprint = "1912.01061",
    archivePrefix = "arXiv",
    primaryClass = "astro-ph.CO",
    doi = "10.1088/1475-7516/2020/04/048",
    journal = "JCAP",
    volume = "04",
    pages = "048",
    year = "2020"
}

@article{Carrilho:2019oqg,
    author = "Carrilho, Pedro and Malik, Karim A. and Mulryne, David J.",
    title = "{Dissecting the growth of the power spectrum for primordial black holes}",
    eprint = "1907.05237",
    archivePrefix = "arXiv",
    primaryClass = "astro-ph.CO",
    doi = "10.1103/PhysRevD.100.103529",
    journal = "Phys. Rev. D",
    volume = "100",
    number = "10",
    pages = "103529",
    year = "2019"
}

@article{Atal:2018neu,
    author = "Atal, Vicente and Germani, Cristiano",
    title = "{The role of non-gaussianities in Primordial Black Hole formation}",
    eprint = "1811.07857",
    archivePrefix = "arXiv",
    primaryClass = "astro-ph.CO",
    reportNumber = "ICCUB-18-022",
    doi = "10.1016/j.dark.2019.100275",
    journal = "Phys. Dark Univ.",
    volume = "24",
    pages = "100275",
    year = "2019"
}

@article{Liu:2020oqe,
    author = "Liu, Jing and Guo, Zong-Kuan and Cai, Rong-Gen",
    title = "{Analytical approximation of the scalar spectrum in the ultraslow-roll inflationary models}",
    eprint = "2003.02075",
    archivePrefix = "arXiv",
    primaryClass = "astro-ph.CO",
    doi = "10.1103/PhysRevD.101.083535",
    journal = "Phys. Rev. D",
    volume = "101",
    number = "8",
    pages = "083535",
    year = "2020"
}

@article{Atal:2021jyo,
    author = "Atal, Vicente and Dom{\`e}nech, Guillem",
    title = "{Probing non-Gaussianities with the high frequency tail of induced gravitational waves}",
    eprint = "2103.01056",
    archivePrefix = "arXiv",
    primaryClass = "astro-ph.CO",
    doi = "10.1088/1475-7516/2021/06/001",
    journal = "JCAP",
    volume = "06",
    pages = "001",
    year = "2021",
    note = "[Erratum: JCAP 10, E01 (2023)]"
}

@article{Xu:2019bdp,
    author = "Xu, Wu-Tao and Liu, Jing and Gao, Tie-Jun and Guo, Zong-Kuan",
    title = "{Gravitational waves from double-inflection-point inflation}",
    eprint = "1907.05213",
    archivePrefix = "arXiv",
    primaryClass = "astro-ph.CO",
    doi = "10.1103/PhysRevD.101.023505",
    journal = "Phys. Rev. D",
    volume = "101",
    number = "2",
    pages = "023505",
    year = "2020"
}

@article{Clesse:2018ogk,
    author = "Clesse, Sebastien and Garc{\'\i}a-Bellido, Juan and Orani, Stefano",
    title = "{Detecting the Stochastic Gravitational Wave Background from Primordial Black Hole Formation}",
    eprint = "1812.11011",
    archivePrefix = "arXiv",
    primaryClass = "astro-ph.CO",
    reportNumber = "IFT--UAM/CSIC--17--119, CERN--TH--2017--259, CERN-TH-2017-259",
    month = "12",
    year = "2018"
}

@article{Riccardi:2021rlf,
    author = "Riccardi, Flavio and Taoso, Marco and Urbano, Alfredo",
    title = "{Solving peak theory in the presence of local non-gaussianities}",
    eprint = "2102.04084",
    archivePrefix = "arXiv",
    primaryClass = "astro-ph.CO",
    doi = "10.1088/1475-7516/2021/08/060",
    journal = "JCAP",
    volume = "08",
    pages = "060",
    year = "2021"
}

@article{Thrane:2013oya,
    author = "Thrane, Eric and Romano, Joseph D.",
    title = "{Sensitivity curves for searches for gravitational-wave backgrounds}",
    eprint = "1310.5300",
    archivePrefix = "arXiv",
    primaryClass = "astro-ph.IM",
    doi = "10.1103/PhysRevD.88.124032",
    journal = "Phys. Rev. D",
    volume = "88",
    number = "12",
    pages = "124032",
    year = "2013"
}

@article{Dicus:2005rh,
    author = "Dicus, Duane A. and Repko, Wayne W.",
    title = "{Comment on damping of tensor modes in cosmology}",
    eprint = "astro-ph/0509096",
    archivePrefix = "arXiv",
    doi = "10.1103/PhysRevD.72.088302",
    journal = "Phys. Rev. D",
    volume = "72",
    pages = "088302",
    year = "2005"
}

@article{Watanabe:2006qe,
    author = "Watanabe, Yuki and Komatsu, Eiichiro",
    title = "{Improved Calculation of the Primordial Gravitational Wave Spectrum in the Standard Model}",
    eprint = "astro-ph/0604176",
    archivePrefix = "arXiv",
    doi = "10.1103/PhysRevD.73.123515",
    journal = "Phys. Rev. D",
    volume = "73",
    pages = "123515",
    year = "2006"
}

@article{Pritchard:2004qp,
    author = "Pritchard, Jonathan R. and Kamionkowski, Marc",
    title = "{Cosmic microwave background fluctuations from gravitational waves: An Analytic approach}",
    eprint = "astro-ph/0412581",
    archivePrefix = "arXiv",
    doi = "10.1016/j.aop.2005.03.005",
    journal = "Annals Phys.",
    volume = "318",
    pages = "2--36",
    year = "2005"
}

@article{Saga:2018ont,
    author = "Saga, Shohei and Tashiro, Hiroyuki and Yokoyama, Shuichiro",
    title = "{Limits on primordial magnetic fields from direct detection experiments of gravitational wave background}",
    eprint = "1807.00561",
    archivePrefix = "arXiv",
    primaryClass = "astro-ph.CO",
    reportNumber = "YITP-18-72, RUP-18-20",
    doi = "10.1103/PhysRevD.98.083518",
    journal = "Phys. Rev. D",
    volume = "98",
    number = "8",
    pages = "083518",
    year = "2018"
}

@article{Yang:2024ntt,
    author = "Yang, Wei and Kang, Yu-Xuan and Ali, Arshad and Sui, Tao-Tao and Wu, Chen-Hao and Hu, Ya-Peng",
    title = "{Primordial black holes and secondary gravitational waves from the inflation potential with a tiny Lorentz function bump}",
    eprint = "2403.15979",
    archivePrefix = "arXiv",
    primaryClass = "astro-ph.CO",
    month = "3",
    year = "2024"
}
\end{document}